\documentclass[aps, pra, 10pt, twocolumn, tightenlines, letterpaper, amsmath, amssymb, preprintnumbers, floatfix, longbibliography, nofootinbib]{revtex4-2}

\usepackage{dsfont}
\usepackage{amsmath}
\usepackage{amssymb}
\usepackage{physics}
\usepackage{tabu}
\usepackage{tabularx}
\usepackage{booktabs}
\usepackage{bm}
\usepackage{tikz}
\usetikzlibrary{shapes.geometric}
\usepackage{graphicx}
\usepackage{placeins}
\usepackage{textgreek}
\usepackage{multirow}
\usepackage{makecell}
\usepackage{soul}
\usepackage{diagbox}
\usepackage{xcolor}
\usepackage[normalem]{ulem}
\usepackage[export]{adjustbox}
\usepackage{float}
\usepackage{wasysym}
\usepackage{fancyvrb}
\usepackage{listings}

\definecolor{darkgreen}{rgb}{0.0, 0.5, 0.0}   

\usepackage[hypertexnames=false]{hyperref}
\hypersetup{
    colorlinks=true,       
    linkcolor=blue,          
    citecolor=blue,        
    filecolor=blue,      
    urlcolor=blue           
}
\usepackage{footnotehyper}

\definecolor{emerald}{HTML}{3EB064}

\newcommand{\Tshot}{T_{\mathrm{shot}}}
\newcommand{\Tacc}{T_{\mathrm{acc}}}
\newcommand{\Gtot}{G_{\mathrm{tot}}}
\newcommand{\Gstep}{G_{\mathrm{step}}}
\newcommand{\Gfixed}{G_{\mathrm{fixed}}}

\newcolumntype{Y}{>{\centering\arraybackslash}X}

\usepackage{orcidlink}

\newlength{\emailcol}\newlength{\emailtmp}
\newcommand{\emailwidest}[1]{%
  \settowidth{\emailtmp}{#1}%
  \ifdim\emailtmp>\emailcol \setlength{\emailcol}{\emailtmp}\fi}

\newcommand{\emailtable}{%
  \begingroup
  \emailwidest{\textcolor{blue}{$^{*}$\,frolandh@uw.edu}}%
  \emailwidest{\textcolor{blue}{$^{\dagger}$\,grabow@uw.edu}}%
  \emailwidest{\textcolor{blue}{$^{\ddagger}$\,segrie@uw.edu}}%
  \emailwidest{\textcolor{blue}{$^{\S}$\,jhartse@uw.edu}}%
  \emailwidest{\textcolor{blue}{$^{\P}$\,alash@uw.edu}}%
  \emailwidest{\textcolor{blue}{$^{**}$\,zhiyaol@uw.edu}}%
  \emailwidest{\textcolor{blue}{$^{\dagger\dagger}$\,ziyuanli@uw.edu}}%
  \emailwidest{\textcolor{blue}{$^{\ddagger\ddagger}$\,spow9@uw.edu}}%
  \emailwidest{\textcolor{blue}{$^{\S\S}$\,mjs5@uw.edu}}%
  \emailwidest{\textcolor{blue}{$^{\P\P}$\,xjyao@uw.edu}}%
  \emailwidest{\textcolor{blue}{$^{***}$\,zemlni@uw.edu}}%

 \begin{tabular}[t]{@{}l l@{}}
    \textcolor{blue}{$^{*}$\,frolandh@uw.edu}      & \textcolor{blue}{$^{\dagger}$\,grabow@uw.edu}        \\
    \textcolor{blue}{$^{\ddagger}$\,segrie@uw.edu}\,\textsuperscript{(c)} & \textcolor{blue}{$^{\S}$\,jhartse@uw.edu} \\
    \textcolor{blue}{$^{\P}$\,alash@uw.edu}        & \textcolor{blue}{$^{**}$\,zhiyaol@uw.edu}            \\
    \textcolor{blue}{$^{\dagger\dagger}$\,ziyuanli@uw.edu} & \textcolor{blue}{$^{\ddagger\ddagger}$\,spow9@uw.edu} \\
    \textcolor{blue}{$^{\S\S}$\,mjs5@uw.edu}\,\textsuperscript{(c)} & \textcolor{blue}{$^{\P\P}$\,xjyao@uw.edu} \\
    \textcolor{blue}{$^{***}$\,zemlni@uw.edu}\, & \\
  \end{tabular}\par
  \smallskip
  {\footnotesize \textsuperscript{(c)} Corresponding author}\par
  \endgroup}

\allowdisplaybreaks

\makeatletter
\newcommand\blfootnote[1]{
  \begingroup
  \renewcommand\thefootnote{}
  \renewcommand\@makefntext[1]{\noindent##1}
  \footnotetext{#1}
  \endgroup
  \addtocounter{footnote}{-1}
}
\makeatother

\begin{document}

\begin{figure}
  \vskip -1.cm
  \leftline{\includegraphics[width=0.15\textwidth]{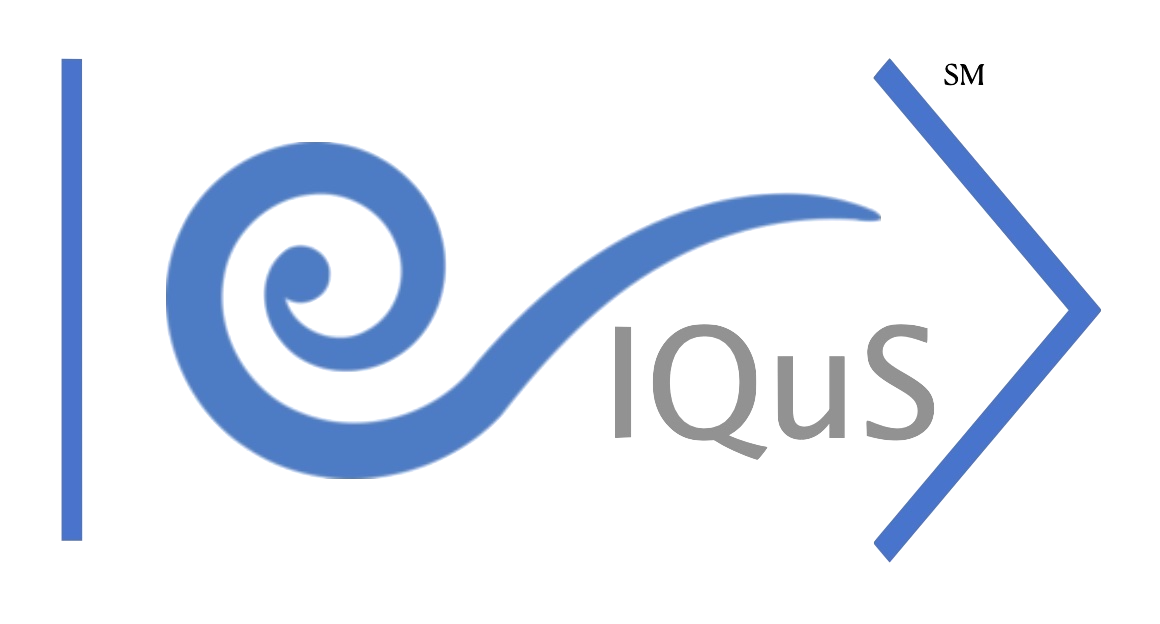}}
  \vskip -1.cm
\end{figure}

\title{The Utility of Sparse Error Detection in Quantum Simulations}

\author{Henry Froland$^{\ast}$\,\orcidlink{0009-0008-4356-0602}}
\author{Dorota M. Grabowska$^{\dagger}$\,\orcidlink{0000-0002-0760-4734}}
\author{Sebastian Grieninger$^{\ddagger}$\,\orcidlink{0000-0002-9523-5819}}
\author{Jeremy Hartse$^{\S}$\,\orcidlink{0009-0004-1943-421X}}
\author{Anne L. Lashbrook$^{\P}$\,\orcidlink{0009-0002-9120-7738}}
\author{Zhiyao Li$^{\ast\ast}$\,\orcidlink{0000-0002-7614-8496}}
\author{Ziyuan Li$^{\dagger\dagger}$\,\orcidlink{0009-0005-2034-1499}}
\author{Sarah J.~M.~Powell$^{\ddagger\ddagger}$\,\orcidlink{0000-0002-5228-8291}}
\author{Martin J.~Savage$^{\S\S}$\,\orcidlink{0000-0001-6502-7106}}
\author{Xiaojun Yao$^{\P\P}$\,\orcidlink{0000-0002-8377-2203}}
\author{Nikita A. Zemlevskiy$^{\ast\ast\ast}$\,\orcidlink{0000-0002-0794-2389}}

\preprint{IQuS@UW-21-132, NTG@UW-26-19}
\date{\today}

\begin{abstract}
\noindent
The recent success of error detecting codes points toward their potential application to fault-tolerant simulations of nature. In this work, we examine the utility of sparse error detection for simulating lattice gauge theories using quantum computers. In particular, we study the time evolution of the lattice Schwinger model embedded into the Iceberg code family, $[[N+2, N, 2]]$, as well as the Hypercube code family, $[[2^N, N, 2]]$. The lattice of electrons and positrons in the axial gauge is embedded into a single code block or into multiple code blocks, and this work finds that large codeblocks are advantageous in the absence of connectivity constraints. Noisy classical simulations with realistic near-term error rates, infrequent syndrome measurements and physics-aware postselection are found to improve observable estimation. Under realistic noise rates for near-term quantum computers,  this work finds that sparse error detection in quantum simulations has the potential to improve accuracy of observable estimation. Additional rounds of error detection are found to systematically drive errors in observables to the noise floor set by the code. These findings suggest that incorporating minimal implementations of fault tolerance in the near-term will enhance the performance of quantum simulations in nuclear physics and high-energy physics.
\end{abstract}
\maketitle
\blfootnote{\emailtable}

\tableofcontents

\section{Introduction}
\label{sec:intro}
\noindent
Recently, there have been several demonstrations of how logical qubits can be
embedded onto quantum computers. This includes 
IBM using superconducting qubits with LDPC codes~\cite{Bravyi_2024},
Google Quantum AI, ETH Z\"urich, and USTC using superconducting qubits with surface
codes~\cite{GoogleQuantumAI2025,Krinner2022,Zhao2022},
Quantinuum and the University of Innsbruck using trapped
ions~\cite{reichardt2024MicroQuant,QuantMicro,MicrosoftQuantinuum2026,Dasu:2026dwm,Postler2022},
the Harvard--MIT--QuEra collaboration using neutral
atoms~\cite{bluvstein2024logical,Rodriguez2025,Bluvstein2026},
the Atom Computing--Microsoft collaboration using cold atoms~\cite{reichardt2025MicroAtom},
and Yale, Amazon (AWS), and Alice \&\ Bob using bosonic
codes~\cite{Sivak2023,Putterman2025Bosonic,Reglade2024}.
These demonstrations, carried out on a variety of qubit architectures and encodings, 
indicate that quantum computers with some degree of fault tolerance (FT)
and quantum error correction (QEC) ~\cite{shor1996fault,Preskill_1998,gottesman1997stabilizer,gottesman1998theory,gottesman2000local} 
will become available in the near future. 
Before universal error-corrected simulations are achieved, heterogeneous simulations, where logical qubits are protected to differing degrees by error detection or error correction, 
can be performed to potentially extend the scientific reach of quantum computers.
For fixed-resource simulations of observables with a target precision, including both systematic errors and statistical errors, an optimization of how resources are distributed within the simulation and workflow needs to be performed.
This includes optimizing with respect to the distribution of fidelities, and optimizing the average fidelity of logical qubits versus the number of logical qubits supporting the physical system.
Further, optimizations in the distribution of error detecting 
and error correcting codes utilized in a given simulation are also necessary. Compared to conventional QEC, lightweight error-detection codes offer a reduction in the number of required qubits and gates, as well as device coherence time, required to implement a FT quantum simulation.
Together with extrapolations in simulation parameters to connect to observables, this reduction in resources will enable simulations of physical systems in the near term.

In fundamental physics, important quantum many-body systems require simulations that lie beyond the capabilities of classical computing alone, and near-term simulations~\cite{Bauer:2022hpo,Bauer:2023qgm,Davoudi:2022bnl,Beck:2023xhh} 
are expected to require hundreds or more high-fidelity qubits in order to be of scientific utility.
Target simulations include establishing the properties of ground and excited states of strongly-interacting matter as well as the dynamics of systems that are far from equilibrium.
Specific examples of such processes are interactions among short-lived particles, high-multiplicity collisions created in accelerators, the preferential generation of matter over anti-matter in the early universe, and the creation of new phases of matter in the high-energy collisions of nuclei.
Results of such simulations will play key roles in extending predictive capabilities into new regimes, required for discovering new fundamental particles and laws of nature. 
Since the formulation of the Standard Model of particle physics~\cite{Glashow:1961tr,Higgs:1964pj,Weinberg:1967tq,Salam:1968rm,Politzer:1973fx,Gross:1973id}, which describes the strong and electroweak interactions in terms of quantum field theories and gauge interactions, numerical simulations of classes of low-energy observables using Monte Carlo sampling techniques in Euclidean space have found enormous success~\cite{Wilson:1974sk,Kogut:1974ag,Creutz:1979zg,Kogut:1979wt,Creutz:1979kf,Creutz:1980zw}.
Furthermore, lattice gauge theories (LGTs) are the only known formulations that provide systematically improvable, first-principle calculations of the properties of the strong interactions, crucial for comparing theoretical predictions of the Standard Model to experimental results.
Rapid progress is being made toward using quantum computers to simulate complementary sets of observables in these theories in the real-time Hamiltonian formulation~\cite{Kogut:1974ag}, 
focused on the dynamics of systems in highly-energetic initial states, with an expectation of near-term quantum advantages.
Part of the reason for this focus is that 
time-evolution from a given initial state prepared on a quantum computer is known to 
be efficient for quantum computers at scale, lying within the bounded-error quantum polynomial time (BQP) complexity class~\cite{Lloyd:1996aai}.

There is an increasing effort to use the local gauge symmetries underlying LGTs as part of QEC protocols not only for simulation, but also as error correcting codes themselves.
One of the important implications of local gauge symmetry is charge conservation.
In the context of U(1) gauge theory (quantum electrodynamics), quantum circuits were designed to perform Gauss's law checks for implementation in simulations in Refs.~\cite{Kaplan:2018vnj,Stryker:2018efp}.
Their circuits provide  error detection using ancilla qubits, and hence a mechanism for post-selecting on states satisfying local charge conservation.
Integrating local symmetries into QEC is a crucial direction
and it has been shown that, in fact, overheads associated with QEC can be reduced 
in U(1) theories by 
including local charge conservation as one of the error detection  
checks~\cite{Rajput:2021trn,Spagnoli:2024mib,Cobos:2025krn}.
A detailed analysis of implementation costs, along with a more general analysis of implementing Gauss's law checks 
has been presented in Ref.~\cite{Pato:2026wow}.
In addition, efforts are being made to promote LGTs, specifically U(1) theories and the connections to rotor codes, to QEC codes, {\it e.g.}, Refs.~\cite{Rothlin:2025thesis,lacambra2026GLCVa}.
These analyses encode all dynamical elements contributing to Gauss's law as dynamical degrees of freedom in the lattice simulation.
As a result, the choice of gauge used in mapping the theory to qubits 
affects QEC and error detection designs.
With constraints on the number of qubits, other gauge-fixing protocols have been fruitfully explored.
This has recently been  extended to SU(2) theories  for both truncated simulations~\cite{Yao:2025cxs} 
and with a generalized approximate formulation for pure SU(2)~\cite{Bradshaw:2026approximate}.
Important progress is being made in 
studying the error-correction properties of discrete group representations of gauge fields, including associated error thresholds and variations with regard to mappings to physical qubits and qudits, {\it e.g.}, Refs.~\cite{Gustafson:2023robustness,PhysRevD.110.054516}.
These works have observed that increased redundancy associated with gauge symmetry is beneficial for identifying and correcting errors.  In the context of classical and QEC codes, this is not surprising, and is an important point to consider when designing and optimizing simulations of LGTs.

It is well known that the utility of QEC and error detection codes are contingent on noise thresholds~\cite{shor1996fault,Aharonov:1999ei,Kitaev1997aa,KnillLaflammeZurek1998ProcRoySoc,Preskill_1998,Knill2005,AliferisGottesmanPreskill2006}.
When the noise falls below the threshold for a given code, the fidelity of logical computation rapidly improves.
Error thresholds are well studied for many QEC codes and have set fabrication targets for physical qubits,
and 
fidelity improvements of a small number of codes have been demonstrated~\cite{Ofek:2016extending,Chen:2021exponential,RyanAnderson:2021realtime,GoogleQuantumAI2025,Krinner2022,Zhao2022,reichardt2024MicroQuant,QuantMicro,MicrosoftQuantinuum2026,Dasu:2026dwm,Postler2022,bluvstein2024logical,Rodriguez2025,Bluvstein2026,reichardt2025MicroAtom,Sivak2023,Putterman2025Bosonic,Reglade2024}.
In addition, 
the structure of physical systems in simulations allows for fine-tuned optimizations in the 
FT design.
During the present period in which error detection may have advantages over QEC, 
dictated by the available quantum computers, optimizations of the non-universal error detection implementations are desirable for systems being simulated.

In this work, we examine the potential of sparse error detection in 
distance-two
$[[N+2, N, 2]]$ 
codes, with $N$ even,~\cite{Vaidman:1996error,Grassl:1997erasure,Rains:1999distance2,Steane:1996simple,Gottesman:1998theory,Viola:1999dynamical}
for lattice gauge theories.  This family of codes is often referred to as Iceberg codes.
We focus on embedding small-spatial volumes of the Schwinger model using one or more combinations of 
$[[N+2, N, 2]]$ with small $N$. 
Axial gauge is chosen, as the lack of explicit gauge degrees of freedom means that the Hamiltonian construction scales to large spatial volumes more readily (made possible by confinement~\cite{farrell2023qcd1,farrell2023qcd2,Farrell:2023fgd,Farrell:2024fit}).
In axial gauge, local checks of Gauss's law are not possible, but violations appear as energy penalties via the Hamiltonian. 
Integral to this analysis is the use of postselection of measurements performed on the final state.
If the initial state of the evolution is an eigenstate of the charge operator, then ideal time evolution will leave the state in the same charge sector. 
Post-selecting measurements of the final state, limited to the space of codewords (codespace) of that charge sector, removes all combinations of errors that change the charge sector.  
This typically eliminates leading order bit-flip errors after the simulation has concluded, but at the cost of a reduced ensemble size.
Adding layers of stabilizer checks with projective measurements throughout the simulations further removes leading-order Pauli errors. 
The size of the uncertainties can be optimized by adjusting the sparsity of the layers, determined by the overall
errors induced by the evolution and by the stabilizer checks.
Specifically, we compare the $[[4,2,2]]^{\otimes 2}$ and $[[6,4,2]]$ embeddings of the Schwinger model with two lattice sites, and a selection of embeddings for four lattice sites, to identify potential advantages of such embeddings for error-detected quantum simulations of quantum field theories. Embedding the Schwinger model into various combinations of Hypercube code families, $[[2^N, N, 2]]$, is also explored, though due to the use of Trotter time-evolution, Iceberg code families are found to carry more benefit. 
Note that  within this work, we will use the notation $[[k,n,d]]^{\otimes{m}}$ to indicate 
$m$ copies of the $[[k,n,d]]$ code, encoding different code blocks.

While simulations of the Schwinger model have been performed on more than 
100 physical qubits~\cite{Farrell:2023fgd,Farrell:2024fit}
(as well as analogous systems of Yang-Mills theories in comparable~\cite{Chen:2026tvd} and smaller spatial volumes~\cite{mezzacapo2015su2circuits,ciavarella2021trailhead,atas2021su2hadrons,rahman2022selfmitigating,ciavarella2022su3vacuum,farrell2023qcd2,farrell2023qcd1,atas2023qcd1d,ciavarella2023improved,calajo2024ionqudits,li2025energyloss,than2025phasediagram, Froland:2025bqf,Cogburn:2026aqy,chernyshev2026pathfinding,Raychowdhury2026a,Froland:2026aff}) and local expectation values have been successfully extracted using error-mitigation techniques~\cite{Urbanek_2021,ARahman:2022tkr,Berg:2022ugn,Farrell:2023fgd,Farrell:2024fit}, simulations necessary for late-time dynamics will require FT, error detection and QEC.
Understanding how to optimally include these protocols requires detailed studies of their behavior when integrated into LGTs.

\subsection{Road Map and Result Summary}
\label{sec:road}
\noindent
With an eye to retaining the underlying symmetries of lattice theories,
including discrete spatial translation invariance that smoothly recovers the continuum limit, we start with a general discussion of FT and error detection (Section~\ref{sec:FTEC}), 
which includes, as a starting point, a detailed analysis of a single code block of the $[[4,2,2]]$ code. 
The reason for choosing  the $[[4,2,2]]$   code is that a single spatial site of 1D quantum electrodynamics (the Schwinger model) defined in axial gauge maps to two qubits using the Jordan-Wigner mapping.
Further, the techniques and mappings generalize to systems with arbitrary numbers of 
spatial sites.
The $[[4,2,2]]$ encoding provides error-detection capabilities for a single site, 
and, while it is not a unique encoding, it is the one with the smallest quantum resource footprint for error detection.
The encoding of two logical qubits into four physical qubits is presented, 
along with the logical state space, 
the stabilizers that identify single bit flips and phase flips which are fault-tolerantly implemented using two ancillas,
and a discussion of the lack of fault-tolerance in implementing general unitary transformations.
The contents of this section can be found in the existing literature, and we review them for context.

Following the discussion of the single $[[4,2,2]]$ code block, we consider the encoded dynamics of two spatial sites ($L=2$) 
of the Schwinger model, requiring four qubits, one for each of the two electron and two positron sites.
The impact of sparse error detection combined with global charge post selection is studied in detail, 
including the approach to recovering noiseless results with increasing number of detection layers.
For the sake of comparison, we  choose to 
focus mainly on two-qubit noise with $p_2=0.001$ and $0.003$.
The results of this work reinforce the role of postselection 
(into the Gauss's law preserving charge sector)
in eliminating ${\cal O}(p_2)$ bit-flip errors 
in both the unencoded and encoded simulations,
and the suppression of 
${\cal O}(p_2^2)$ errors by mid-circuit measurement or postselection on the results of stabilizer measurements.
Further, it shows the importance of embedding the logical states into a larger Hilbert space in order to reduce the 
${\cal O}(p_2^2)$ 
from  several ${\cal O}(p_2)$ errors conspiring to return the system into the codespace, generating an undetectable logical error.
For resource-constrained simulations, with the number of available shots fixed to be $10^4$, a small number of rounds of FT error detection is shown to significantly reduce the systematic error over a large time interval, out to such times that remaining ensemble sizes are small enough for
statistical uncertainty to become comparable to the systematic error. 
However, beyond a small number, increasing the density of error detection does not continue to reduce the systematic error. 
By fitting a simple functional form to the $L=2$ shot-acceptance results, 
predictions for shot-acceptance in larger systems are made
from the ratio of Hilbert-space dimensions.
Finally, we  verify that given the limited number of operations performed on the error-detection ancilla qubits, 
they can be of lower fidelity without noticeably impacting the quality of the computed observables.
Regarding this, we mainly analyze a $[[4, 2, 2]]^{\otimes 2}$ encoding. In addition, we perform the same simulations using a $[[6,4,2]]$ encoding, 
where there are no cross-block operations required to implement time evolution.  
For this encoding, 
the stabilizers involved $Z^6$ and $X^6$ operations.
The results are consistently better than for the $[[4,2,2]]^{\otimes 2}$ encoding for the studied observables.

To provide confidence that our conclusions  are not specific to the $L=2$ system, 
we  study some of the aspects in larger systems.  
The larger systems are also richer with regard to the nature of the errors.
We  do not use the features related to the spacetime location of errors with respect to the location of local observables, and we  restrict ourselves to global rejection of a given shot if an error is detected.
A main focus is the $L=4$ system encoded into a single $[[10,8,2]]$ code block.
For the same error rates, the shot rejection is parametrically larger than for the $L=2$ system,
and some of the observations that we make require $> 10^6$ shots to clearly identify.
This larger system is used to systematically study the reduction of systematic error as a function of time with increasing numbers of rounds of error detection.
Simulations were performed out to $T=45$, using  up to 
90 Trotter steps under the effect of depolarizing noise with $p_2=0.001$. 
At any given time, it is found that the systematic error is reduced by including error-detection layers, 
up to some number at which the improvement saturates.  
The longer the time interval, the greater number of layers required before saturation.
For a given evolution operator and $p$, there is an optimal density of error-detection layers that can be included
to reduce systematic errors due to noise.  
This number will need to be identified by pre-production run tuning, along with other run parameters.
The larger system  enables different partitions into code blocks from different codes, including
$[[10,8,2]]$, $[[8,6,2]]\otimes [[4,2,2]]$, $[[6,4,2]]^{\otimes 2}$, $[[6,4,2]]\otimes [[4,2,2]]^{\otimes 2}$ and 
$[[4,2,2]]^{\otimes 4}$.   
Further, there is a trade-off between shot rejection rates and the reduction of systematic errors. 
The code that performs the best evolves from $[[10,8,2]]$ for small numbers of shots to $[[6,4,2]]^{\otimes 2}$ for large numbers of shots, with the difference between code performance largest at small shot numbers.

The final aspect of Iceberg codes that we explore is the scaling 
of the ``failure time'' 
of monolithic codes
(in which all of the logical qubits reside in single code block)
of the form $[[2L+2, 2L,2]]$, with the number of spatial sites, $L$, for fixed resources and error rate.
By defining a point of failure of the error detection, 
determined by deviations from the exact result,
these times decrease with increasing system size, 
reducing to $\sim 1$ Trotter step for $L=13$ spatial sites with $p_2=0.003$ and $10^4$ shots.
This indicates that encoding into monolithic Iceberg  codes improves such partially FT simulations over some time interval but  becomes impractical beyond that time with fixed resources.  Given the exponential suppression of the final ensemble, exponentially increasing numbers of shots are required to extend the time interval by an order-one time.
The numerical results from simulations are fit to functional forms
and extrapolated to very large systems, to make resource-requirement predictions for 
$[[102,100,2]]$, as an example.

Another direction to build larger code blocks for larger systems is to utilize the Hypercube code family, $[[2^N, N, 2]]$, as both the Iceberg code family and the Hypercube code family coincide at $N=2$.
Systems comprised of $N=3$ code blocks, $[[8,3,2]]$, are explored in detail.
Because of the structure of the encoding, the acceptance probabilities for $L=2,4,6$ 
are found to be significantly lower than for the comparable embeddings in Iceberg codes.
Consequently, for these simulations, the Hypercube codes are found to 
underperform  compared  to the Iceberg codes in fixed-resource scenarios.

With regard to the error analysis and shot-rejection through error-detection, 
unlike more sophisticated shot-rejection methods based on correlations, 
{\it e.g.}, Ref.~\cite{Froland:2026rzt}, here a shot is rejected if an error is detected anywhere in the system.
This provides lower-bound estimates 
(within the context of the error model) for remaining ensemble sizes, as it does not include spacetime localization (backward light-cones for local operators) or noise that commutes with observables.

\section{Error Detection and Fault-Tolerance}
\label{sec:FTEC}
\noindent
Error detection is designed to eliminate contributions from leading-order (LO) errors and partial higher-order errors,
occurring with a two-qubit rate that is much greater than the single-qubit error rate.
One way that is already used in quantum simulations is to only accept final quantum states that are permitted by physics.
For example, charge-neutral states under charge-conserving dynamics should produce final states with vanishing total electric charge, and violations of this detected in final measurements can be removed with postselection.
This eliminates all ${\cal O}(p_2)$ charge-violating errors and the corresponding ${\cal O}(p_2^n)$ errors.~\footnote{This does not preclude charge-violating errors from conspiring to produce a charge-neutral final state; these errors contribute at $\mathcal O(p_2^2)$ and higher.} 
When measuring observables commuting with electric charge, 
this physics-aware postselection is ``free'' and should be implemented wherever practical.

Introducing a number of stabilizer measurements during the time evolution increases the circuit depth and requires ancilla qubits.
The time evolution of a system is simulated using an amount of machine time that depends on the depth of Trotterized circuits.
Within this machine time, the total error rate, $p$, is determined by the error rate per unit evolution time, $\overline{p}$.  
With the sparsely interspersed layers of stabilizer measurements, 
the error rate per unit evolution time is increased
$\overline{p}_s > \overline{p}$.
The total error rate becomes $p=1-(1-\bar{T}\ \bar{p})^m$ for $m$ layers, where $\bar{T}$ is the time interval of evolution between layers.
Without postselection, encoded circuits produce inferior logical error rates compared to unencoded circuits because of the overhead associated with error detection.~\footnote{This assumes the overhead of mapping the system degrees of freedom to qubits is equal in encoded and unencoded implementations.}
The code's benefit comes from postselection, which suppresses 
logical errors to 
${\cal O}(p^2)$
but discards an ${\cal O}(p)$ fraction of runs — turning a single-metric comparison into a tradeoff between logical fidelity and acceptance rate.
With the stabilizers eliminating ${\cal O}(\overline{p})$ 
errors at each application, the total ${\cal O}(p^2)$ is minimized when
the extra machine time required for implementation of each stabilizer layer is equal to the Trotter step size.
However, for a sufficiently large number of Trotter steps the quantum computer will decohere, and for a  small number of Trotter steps, the systematic Trotter errors will likely dominate measured observable expectation values.  
Therefore, the inclusion of error detection into simulations requires further 
{\it in-situ} optimizations with regard to frequency to minimize the errors in the target observables.

The discussion surrounding error detection is only partially decoupled from  the 
discussion of FT.
The stabilizers and postselection identify errors  that take the system out of the 
codespace.
A fully FT circuit would ensure that any ${\cal O}(p)$ error would not  propagate coherently, for instance, through the time evolution operator, into multiple errors, some of which may place the system back into the codespace 
(a discussion of  error propagation through circuit elements can be found in App.~\ref{app:CEP}).
Without FT, such coherent errors are present and  may escape detection, 
providing undetectable unphysical contributions to observables at ${\cal O}(p)$.
A fully FT simulation with error detection parametrically suppresses 
such contributions to observables to ${\cal O}(p^2)$,
while partial fault-tolerance includes ${\cal O}(p)$ errors with a reduced coefficient.

There are a large number of ways to embed physical qubits into logical qubits. 
Compelling configurations for error detection 
in systems with regularly repeating units, 
such as lattice gauge theories,
are the
$[[N+2, N, 2]]^{\otimes q}$ codes that furnish
$q N$  logical qubits from $q (N+2)$  physical qubits
with Hamming distance 2.
With such embeddings,
errors that take the system out of the codespace can be detected 
with each layer of stabilizer checks,
but not corrected.

\subsection{A Single Block of [[4,2,2]]}
\label{sec:422}
\begin{figure}[ht!]
    \centering
\includegraphics[width=0.95\linewidth,alt={The connectivity diagram for the [[4,2,2]] code.}]{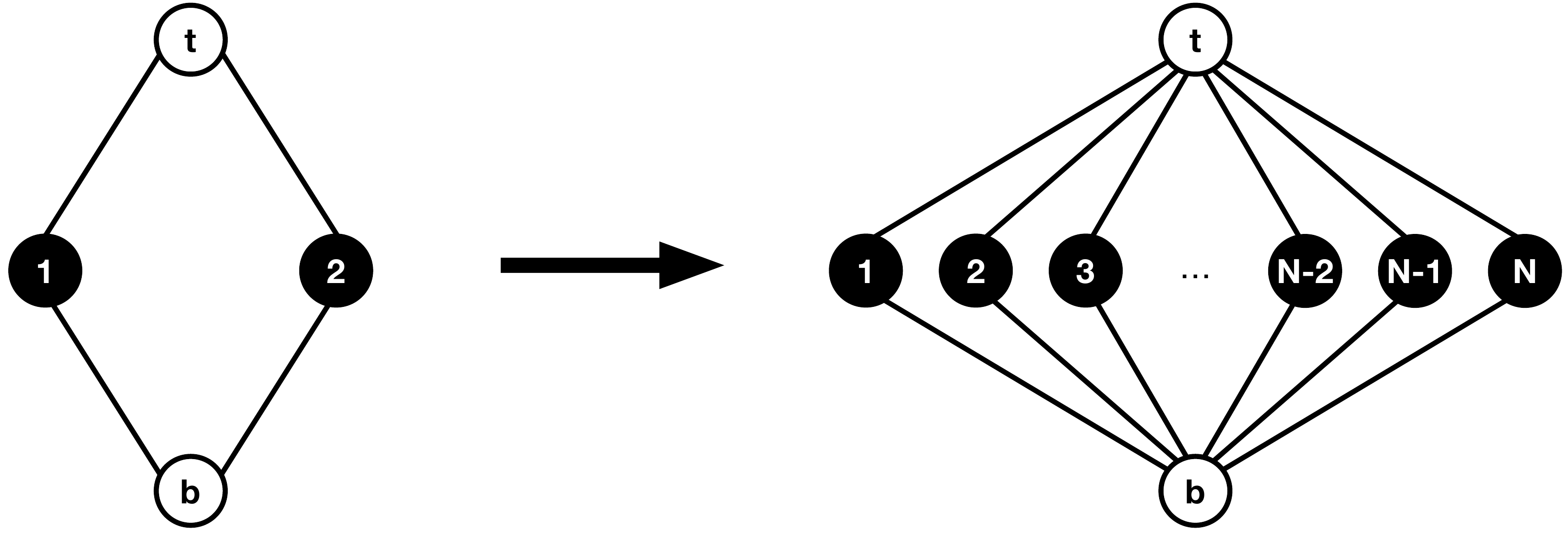}
\caption{The connectivity diagram for the $[[4,2,2]]$ code (left)
and the generalization to the Iceberg code $[[N+2, N, 2]]$ (right).  
The top and bottom qubits are used to stabilize the $N$ logical qubits 
in the Iceberg code family. }
    \label{fig:422connect}
\end{figure}

\noindent
The [[4,2,2]] code (also known as the $C_4$ code)
is the simplest code with which to establish logical qubits from physical qubits and to include FT error detection.
Four physical qubits are used to build two logical qubits with codewords separated by two bit flips.
Fig.~\ref{fig:422connect} shows the connectivity map of the [[4,2,2]] code, as well as its generalization to larger numbers of logical qubits.
The physical qubits shown in Fig.~\ref{fig:422connect} are laid out in the order: $t, 1, 2, b$,
and the logical qubits are entangled states of these physical qubits.
The $|\bar{0} \bar{0}\rangle$ state in the [[4,2,2]] code, 
corresponding to the two logical qubits in the $|0\rangle$ state, 
is simply the $\ket{\text{GHZ}(4)}$ state,
\begin{align}
|\bar{0} \bar{0}\rangle & =  
\frac{1}{\sqrt{2}} 
\left[\ 
|0000\rangle + |1111\rangle
\ \right]
\ .
\end{align}

The logical operations are defined by~\footnote{
The logical $\widehat{\overline{Y}}_i$ are derived from the combined action
$\widehat{\overline{X}}_i \widehat{\overline{Z}}_i = -i \widehat{\overline{Y}}_i$.
}
\begin{align}
    \widehat{\overline{X}}_1 & =  \hat X\hat X\hat I\hat I
    \ ,& \widehat{\overline{X}}_2 \ &=\  \hat X\hat I\hat X\hat I \ ,
    \nonumber\\
    \widehat{\overline{Z}}_1 & =  \hat I\hat Z\hat I\hat Z
    \ ,& \widehat{\overline{Z}}_2 \ &=\  \hat I\hat I\hat Z\hat Z
\ ,
\label{eq:422logicals}
\end{align}
which act on the $|\bar{0} \bar{0}\rangle$ state to give the full logical basis
\begin{align}
|\bar{0} \bar{0}\rangle & =  {1\over\sqrt{2}} 
\left[\ 
|0000\rangle + |1111\rangle
\ \right]
\ \ ,
\nonumber\\
|\bar{1} \bar{0}\rangle & = {1\over\sqrt{2}} 
\left[\ 
|1100\rangle + |0011\rangle
\ \right]
\ \ ,
\nonumber\\
|\bar{0} \bar{1}\rangle & =  {1\over\sqrt{2}} 
\left[\ 
|1010\rangle + |0101\rangle
\ \right]
\ \ ,
\nonumber\\
|\bar{1} \bar{1}\rangle & = {1\over\sqrt{2}} 
\left[\ 
|0110\rangle + |1001\rangle
\ \right]
\ .
\label{eq:422codewords}
\end{align}
These are  Hamming-distance 2 states, which permit error detection, as a single (physical qubit) error takes any of the states out of the code space to a state that could have originated from more than one codeword.

The stabilizers for the logical states are 
\begin{align}
\hat S_z & =  \hat Z  \hat Z  \hat Z  \hat Z 
\ \ \text{and}\ \ 
\hat S_x \ =\  \hat X  \hat X  \hat X  \hat X 
\ ,
\label{422stabs}
\end{align}
for which the logical states are eigenstates with eigenvalues $+1$.
For this to be the case, the states have a well-defined  Hamming weight 
({\it i.e.}, the number of ``1''s is even).
These stabilizers, by definition, can be measured during a computation 
(or period of sensing)
and do not change a wavefunction in the logical space, as the measurement will always produce $+1$.
In contrast, a measurement of either stabilizer producing a result that is different from $+1$ indicates that an error 
is present somewhere within the 4 physical qubits.
While the ${\cal O}(p)$ errors are detected by these stabilizers, as well as with post selection, they also detect some of the 
${\cal O}(p^2)$ errors. 
While order-2 products of $\hat X$, $\hat Y$ and $\hat Z$ with one flip can be detected, products
$\hat X_i\hat X_j$, $\hat Y_i\hat Y_j$ and $\hat Z_i\hat Z_j$ with an even number of flips remain undetectable. 
This is true for the general $[[N+2, N, 2]]$ codes.

\begin{figure}[ht!]
    \centering
\includegraphics[width=0.41\linewidth,alt={FT Neel}]{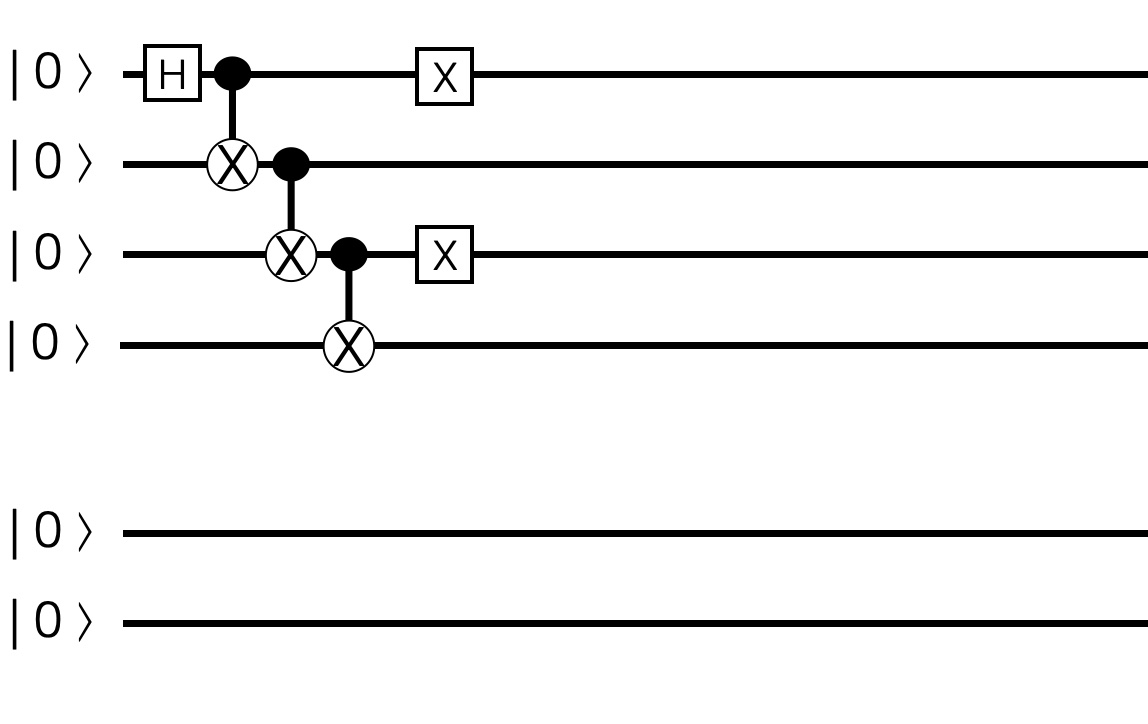}
\includegraphics[width=0.44\linewidth,alt={FT Neel}]{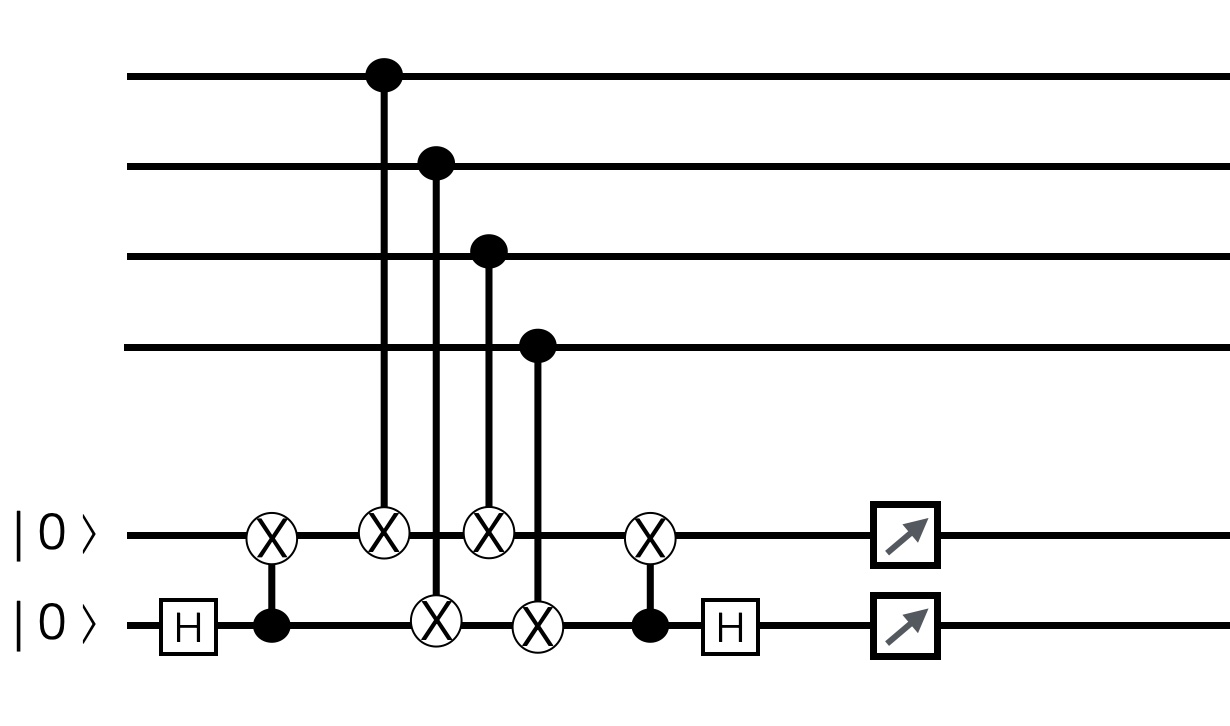}
\includegraphics[width=0.42\linewidth,alt={FT Neel}]{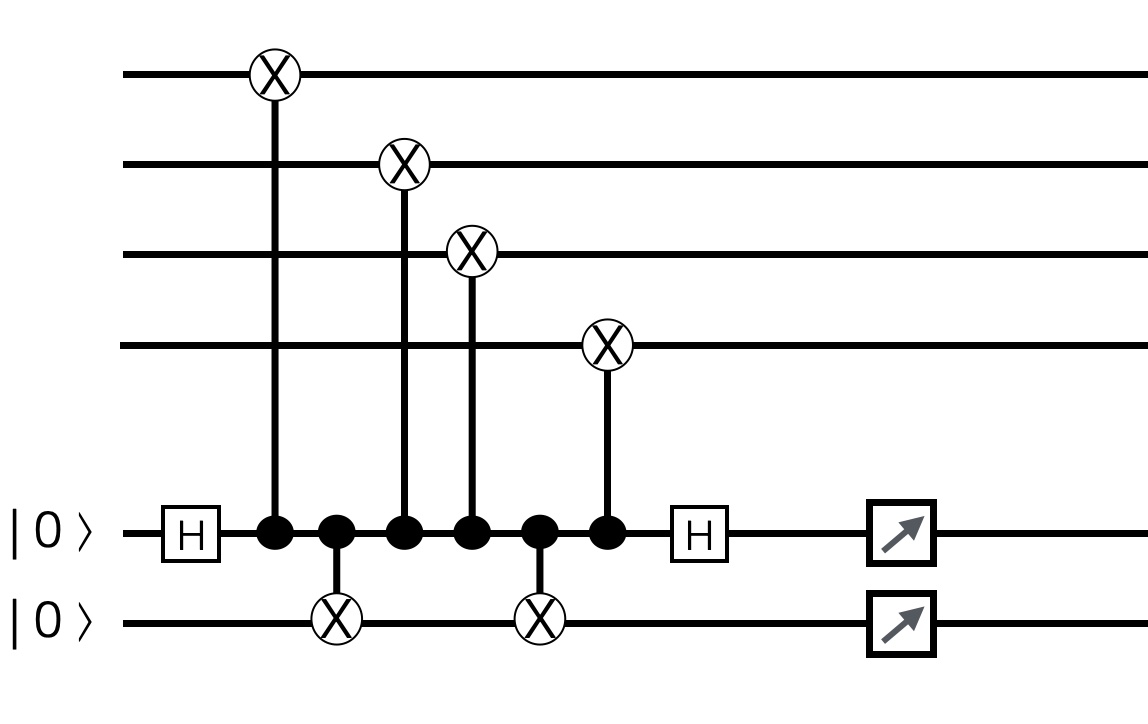}
\includegraphics[width=0.42\linewidth,alt={FT Neel}]{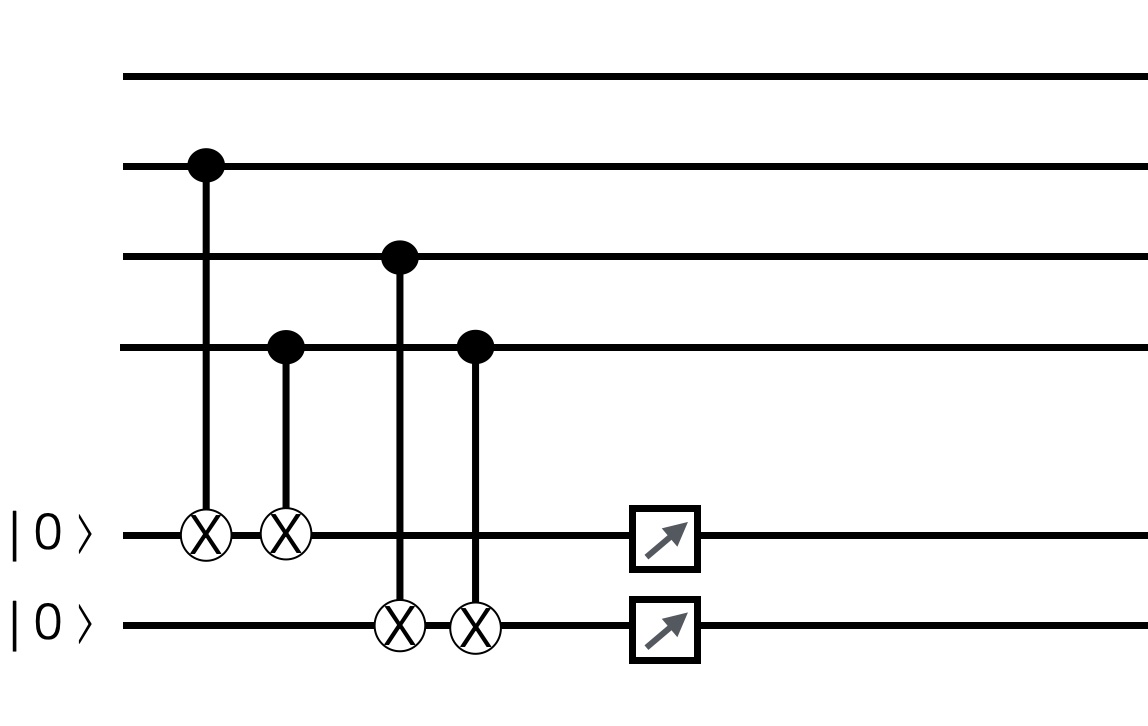}
\caption{
Quantum circuits for a  FT preparation of the Neel state $|\bar{0} \bar{1}\rangle$.
The upper left circuit furnishes a non-FT preparation of the Neel state (Neel0).
The upper right circuit (the $\hat S_z$ given in Eq.~(\ref{422stabs})) provides a FT stabilizer check for bit-flips ($\hat S_z$).
The lower-left circuit (the $\hat S_x$ given in Eq.~(\ref{422stabs}))
provides a FT stabilizer check for phase-flips ($\hat S_x$).
The lower-right circuit measures $\widehat{\overline{Z}}$ for each logical qubit without undetectable errors.}
    \label{fig:422NeelFT}
\end{figure}

A FT preparation of
the Neel state 
(the strong-coupling vacuum in the Schwinger model)
in this block, $|\bar{0} \bar{1}\rangle$, can be accomplished with the circuits shown in Fig.~\ref{fig:422NeelFT}.
The circuits in Fig.~\ref{fig:422NeelFT} are applied in the sequence
\begin{align}
{\rm Neel0}\  ,\  \hat S_z\  ,\  \hat S_x\  ,\  \widehat{\overline{Z}}\  ,\  \hat S_z\  ,\  \hat S_x 
\end{align}
to obtain the logical Neel state after measurements on the ancilla such that
$\hat S_z$ gives $00$,
$\hat S_x$ gives $00$ and 
$\widehat{\overline{Z}}$ gives $01$.
The Neel0 preparation
(see Fig.~\ref{fig:422NeelFT})
is not FT and can have errors that are detected and selected against with the subsequent stabilizer and $\widehat{\overline{Z}}$ measurements.  
After the $\widehat{\overline{Z}}$ measurements, 
there can remain detectable errors only, which are then selected against with the two following stabilizer measurements.
To create a chain of such Neel states, associated with the strong-coupling ground state of the Schwinger model, this preparation is repeated for each logical block.~\footnote{For larger systems we found it to be beneficial to skip the $\widehat{\overline Z}$ measurement which adds $4L$ CNOT gates.}

\begin{figure}[ht!]
    \centering
\includegraphics[width=0.45\linewidth,alt={FT rotations}]{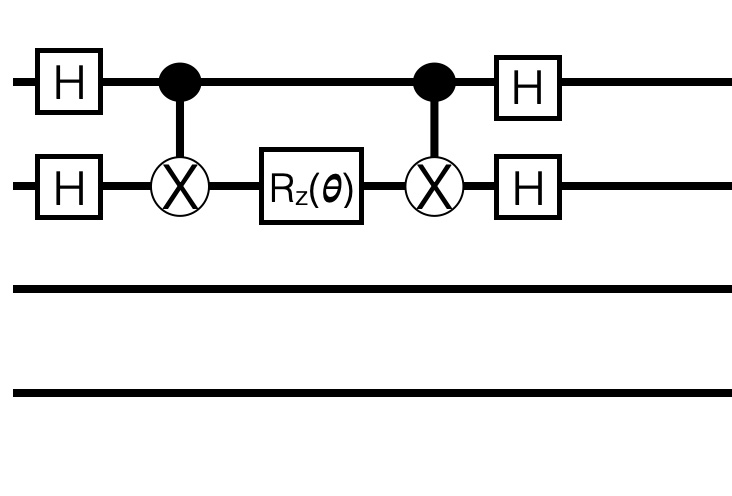}
\includegraphics[width=0.45\linewidth,alt={FT rotations}]{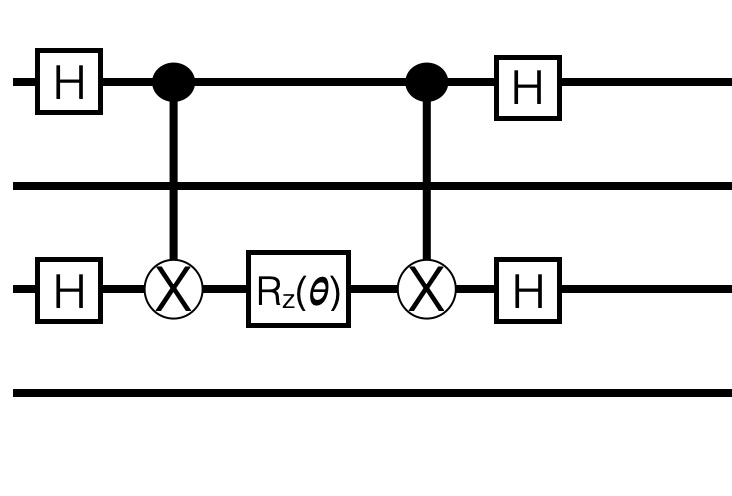}
\caption{
Non-FT circuits for rotating logical states in the [[4,2,2]] code through
$e^{-i\frac{\theta}{2} \widehat{\overline{X}}_j} = e^{-i \frac{\theta}{2} \hat X_0\hat X_j}$, with $j=1$ (left) and $j=2$ (right).}
    \label{fig:422Rots}
\end{figure}

While the Hadamard, phase gate and CNOT gate are transversal, the T-gate is not, and as such, continuous rotations are not naively FT. 
For example, a  non-FT rotation of the form 
$e^{-i\frac{\theta}{2} \widehat{\overline{X}}_1} = e^{-i \frac{\theta}{2} \hat X_0\hat X_1}$ 
is implemented with entangling operations between qubits 0 and 1, and generates undetectable correlated two-qubit errors of the form of a logical operation, see Fig.~\ref{fig:422Rots}.
The  inter-block and intra-block logical rotations cannot be made FT  because of the CNOT gates between physical (data) qubits.~\footnote{Propagation of errors can be minimized by incorporating ancilla qubits into logical rotation gadgets, see {\it e.g.}, Refs.~\cite{Froland:2026rzt,Zhong:2026jps,Gerhard:2024peb}.} 
This can only be overcome using magic state injection via transversal CNOT gates from ``magic state factories''~\cite{shor1996fault,Gottesman1998,Gottesman1999,Zhou2000,Bravyi2005,Eastin2009,Paetznick2013,Anderson2014,Campbell2017}
and utilizing one logical qubit per code block, or by code switching~\cite{Paetznick2013,Anderson2014,Bombin2015,Kubica2015,Beverland2021,Pogorelov2025}.
Without  these techniques, 
the time evolution of systems embedded in one or more $[[4,2,2]]$ error-detecting codes 
will not be  FT,  and errors 
from logical rotations could escape into the codespace.
The use of non-FT rotations with other FT gadgets (such as syndrome extraction) can still suppress errors provided the error propagation from non-FT gadgets is limited. 
Recent work has used this approach to demonstrate improvements of encoded circuits over unencoded ones~\cite{Dasu:2026dwm,Froland:2026rzt}, and is the approach taken in this work.

The practical implementation of the stabilizer and logical checks requires a small number of ancilla qubits in each block, along with ancilla resets.
The blockwise stabilizer and $\widehat{\overline Z}$ checks shown in Fig.~\ref{fig:422NeelFT} produce four classical flags per block in each error detection round.
Following this, runs that detect any errors are discarded.
For sufficiently long runs with a fixed error rate, the number of surviving members of the ensemble becomes exponentially small (until the code floor is reached),
scaling as $\sim \exp(-p_2\ \# {\rm gates})$.
The practical upper limit on the length of any given run is reached when the statistical error becomes larger than the target uncertainty.

For post-processing it can be helpful to  map
the physical states defining the logical states into one basis state (that is not an eigenstate of the stabilizers).
For example, 
$|\bar{0} \bar{1}\rangle  =  {1\over\sqrt{2}} \left[\ |1010\rangle + |0101\rangle\ \right]$
$\rightarrow |0101\rangle$.
\begin{figure}[ht!]
    \centering
\includegraphics[width=0.4\linewidth,alt={FT rotations}]{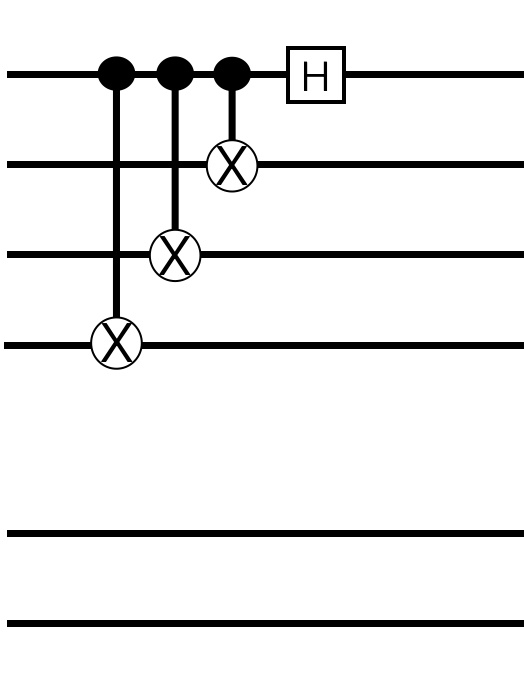}
\caption{
A quantum circuit to  map the physical states contributing to each logical state into a single basis state. 
}
    \label{fig:422Final}
\end{figure}
The quantum circuit to achieve this in a single  code block is shown in Fig.~\ref{fig:422Final}.

\section{The $L=2$ Schwinger Model}
\label{sec:SM}
\noindent
The lattice Schwinger model provides a well-explored framework in which to examine error detection, 
providing an entry point toward studies in more complex theories describing fundamental physics.

The Hamiltonian is most easily constructed in axial-gauge, where there are no space-like gauge links, and the time-like gauge links can be replaced by a non-local ``Coulomb'' interaction across the lattice,
\begin{eqnarray}
\hat H & = &   
\hat H_m  + \hat H_{el} + \hat H_{kin} \ ,
\nonumber\\
& = & 
\frac{m}{ 2}\sum_{j=0}^{2L-1}\ \left[ (-1)^{j+1} \hat Z_j + \hat{I} \right] 
\ + \ \frac{g^2}{ 2}\sum_{j=0}^{2L-2}\bigg (\sum_{k\leq j} \hat Q_k \bigg )^2 
\nonumber \\ 
&& 
\ + \ 
\frac{1}{2}\sum_{j=0}^{2L-2}\ \left( \hat \sigma^+_j \hat\sigma^-_{j+1} + {\rm h.c.} \right) 
\ ,
\nonumber \\ 
\hat Q_k & = &  -\frac{1}{2}\left[ \hat Z_k - (-1)^k\hat{I} \right] \ .
\label{eq:Hgf}
\end{eqnarray}
There are three distinct contributions: the mass term, the gauge term and the kinetic term. Here, $L$ is the number of  spatial lattice sites, 
corresponding to $2L$ staggered (fermion) sites, $m$ and $g$ are the (bare) electron mass and charge, respectively, and the staggered lattice spacing $a$ has been set to one.
Starting from the left-most lattice site, with an embedding, 
$e^+, e^-, e^+, e^-, e^+, ..., e^-$,
the electric field to the right of a given site is the sum of charges up to and including that site,
and vanishing on the left  and right hand sides of the lattice in the $\sum Q_i=0$ sector.
For example, $E_0=Q_0$, $E_1 = Q_0+Q_1$, and so forth,
where we are labeling sites $n=0,1,2,.., 2L-1$.
An alternate (symmetric) construction in $\sum Q_i=0$ sector involves forming the electric field from either end of the lattice.

\subsection{Physical Qubits: Unencoded}
\label{sec:SMLeq2phys}
\noindent
Quantum simulations of the Schwinger model using physical qubits are well-established, 
particularly for small systems.
The Hamiltonian in Eq.~(\ref{eq:Hgf}) reduces to 
\begin{eqnarray}
\hat H_{kin} & = & 
{1\over 4}
\left[\ 
\hat X_0\hat X_1+\hat Y_0\hat Y_1
+\hat X_2\hat X_3+\hat Y_2\hat Y_3
\right.\nonumber\\
&&\left.
+ \hat X_1 \hat X_2
+  \hat Y_1 \hat Y_2 
\ \right]
\nonumber\\
\hat H_{mass} & = & 
{m\over 2}
\left[\ 
4 \hat I 
-\hat Z_{0}  +\hat Z_{1}  
-\hat Z_{2}  +\hat Z_{3} 
\ \right]
\nonumber\\
\hat H_{gauge} & = & 
{g^2\over 2}
\left[
2 \hat I +
\hat Z_{0}\hat Z_{1}
+{1\over 2}
\hat Z_{0} \hat Z_{2} 
+{1\over 2}
\hat Z_{1} \hat Z_{2} 
\right. \nonumber\\
& & 
\left.
\qquad
-\hat Z_{0} 
-{1\over 2} \hat Z_{1} 
-{1\over 2} \hat Z_{2} 
\ \right]
\ ,
\label{eq:HamLeq2LOG}
\end{eqnarray}
after restricting the lattice to 
two spatial sites, $L=2$.
We perform classical simulations using {\tt NoiseModel()} in the {\tt qiskit} {\tt AerSimulator}~\cite{Qiskit2024} 
with isotropic depolarizing noise. 
The single-qubit error rate is one-tenth ($1/10$) of $p_2$.~\footnote{This is implemented via 
{\tt noise\_model.add\_all\_qubit\_quantum}
{\tt \_error(depolarizing\_error($p_2$, 2), ['cx'])}. 
The transpiled basis set of gates is {\tt ['h', 'x', 'rz', 'sx', 'cx']},
and 
as $rz$ gates are generally implemented in software, those operations have been assigned $p_{rz}=0$.
}
We compare sets of results obtained with $p_2=0.001$ and $p_2=0.003$.
The former is consistent with present day hardware error rates, 
while the latter reflects error rates of the recent past.
Working with a fixed number of shots is a realistic constraint for production  runs on accessible quantum computers, and for $L=2$ we restrict our analysis to simulations using $10^4$ shots, and for larger lattices consider simulations with  greater than $10^5$  shots.

As the focus of this work is on evaluating the utility of error detection,
we perform simulations without implementing post-processing error mitigation, such as Decoherence Renormalization (DR)~\cite{Urbanek:2021oej,ARahman:2022tkr}, Operator Decoherence renormalization (ODR)~\cite{Farrell:2023fgd,Farrell:2024fit} 
and Zero-Noise Extrapolation (ZNE)~\cite{ZNE1,ZNE2,Klco:2018kyo,ZNE3}. 
Improvements or degradations resulting from error-detection will likely
lead to the same trends after 
error mitigation, and 
quantifying the interplay between them is a study for the near future.
Further, given some of our results, extrapolations in the number of detection layers may also prove fruitful.

The quench dynamics from the  strong-coupling vacuum (Neel) state is studied.~\footnote{Quenching means evolving a pure state forward in time under dynamics defined by a Hamiltonian for which the state is not an eigenstate.}
While this process lies outside of the regime of applicability of the low-energy effective lattice theory,
it enables direct comparisons between different methods of error suppression.
As discussed above, to include the impact of fixed quantum resources,
the total number of shots for each set of measurements is limited to $10^4$. 
At each time step, 
performed using leading-order Trotterized time evolution,
the energy in the electric field, 
the chiral condensate, $\Sigma$,
and the number of $e^+e^-$ pairs are computed.
This corresponds to performing ensemble measurements after each $n<N_{\rm max}$ Trotter steps.
The Trotter steps are chosen to furnish small systematic Trotter errors across the time interval of simulation, which depends on the selected parameters defining the Hamiltonian in  Eq.~(\ref{eq:HamLeq2LOG}). 
Given that the system with $L=2$ is too small to reflect continuum, infinite-volume physics, 
we use parameters $m=0.1$ and $g=0.3$
that result in a confinement length that is contained in near-term larger-scale simulations.   Simulations are performed over a time interval from $t=0$ to $t=T=30$, 
with a  Trotter step size of $\Delta T=0.5$. 

\begin{figure}[ht!]
    \centering
\includegraphics[width=0.95\linewidth,alt={FT rotations}]{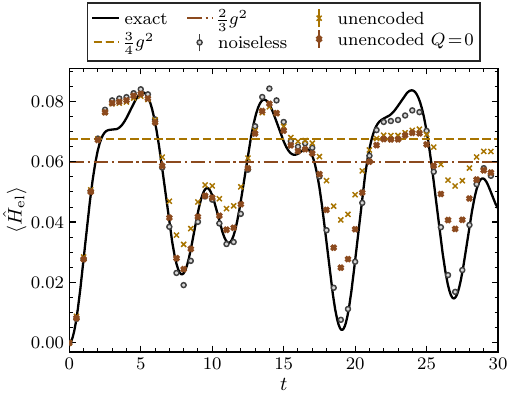}
\caption{
The total energy in the electric field in the $L=2$ Schwinger model as a function of time starting from the Neel state.  The solid black line corresponds to exact unitary evolution.
The black circles correspond to noiseless Trotterized time evolution.
The golden points correspond to noisy Trotterized time evolution with $p_2=0.003$ depolarizing noise without postselection into the physical space, while the brown points are the results obtained with postselection.
The golden dashed line corresponds to the value in the limit of complete depolarization of $g^2 L (L+1)/8=0.0675$
(determined by the matrix elements of the identity operator of the symmetric gauge term). 
The postselection to the $Q=0$ sector restricts this to  $g^2L(L+2)/12=0.06$ (brown dashed line).
The Trotterized simulations were performed using $10^4$ shots. 
}
    \label{fig:Leq2physELEC}
\end{figure}

\begin{figure}[ht!]
    \centering
\includegraphics[width=0.95\linewidth,alt={FT rotations}]{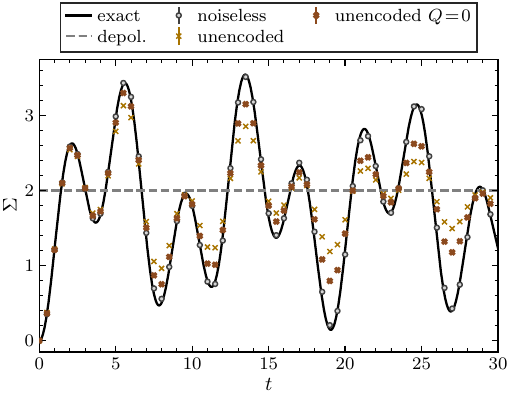}
\caption{
The chiral condensate in the $L=2$ Schwinger model as a function of time, starting from the Neel state.  The solid black line corresponds to exact unitary evolution.
The black circles correspond to noiseless Trotterized time evolution.
The golden points correspond to noisy Trotterized time evolution with $p_2=0.003$ depolarizing noise without postselection into the physical space, while the brown points are the results obtained with postselection.
The gray dashed line corresponds to the value in the limit of complete depolarization of $2$
(determined by the matrix elements of the identity operator).
The Trotterized simulations were performed using $10^4$ shots.
}
    \label{fig:Leq2physChiral}
\end{figure}

The different evolution scenarios we consider are (all performed using 4 qubits): 
\begin{enumerate}
\item Exact unitary time evolution without noise; 
\item  Trotterized time evolution without noise;
\item  Trotterized time evolution with noise ($p_2$) and without postselection into the codespace;
\item  Trotterized time evolution with noise and with postselection into the codespace; 
\end{enumerate}
Note that we have not included measurement errors in this analysis.
Figures~\ref{fig:Leq2physELEC} and \ref{fig:Leq2physChiral} show the 
time evolution of the total energy in the electric field and the chiral condensate, 
respectively, for these scenarios,
starting from the Neel state.

\begin{figure}[ht!]
    \centering
\includegraphics[width=0.95\linewidth,alt={FT rotations}]{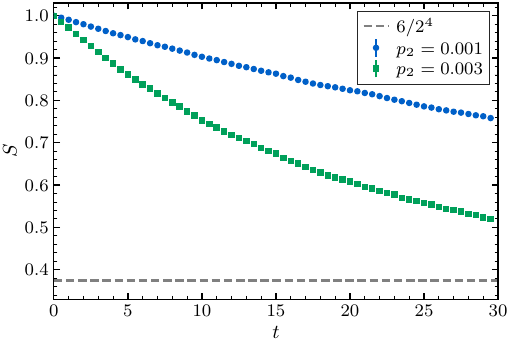}
\caption{
The postselection acceptance rate in the $L=2$ Schwinger model as a function of time starting from the Neel state.  The markers correspond to the acceptance rate $S$ for post-selected noisy Trotterized time evolution with $p_2 = 0.001$
(blue) and $0.003$ (green),
determined by the number of events in the $Q=0$-sector after the evolution normalized to the number of shots.
The Trotterized simulations were performed using $10^4$ shots.
The dashed gray line corresponds asymptotic acceptance rate, corresponding to the number of $Q=0$ states divided by the total number of four-qubit states, $6/16$.
}
    \label{fig:Leq2physAccept}
\end{figure}

Postselection into the physical space (after measurements) for this system is an effective way to significantly reduce the systematic errors induced by the noise.
This is equivalent to only keeping  probabilities for states in the $Q=0$ sector, 
renormalizing to the total number of events in these states,
and using these quantities to determine observables of interest.
Thus, this is a global Gauss's law check on the final state. 
While reducing the systematic errors, postselection increases the statistical errors because of the reduced size of the surviving ensemble, as shown in Fig.~\ref{fig:Leq2physAccept}.
The large-time asymptotic value of the postselection acceptance rate is determined by the fraction of $Q=0$ states compared with the total number of states in the Hilbert space.
This is because complete depolarization results in all states in the Hilbert space being equally probable.
The form of the decay is consistent with a single exponential shifted by a constant, from which estimates about the behavior of larger systems can be made.

\subsection{Encoding in $[[4,2,2]]^{\otimes 2}$}
\label{sec:SMLeq2422}
\noindent
We examine the potential of error detection in the Schwinger model using a small system with two spatial sites,
mapped to four staggered fermion sites and embedded in a $[[4,2,2]]^{\otimes 2}$ error-detection code. Each spatial site is embedded into a $[[4,2,2]]$ block.

\subsubsection{The Encoding and Quantum Circuits}
\label{sec:SMLeq2422embed}
\noindent
The $L=2$ Hamiltonian given in Eq.~(\ref{eq:HamLeq2LOG}),
written in terms of physical qubits, can be embedded into a 
$[[4,2,2]]^{\otimes 2}$ code with layout
$(t_1,1,2,b_1)(t_2,3,4,b_2)$, 
becoming
\begin{eqnarray}
\hat H_{kin} & = & 
{1\over 4}
\left[\ 
\hat X_1\hat X_2+\hat Y_1\hat Y_2
+\hat X_3\hat X_4+\hat Y_3\hat Y_4
\right.\nonumber\\
&&\left.
\qquad
+ \hat X_{t_1}\hat X_2\hat X_{t_2}\hat X_3
+ \hat X_{t_1} \hat Y_2\hat Z_{b_1}\ \hat X_{t_2}\hat Y_3 \hat Z_{b_2}
\ \right]
\nonumber\\
\hat H_{mass} & = & 
{m\over 2}
\left[\ 
4 \hat I 
-\hat Z_{1} \hat Z_{b_1}+\hat Z_{2} \hat Z_{b_1}
-\hat Z_{3} \hat Z_{b_2}+\hat Z_{4} \hat Z_{b_2}
\ \right]
\nonumber\\
\hat H_{gauge} & = & 
{g^2\over 2}
\left[
\hat Z_{1}\hat Z_{2}
+{1\over 2}
\hat Z_{1}\hat Z_{b_1}\hat Z_{3}\hat Z_{b_2}
+{1\over 2}
\hat Z_{2}\hat Z_{b_1}\hat Z_{3}\hat Z_{b_2}
\right. \nonumber\\
& & 
\left.
\ 
-\hat Z_{1}\hat Z_{b_1}
-{1\over 2} \hat Z_{2}\hat Z_{b_1}
-{1\over 2} \hat Z_{3}\hat Z_{b_2}
\ +\ 2 \hat I 
\ \right]
\ .
\label{eq:HamLeq2phys2}
\end{eqnarray}
The 6 codewords for the allowed states ($Q=0$), formed from those of a single 
$[[4,2,2]]$ block in Eq.~(\ref{eq:422codewords}),
are 
\begin{align}
    |\_\ \_\ \_\ \_\ \rangle & = |\bar{0} \bar{1}\rangle\otimes |\bar{0} \bar{1}\rangle\ ,
    \nonumber\\
     |e^+e^-\_\ \_\rangle & = |\bar{1} \bar{0}\rangle\otimes |\bar{0} \bar{1}\rangle\ ,
    \nonumber\\
    |\ \_\ \_\ e^+e^-\rangle & = |\bar{0} \bar{1}\rangle\otimes |\bar{1} \bar{0}\rangle\ ,
    \nonumber\\
    |\ \_\ e^-e^+\ \_\ \rangle & = |\bar{0} \bar{0}\rangle\otimes |\bar{1} \bar{1}\rangle\ ,
    \nonumber\\
    |e^+\ \_\ \_\ e^-\rangle & = |\bar{1} \bar{1}\rangle\otimes |\bar{0} \bar{0}\rangle\ ,
    \nonumber\\
    |e^+e^-e^+e^-\rangle & = |\bar{1} \bar{0}\rangle\otimes |\bar{1} \bar{0}\rangle
    \ ,
\end{align}
where a `` $\_$ '' denotes an unoccupied staggered site. 
See App.~\ref{app:422sq} for more details.
These states are eigenstates of the stabilizers given in Eq.~(\ref{422stabs}) acting on both sites, 
and, as such, the Hamiltonian commutes with the stabilizers, as desired.

\begin{figure}[ht!]
    \centering
\includegraphics[width=0.55\linewidth]{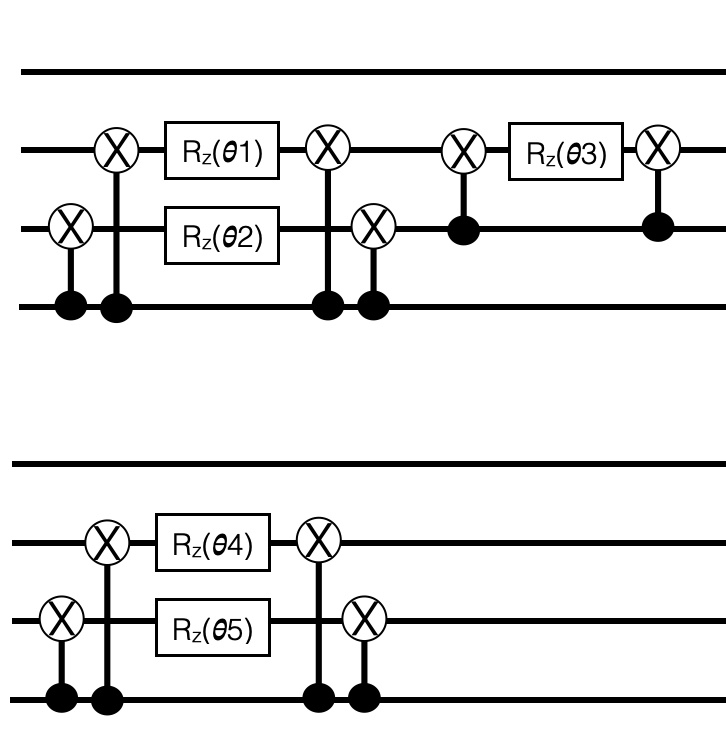}
\caption{
A  non-FT quantum circuit 
implementing time evolution of the mass and gauge terms in
Eq.~(\ref{eq:HamLeq2phys2}) for the  $[[4,2,2]]^{\otimes 2}$ with $L=2$ spatial sites.
The angles are :
$\theta_1=-t(m+\frac{g^2}{2})$,
$\theta_2=+t m$,
$\theta_3=+t\frac{g^2}{2}$,
$\theta_4=-t m$,
$\theta_5=+t(m+\frac{g^2}{2})$.
}
    \label{fig:422sqUmg}
\end{figure}

\begin{figure}[ht!]
    \centering
\includegraphics[width=0.65\linewidth]{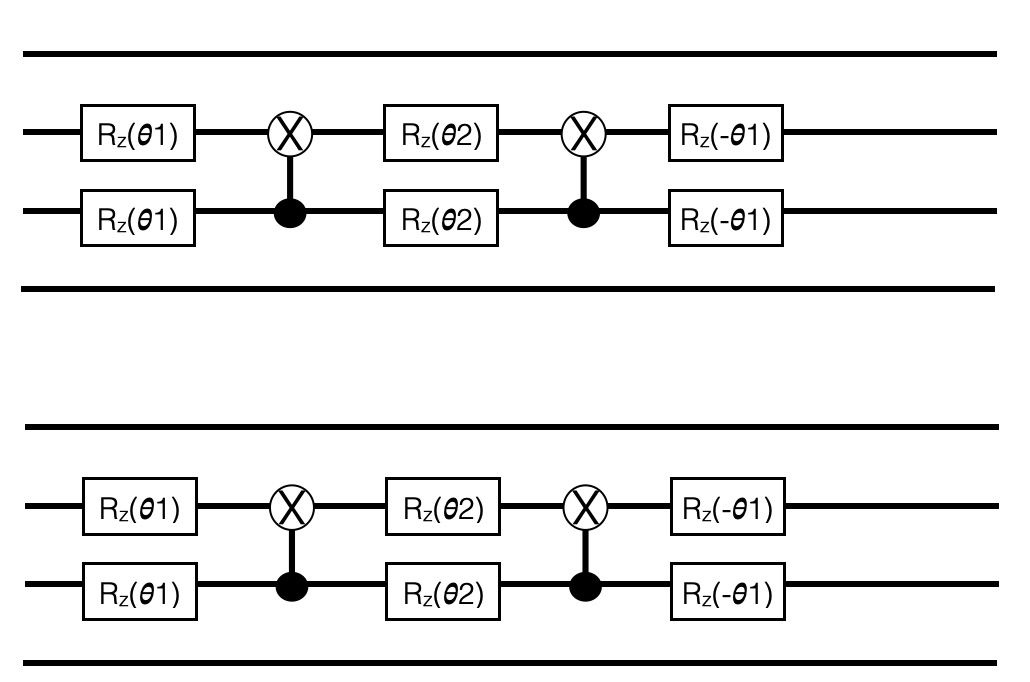}

\vspace{2em}
\includegraphics[width=0.65\linewidth]{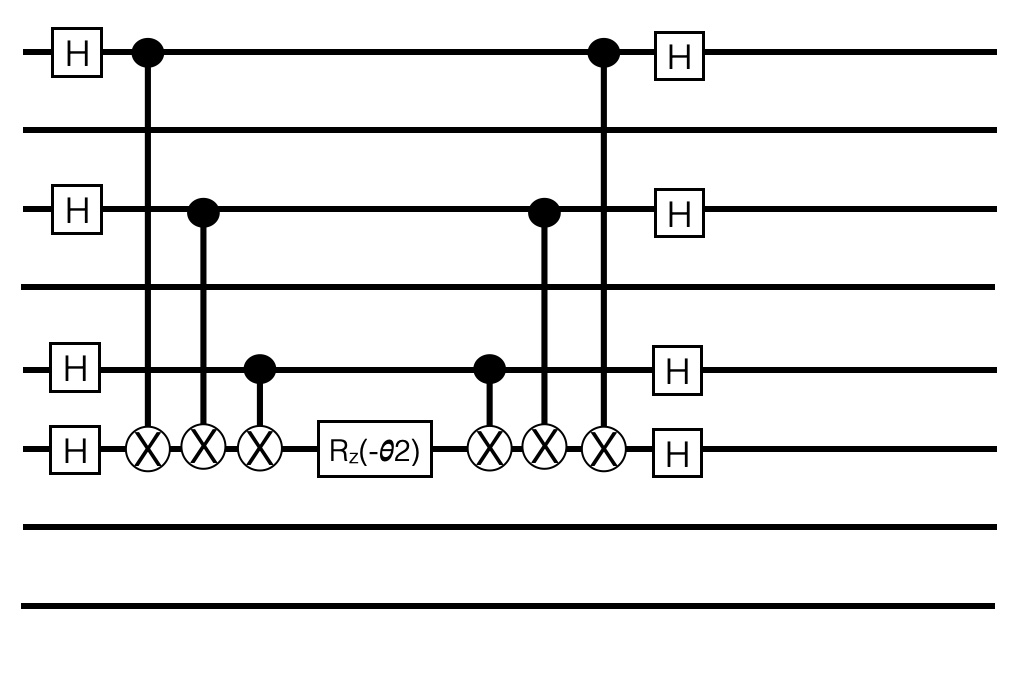}

\vspace{2em}
\includegraphics[width=0.65\linewidth]{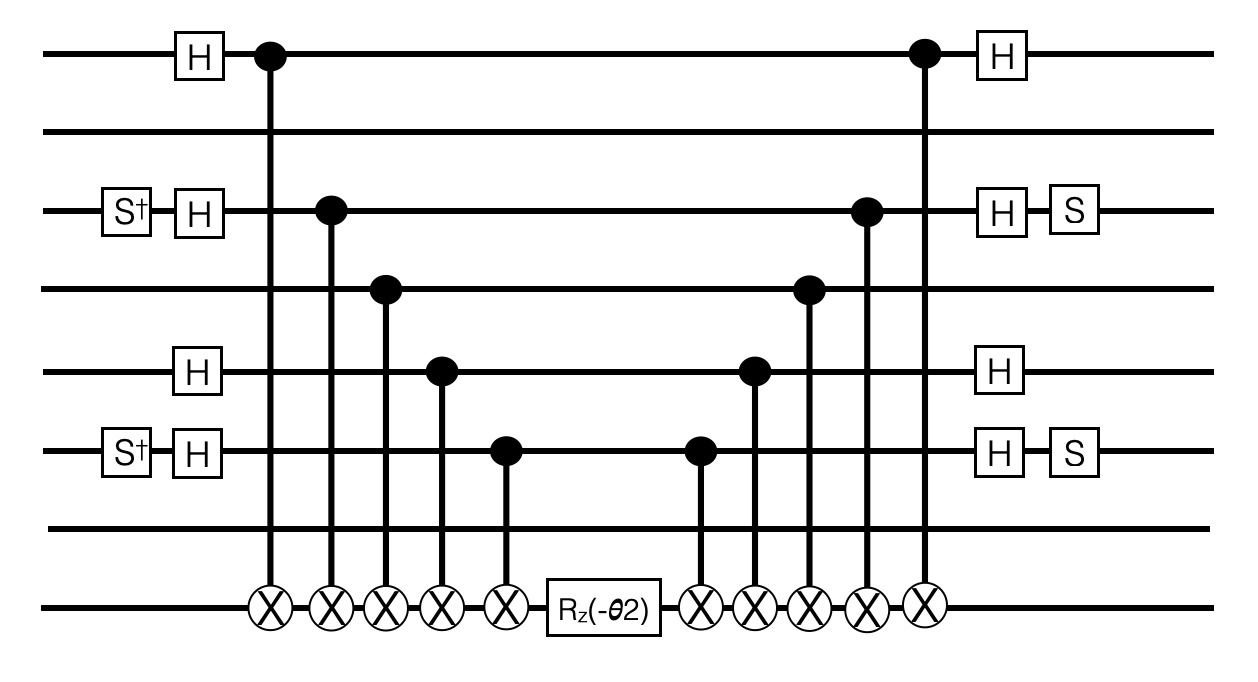}
\caption{
Quantum circuits implementing Trotterized time evolution of the kinetic term in
Eq.~(\ref{eq:HamLeq2phys2}) for the  $[[4,2,2]]^{\otimes 2}$  code with $L=2$ spatial sites.
The angles are :
$\theta_1=\frac{\pi}{2}$ and 
$\theta_2= -\frac{t}{2}$.
The upper circuit implements
$e^{-i t \frac{1}{4}
\left[\ 
\hat X_{1}\hat X_{2} + \hat Y_{1}\hat Y_{2}
+
\hat X_{5}\hat X_{6} + \hat Y_{5}\hat Y_{6}
\right] }$,
the center circuit implements
$e^{-i t \frac{1}{4} 
\hat X_{0}\hat X_{2} \hat X_{4}\hat X_{5} }$,
while the third implements
$e^{-i t \frac{1}{4} 
\hat X_{0}\hat Y_{2} \hat Z_{3}\hat X_{4} \hat Y_{5}\hat Z_{7}  }$. 
The qubit ordering from top to bottom is: 
$t_1,1,2,b_1,t_2,3,4,b_2 \equiv q_0,q_1,q_2,q_3,q_4,q_5,q_6,q_7$,
and the ancilla used for error detection are not shown.
}
    \label{fig:422sqUk}
\end{figure}

As mentioned above, the circuit implementing time-evolution cannot be made FT 
when both logical qubits per code block are used in the mapping. 
If only one logical qubit of each block is used, transversal operations provide
a fault-tolerant Clifford gate set~\cite{Gottesman2016,Linke2017,Vuillot2018,HarperFlammia2019},
which can in principle be lifted to universal FT circuits through verified
magic-state injection~\cite{Bravyi2005,Knill2005}, albeit with a
substantial overhead in both qubits and post-selection.
In mapping two spatial sites to the $[[4,2,2]]^{\otimes 2}$ code, individually addressed
logical operations are no longer transversal~\cite{Gottesman2016,Eastin2009},
and the arbitrary-angle rotations required for the time evolution are
necessarily non-FT.
Instead, this work uses non-FT operations to implement time evolution ({\it e.g.}, according to the Hamiltonian in Eq.~(\ref{eq:HamLeq2phys2})) 
by explicitly constructing the corresponding logical rotations in terms of physical gates.
However, Trotterization of the evolution operator can be optimized with regard to the number of 
(coherent) errors escaping into the codeword space.
The time evolution from the mass term and gauge term can be implemented together, 
as shown in Fig.~\ref{fig:422sqUmg},
which can also be implemented parallel across the spatial sites.
The time evolution induced by the kinetic terms can be implemented in three layers, 
as shown in Fig.~\ref{fig:422sqUk},
with the operations confined to a single site performed in parallel.

The encoded simulations proceed in several steps:
\begin{enumerate}
    \item FT preparation of the Neel state, using the circuits in Fig.~\ref{fig:422NeelFT}.
    \item Non-FT time evolution using the Trotter steps shown in Figs.~\ref{fig:422sqUmg} and~\ref{fig:422sqUk}.
    \item Sparse error detection using the FT stabilizer circuits in Fig.~\ref{fig:422NeelFT}.
    \item Final measurements implemented on the states produced by the non-FT circuit shown in Fig.~\ref{fig:422Final}.
\end{enumerate}

\subsubsection{Results from Classical Simulations}
\label{sec:SMLeq2422results}
\begin{figure}[ht!]
    \centering
\includegraphics[width=0.95\linewidth,alt={FT rotations}]{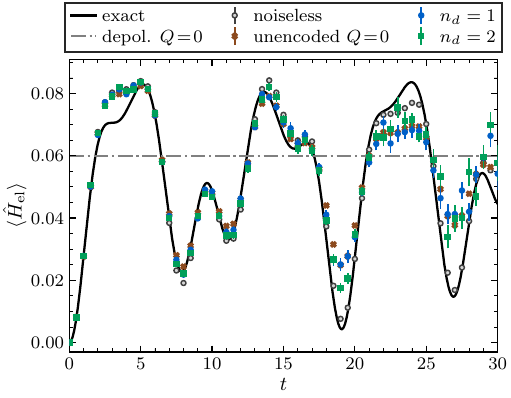}
\caption{
The total energy in the electric field in the $L=2$ Schwinger model 
embedded into the  $[[4,2,2]]^{\otimes 2}$ code
as a function of time starting from the Neel state.  
The solid black line corresponds to exact unitary evolution.
The black circles correspond to noiseless Trotterized time evolution.
The brown points correspond to noisy Trotterized time evolution using physical qubits with $p_2=0.003$ depolarizing noise with postselection. 
The blue points correspond to a 12-qubit simulation using a single layer of FT error detection at the mid-point of the evolution,
while the green points correspond to error detection at the $1/3$ and $2/3$ points during the evolution.
The Trotterized simulations were performed using $10^4$ shots.
}    
\label{fig:Leq2422422ELEC}
\end{figure}

\begin{figure}[ht!]
    \centering
\includegraphics[width=0.95\linewidth,alt={FT rotations}]{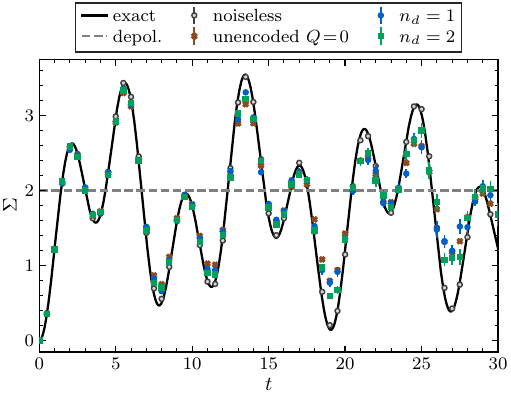}
\caption{
The chiral condensate in the $L=2$ Schwinger model 
    embedded into  the $[[4,2,2]]^{\otimes 2}$ code
as a function of time starting from the Neel state.  The solid black line corresponds to exact unitary evolution.
The black circles correspond to noiseless Trotterized time evolution.
The green points correspond to noisy Trotterized time evolution using physical qubits with $p_2=0.003$ depolarizing noise with postselection. 
The blue points correspond to a 12-qubit simulation using a single layer of FT error detection at the mid-point of the evolution,
while the light-red points correspond to error detection at the $1/3$ and $2/3$ points during the evolution.
The Trotterized simulations were performed using $10^4$ shots.}
    \label{fig:Leq2422422Chiral}
\end{figure}

\noindent
In this section, 
we consider noisy Trotterized time evolution with postselection  
with sets of FT error detection at the 
$1/(n_d+1), 2/(n_d+1), ..., n_d/(n_d+1) $
fractions of the evolution
for a small number of layers of error detection $n_d\ge 1$.
The results are shown in Figs.~\ref{fig:Leq2422422ELEC} and 
\ref{fig:Leq2422422Chiral},
for a two-qubit error rate of $p_2=0.003$. 
One observes that the single-round of error-detection reduces the systematic error in both the 
energy in the gauge field and the chiral condensate for each evaluation time.
However, after $t\sim 20$, the surviving sample size is substantially reduced from $10^4$,
resulting in noticeable statistical error bars.
Therefore, for this small system,  tripling of the number of active physical qubits leads to an improvement in the overall post-diction of the considered observables.

\begin{figure}[ht!]
    \centering
\includegraphics[width=0.95\linewidth,alt={FT rotations}]{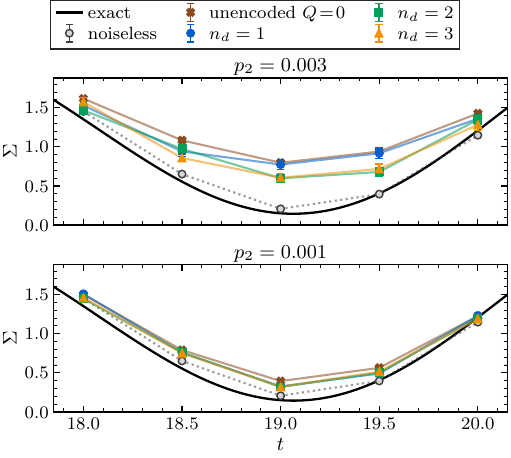}
\caption{
The chiral condensate in the $L=2$ Schwinger model 
embedded into the $[[4,2,2]]^{\otimes 2}$ code
as a function of time starting from the Neel state.  
The solid black line corresponds to exact unitary evolution.
The black circles correspond to noiseless Trotterized time evolution.
The brown points correspond to noisy Trotterized time evolution using physical qubits with 
with $p_2=0.003$ (upper panel)
and $p_2=0.001$  (lower panel)
depolarizing noise with postselection. 
The dark-blue points correspond to a 12-qubit simulation using a single layer of FT error detection at the mid-point of the evolution,
the green points correspond to error detection at the $1/3$ and $2/3$ points during the evolution,
and the orange points correspond to error detection at the $1/4$, $1/2$ and $3/4$ points during the evolution.
The Trotterized simulations were performed using  $10^4$ shots.
}
    \label{fig:Leq2422422Chiralinset}
\end{figure}

Zooming in on the convergence behavior in a given time interval,
Fig.~\ref{fig:Leq2422422Chiralinset} shows the chiral condensate around $t=19$ for $p_2=0.003$. 
Including one layer of error detection at the mid-point, and two layers at 
the $1/3, 2/3$ points both improve upon the prediction
obtained from unencoded simulations with postselection, 
while including another layer, (corresponding to syndrome measurements at $1/4, 1/2$ and $3/4$) does not noticeably improve upon the  $1/3, 2/3$ results. 
Further, the results have not converged to their noiseless values.

\begin{figure}[ht!]
    \centering
\includegraphics[width=0.95\linewidth,alt={FT rotations}]{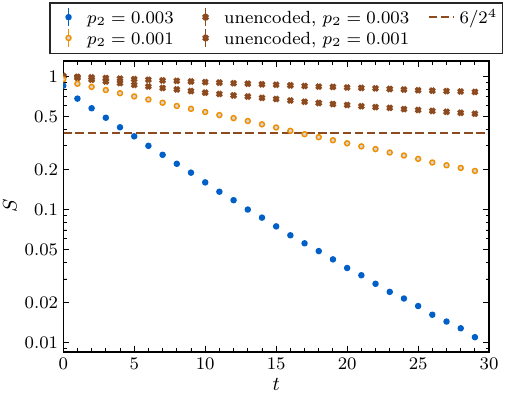}
\caption{
The postselection acceptance rate in the $L=2$ Schwinger model 
embedded into  the $[[4,2,2]]^{\otimes 2}$ code
as a function of time starting from the Neel state with a single layer of stabilizer projective measurements at the mid-point of the evolution.
The orange points correspond to post-selected noisy Trotterized time evolution with 
$p_2 = 0.001$ and the blue points to $p_2=0.003$,
determined by the number of events in the $Q=0$ sector after the evolution normalized to the number of shots. The brown curves are the corresponding unencoded noisy simulations using Gauss law postselection only.
The Trotterized simulations were performed using $10^4$ shots.
}

    \label{fig:Leq2422422LOGAccept}
\end{figure}
The rejection rate after some time is shown in Fig.~\ref{fig:Leq2422422LOGAccept}.
This is to be contrasted with the postselection-only acceptance rate of the evolution of physical qubits,
shown in Fig.~\ref{fig:Leq2physAccept}.
It is unsurprising that the rejection rate for 12 qubits with error detection is substantially greater than that for 4 qubits with only postselection.

\subsubsection{Rejection Rates and Scaling}
\label{sec:SMLeq2422scaling}
\noindent
The results we have obtained for rejection rates as a function of time permit an 
evaluation of scaling
with system size and error rates, and a consideration of potential
time intervals for physics extraction.

First, it is no surprise that the acceptance rate scales exponentially 
with the product of two-qubit error rate and time interval with sparse error detection,
$\sim \exp(-p_2 \ \ \# \text{gates})$.
Therefore, for simulations of observables (in fixed volume) over a given time interval, 
improving two-qubit gates can reduce statistical errors substantially, with a globally reduced systematic error.

\begin{figure}[ht!]
    \centering
\includegraphics[width=0.95\linewidth,alt={FT rotations}]{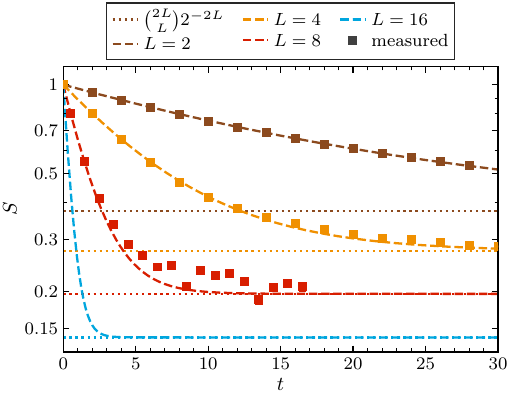}
\caption{
The postselection acceptance rate $S$ in the unencoded $L=2, 4, 8, 16$ Schwinger model 
as a function of time starting from the Neel state with only postselection. 
The filled markers correspond to post-selected noisy Trotterized time evolution with 
$p_2 = 0.003$,
determined by the number of events in the $Q=0$ sector after the evolution normalized to the number of shots. The dashed curves are $S_L(t) = f_L + (1-f_L) e^{-\beta_L\, t}$; dotted lines are the exact charge-sector floors $f_L = \binom{2L}{L}/2^{2L}$. Using the $L=2$ data, we fitted $\beta_2 = 0.050$ which determines the remaining decay constants through $\beta_L = \beta_2 (g_L/g_2)(1-f_2)/(1-f_L)$. The factor $g_L = 2L^2 + 0.0955(8L-4)$ is the weighted noisy-gate count of one Trotter step. The data for $L=4$ (yellow) and $L=8$ (red) confirm the prediction.
}
    \label{fig:Leq422422predict003}
\end{figure}

While imperfect, an approximate scaling of the rejection rates as a function of system size can be determined.
Assuming that the number of entangling gates scales with system size 
(which is only approximately true, modified by the confinement scale and boundary effects),
the predicted rejection rates for increasing system sizes to $L=4, 8, 16$ respectively, are shown in Fig.~\ref{fig:Leq422422predict003} for $p_2=0.003 $.
Fig.~\ref{fig:Leq422422predict003} shows the maximum time over which evolution can be fruitfully performed before all quantum states in the codespace are equally probable, 
with the potential to predict observables via error-mitigation techniques limited to $t\sim 10$ 
(20 Trotter steps)
for 
$L=16$  with postselection alone.   
As the argument of the exponential depends on $p_2 t$, this time interval can be extended by a reduction in $p_2$.  The overall rejection rate is such that an increase in statistics of less than two orders of magnitude ($10^6$ shots) is sufficient to recover  a statistical uncertainty
in the $L=16$ system comparable to that in the $L=2$ system 
(modulo the caveats related to circuit depth in larger systems).
It is interesting to note that even with a large increase in the number of shots (above $10^4$), 
the Hilbert space is not large enough for postselection to be effective because the maximally-mixed state
provides a high probability of being in the $Q=0$ sector.

\begin{figure}[ht!]
    \centering
\includegraphics[width=0.95\linewidth,alt={FT rotations}]{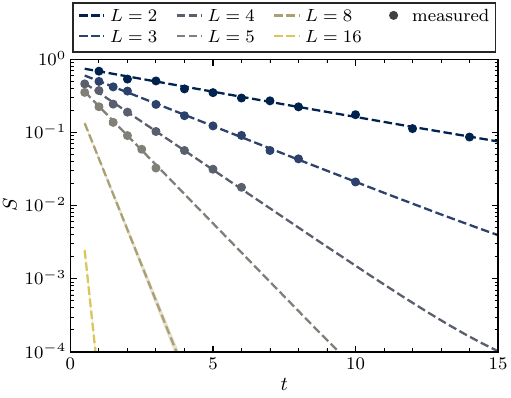}
\caption{
The postselection acceptance rate $S$ in the $L=2$ Schwinger model 
embedded into   the $[[4,2,2]]^{\otimes 2}$ code
as a function of time starting from the Neel state with postselection and error detection at the mid-point with  $p_2 = 0.003$,
determined by the number of events in the $Q=0$ sector after the evolution normalized to the number of shots.
The Trotterized simulations were performed using up to $4\cdot 10^4$ shots.
The dashed lines are $S_L(t) = f_L +(1-f_L) e^{-\lambda g_{\rm tot}^L(t)}$, where $g_{\rm tot}^L(t)$ 
 is the weighted noisy-gate count of the built circuit (in analogy with Fig.~\ref{fig:Leq422422predict003}, see main text for functional form).  The floors $f_L$ are fitted (order $10^{-3}$ for $L=2,3$, $10^{-4}$ for $L=4$) and $\lambda = 2.56\cdot 10^{-3}\sim p_2$.
}

    \label{fig:Leq422422Q12predict003}
\end{figure}

Fig.~\ref{fig:Leq422422Q12predict003} shows the corresponding acceptance probability as a function of time in the
$[[4,2,2]]^{\otimes 2}$ embedding with $p_2=0.003$.~\footnote{The data in this figure used {\tt qiskit} {\tt AerSimulator} with \texttt{optimization\_level=0}.} The functional form used for the dashed lines consists of the gate count $g_{\rm tot}^L(t)=g_{\rm fixed}(L,n_d) + g_{\rm step}(L) k$, where the first part is the gate count of encoding, decoding and syndrome measurements and the second part the gate count per Trotter step ($t=k\, dt)$.~\footnote{As explained in more detail in Section~\ref{sec:SMice}: $g_{\rm fixed}  = 6L(5 + 2 n_d) + 0.0955 \, 4L(3 + n_d);
    g_{\rm step} = 6L^2 + 6L - 6 + 0.0955\, (14L - 10)$. }  The  larger Hilbert space significantly reduces 
the maximally-mixed state probability of being in the $Q=0$ sector, and being selected during postselection.
This means that a very high-statistics set of measurements, $\sim 10^{10}$, could provide an improved set of predictions out to times $t\sim 15$ for $L=8$, but a much larger set would be required for comparable errors at $L=16$.

In order to have a high-statistics result ($\sim 10^8$ shots) 
with reduced systematic errors 
for $t\leq 30$ (using a Trotter step size of $\Delta t=0.5$)
through sparse error detection for systems with $N=100$ logical qubits, 
we find that the two-qubit error rate (scaling the above results) should satisfy
$p_2 \leq 10^{-4}$.

\subsubsection{Noisy Ancillas}
\label{sec:SMLeq2422DA}
\noindent
Previously, it has been suggested that a heterogeneous distribution of qubit fidelities could be optimized in simulations.
This is in standard practice for simulations using available machines in NISQ-era simulations, but has been suggested as a potentially fruitful avenue to explore for systems of logical qubits~\cite{Klco:2021jxl}.
Since ancilla qubits are only used for sparsely-spaced syndrome measurements, a natural simplification is to dedicate the noisy device qubits to these ancillas.
We have studied the impact of using ancilla qubits that have larger error rates 
than data qubits.  
With fidelity asymmetries 
among the ancilla and  data qubits in the $[[4,2,2]]$ blocks,
we find that, for these circuits implemented, 
the error rates in the ancilla qubits
can be up to three times as large as in the 
$[[4,2,2]]$ blocks before noticeable degradation in the quality of the simulation results is observed.

\subsection{Encoding into $[[6,4,2]]$}
\label{sec:SMLeq2642}
\noindent
The $L=2$ system above can also be embedded into a single block of the $[[6,4,2]]$ code.
The Hamiltonian given in Eq.~(\ref{eq:HamLeq2LOG}),
written in terms of physical qubits, can be embedded into a 
$[[6,4,2]]$ code with layout
$(t,1,2,3,4,b)$, 
becoming
\begin{eqnarray}
\hat H_{kin} & = & 
{1\over 4}
\left[\ 
\hat X_1\hat X_2+\hat Y_1\hat Y_2
+\hat X_3\hat X_4+\hat Y_3\hat Y_4
\right.\nonumber\\
&&\left.
\qquad
+  \hat X_2 \hat X_3
+  \hat Y_2 \hat Y_3
\ \right]
\nonumber\\
\hat H_{mass} & = & 
{m\over 2}
\left[\ 
4\ \hat I 
-\hat Z_{1} \hat Z_{b}+\hat Z_{2} \hat Z_{b}
-\hat Z_{3} \hat Z_{b}+\hat Z_{4} \hat Z_{b}
\ \right]
\nonumber\\
\hat H_{gauge} & = & 
{g^2\over 2}
\left[
2 \hat I\ +\ 
\hat Z_{1}\hat Z_{2}
+{1\over 2}
\hat Z_{1} \hat Z_{3} 
+{1\over 2}
\hat Z_{2} \hat Z_{3} 
\right. \nonumber\\
& & 
\left.
\qquad
-\hat Z_{1}\hat Z_{b}
-{1\over 2} \hat Z_{2}\hat Z_{b}
-{1\over 2} \hat Z_{3}\hat Z_{b}
\ \right]
\ .
\label{eq:HamLeq2phys}
\end{eqnarray}
The logical states are found by acting with the logical Pauli operators on the 
$|\bar{0} \bar{0}\bar{0} \bar{0}\rangle$ state,
{\it e.g.}, 
\begin{eqnarray}
|\bar{0} \bar{0} \bar{0} \bar{0}\rangle & = & {1\over\sqrt{2}} 
\left[\ 
|000000\rangle + |111111\rangle
\ \right]
\ \ ,\ \ 
\nonumber\\
|\bar{1} \bar{0}\bar{0} \bar{0}\rangle & = &  {1\over\sqrt{2}} 
\left[\ 
|110000\rangle + |001111\rangle
\ \right]
\ \ ,
\end{eqnarray}
and the stabilizers for these states are 
\begin{eqnarray}
\hat S_z & = & \hat Z^{\otimes 6}
\ \ ,\ \ 
\hat S_x \ =\  \hat X^{\otimes 6} 
\ .
\end{eqnarray}
The circuits required for implementing Trotterized time evolution are straightforward to determine (from those used in previous sections), as are the constructions of the codewords and observables 
(details of which can be found in App.~\ref{app:642}).
The main difference between the 
$[[4,2,2]]^{\otimes 2}$ (using 8+4=12 qubits)
and the $[[6,4,2]]$ code (using 6+2=8 qubits)
embeddings is the fidelity of the stabilizer checks.  
While leading-order errors in the 2 logical qubits 
in each of the $[[4,2,2]]$ blocks are detected,
they are detected only on the 4 logical qubits in the single $[[6,4,2]]$ block.

\begin{figure}[ht!]
    \centering
\includegraphics[width=0.95\linewidth,alt={FT rotations}]{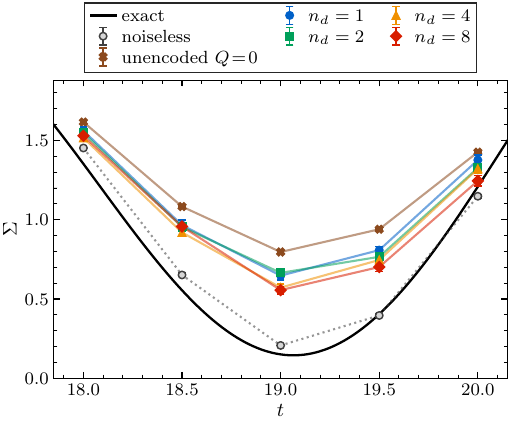}
\caption{
The chiral condensate in the $L=2$ Schwinger model 
embedded into the $[[6,4,2]]$ code
as a function of time starting from the Neel state.  
The solid black line corresponds to exact unitary evolution.
The black circles correspond to noiseless Trotterized time evolution.
The brown points correspond to noisy Trotterized time evolution using physical qubits with 
$p_2=0.003$
depolarizing noise with postselection. 
The dark-blue points correspond to a (6+2) 8-qubit simulation using a single layer of 
$\hat Z^6$ and $\hat X^6$
FT error detection at the mid-point of the evolution,
the green points correspond to error detection at the $1/3$ and $2/3$ points during the evolution,
the orange points correspond to error detection at the $1/5$, $2/5$, $3/5$ and $4/5$ points during the evolution,
and 
the red points correspond to error detection at the  points $1/9$, $2/9$, $3/9$, $4/9$, $5/9$, $6/9$, $7/9$ and $8/9$ during the evolution.
The Trotterized simulations were performed using  $10^4$ shots.
}
    \label{fig:Leq2642Chiralinset}
\end{figure}
Fig.~\ref{fig:Leq2642Chiralinset} displays the behavior of the chiral condensate over a 
small time interval as a function of the number of stabilizer layers of $\hat Z^6$ and $\hat X^6$. 
Comparing these results with those displayed in Fig.~\ref{fig:Leq2422422Chiralinset} 
shows that the coarser error detection for this system is comparably effective in reducing the systematic errors.
This suggests that these expectation values are approximately independent of the density of detectable errors at a fixed distance.

\begin{figure}[ht!]
    \centering
\includegraphics[width=0.95\linewidth,alt={FT rotations}]{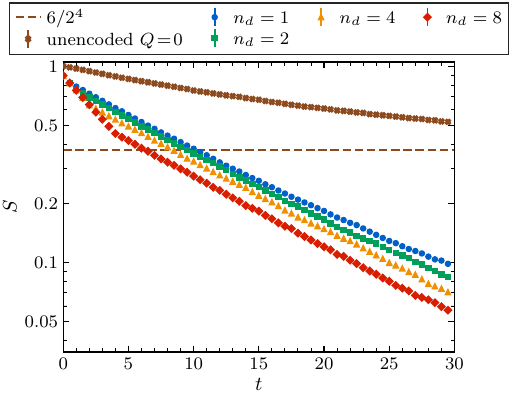}
\caption{
The  postselection acceptance rate in the $L=2$ Schwinger model 
embedded into the $[[6,4,2]]$ code
as a function of time starting from the Neel state with a selection in the number of 
layers of FT error detection during the evolution, followed by postselection onto the $Q=0$ sector.
The Trotterized simulations were performed using $10^5$ shots.
The dashed orange line corresponds asymptotic acceptance rate, corresponding to the number of $Q=0$ states divided by the total number of four-qubit states, $\frac{6}{2^{4}}$.}
    \label{fig:Leq2642LOGAccept}
\end{figure}
Fig.~\ref{fig:Leq2642LOGAccept} shows the acceptance rate for different numbers of uniformly-spaced stabilizer layers, followed by postselection, associated with the results shown in Fig.~\ref{fig:Leq2642Chiralinset}.
Increasing the number  of layers reduces the acceptance rate, as expected, but less  significantly than increasing the number of layers in the $[[4,2,2]]^{\otimes 2}$ encoding.  
This is consistent with the discussion above related to the relatively lower stabilizer density per spatial lattice site.
Some of the curves exhibit changes of slope.  This is due to saturation of stabilizer layers at small numbers of Trotter steps (which are limited to no more than one per step).

\section{The $L=4$ Schwinger Model}
\label{sec:SML4}
\noindent
Doubling the spatial extent of the lattice more than 
doubles the system error rate for the same gate error rate.~\footnote{Without a confinement-guided truncation of the gauge contribution to the Hamiltonian in axial gauge, 
the number of gates grows ${\cal O}(L^2)$, compared with ${\cal O}(\xi L)$ for large systems, 
where $\xi$ is the confinement scale~\cite{Farrell:2023fgd,Farrell:2024fit}.
For small system sizes considered here, this truncation is not applicable.}
The $L=4$ system, mapped to 8 logical qubits, provides a (small) sandbox to begin studying the relative effectiveness of error detection in subregions of the spatial lattice, using both homogeneous and heterogeneous partitions. 
We compare the resulting uncertainties in the chiral condensate and energy in the gauge field as function of time for 
$[[10,8,2]]$, $[[8,6,2]]\otimes [[4,2,2]]$, $[[6,4,2]]^{\otimes 2}$, $[[6,4,2]]\otimes [[4,2,2]]^{\otimes 2}$ and 
$[[4,2,2]]^{\otimes 4}$
encodings with equal resource constraints.

\subsection{Encoding into $[[10,8,2]]$}
\label{sec:SM1082}
\noindent
The quantum circuits used for state preparation, time evolution, final-state consolidation and stabilizer measurements for embedding the $L=4$ Schwinger model in the [[10,8,2]] code are analogous to those discussed in previous sections.~\footnote{The stabilizers $\hat X^8$ and $\hat Z^8$ are implemented using the two-ancilla fault-tolerant circuits used in previous sections.}
The terms in the Hamiltonian describing the $L=8$ system, written using logical operators, are
\begin{align}
\bar{H}_{m} & = \frac{m}{2}\Big[
-\widehat{\overline{Z}}_1 + \widehat{\overline{Z}}_2 - \widehat{\overline{Z}}_3 + \widehat{\overline{Z}}_4
\nonumber\\
& \qquad \qquad
- \widehat{\overline{Z}}_5 + \widehat{\overline{Z}}_6 - \widehat{\overline{Z}}_7 + \widehat{\overline{Z}}_8 + 8\,\hat{I}\Big]
\ ,
\nonumber\\
\bar{H}_{\mathrm{kin}} & = \frac{1}{4}\sum_{i=1}^{7}
      \big(\widehat{\overline{X}}_i\widehat{\overline{X}}_{i+1} + \widehat{\overline{Y}}_i\widehat{\overline{Y}}_{i+1}\big)
      \ ,
\nonumber\\
      \bar{H}_{g} & = \frac{g^{2}}{2}\Bigg[\; 5\,\hat{I}
\;-\;\widehat{\overline{Z}}_1 - \tfrac{1}{2}\widehat{\overline{Z}}_2 - \tfrac{1}{2}\widehat{\overline{Z}}_3
\;+\;\tfrac{1}{2}\widehat{\overline{Z}}_6 + \tfrac{1}{2}\widehat{\overline{Z}}_7 
\nonumber\\
& \qquad \qquad
+ \widehat{\overline{Z}}_8
\;+\;\tfrac{3}{2}\,\widehat{\overline{Z}}_1\widehat{\overline{Z}}_2
   + \widehat{\overline{Z}}_1\widehat{\overline{Z}}_3
   + \tfrac{1}{2}\widehat{\overline{Z}}_1\widehat{\overline{Z}}_4
\nonumber\\
& \qquad \qquad
   + \widehat{\overline{Z}}_2\widehat{\overline{Z}}_3
   + \tfrac{1}{2}\widehat{\overline{Z}}_2\widehat{\overline{Z}}_4
   + \tfrac{1}{2}\widehat{\overline{Z}}_3\widehat{\overline{Z}}_4
\;+\;\tfrac{1}{2}\,\widehat{\overline{Z}}_6\widehat{\overline{Z}}_7
\nonumber\\
& \qquad \qquad
   + \tfrac{1}{2}\widehat{\overline{Z}}_6\widehat{\overline{Z}}_8
   + \widehat{\overline{Z}}_7\widehat{\overline{Z}}_8
\;\Bigg]
\ ,
\label{eq:L8HamiLog}
\end{align}
Where $i$ labels the lattice site, $i=1,2,...,8$.
The symmetric representation of the energy in the gauge field has been used.
As the $L=4$  system is twice the spatial volume of the $L=2$ system, we expect the error rate to be at least twice as large.  
The increased gate depth associated with the gauge interactions means that it will not simply be a factor of two larger.

\begin{figure}[ht!]
    \centering
\includegraphics[width=0.95\linewidth,alt={FT rotations}]{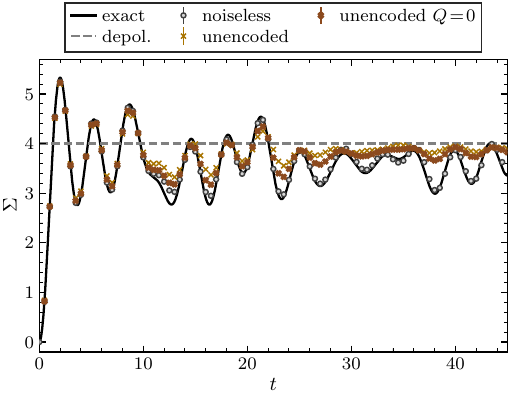}
\caption{
The chiral condensate in the unencoded $L=4$ Schwinger model 
as a function of time starting from the Neel state.  
The solid black line corresponds to exact unitary evolution.
The black circles correspond to noiseless Trotterized time evolution.
The golden points correspond to noisy Trotterized time evolution using 8 physical qubits with $p_2=0.001$ depolarizing noise without postselection, while the brown points correspond to the inclusion of postselection. 
The dashed gray line corresponds to the completely depolarized limit of 4.
The Trotterized simulations were performed using  $1.2\times 10^6$ shots.
}
    \label{fig:Leq41082Chiral}
\end{figure}

\begin{figure}[ht!]
    \centering
\includegraphics[width=0.95\linewidth,alt={FT rotations}]{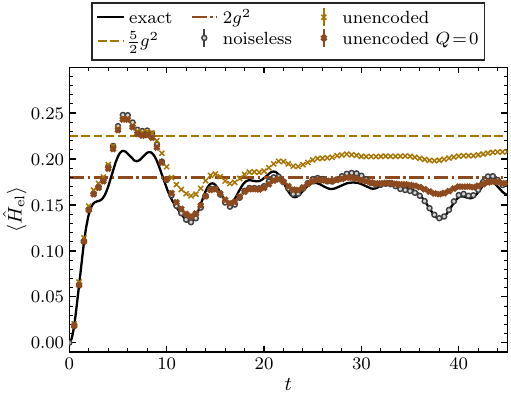}
\caption{
The total energy in the gauge field in the unencoded $L=4$ Schwinger model 
as a function of time starting from the Neel state.  
The solid black line corresponds to exact unitary evolution.
The black circles correspond to noiseless Trotterized time evolution.
The golden points correspond to noisy Trotterized time evolution using 8 physical qubits with $p_2=0.001$ depolarizing noise without postselection, while the brown points correspond to the inclusion of postselection. 
The golden (brown) dashed line corresponds to the depolarized limit without (with) postselection of 0.225 (0.18).
The Trotterized simulations were performed using  $1.2\times 10^6$ shots.
}
    \label{fig:Leq41082Gauge}
\end{figure}

Computing the same observables as above,
the chiral condensates and total energy in the gauge field are displayed in Fig.~\ref{fig:Leq41082Chiral} and 
Fig.~\ref{fig:Leq41082Gauge}, respectively.
Because of the volume of this system, we  use $p_2=0.001$ 
and  $1.2\times10^6$ shots.
The oscillatory behavior of both quantities is smaller than for $L=2$.   
As such, variations between different numbers of layers of error detection will be smaller in magnitude, and hence the need for increased statistical precision.

With this system, we study the scaling of the reduction in 
 systematic error of the chiral condensate
resulting from one-qubit and two-qubit errors as a function of time and of 
the number of error-detection layers.
\begin{figure}[ht!]
    \centering
\includegraphics[width=0.95\linewidth,alt={FT rotations}]{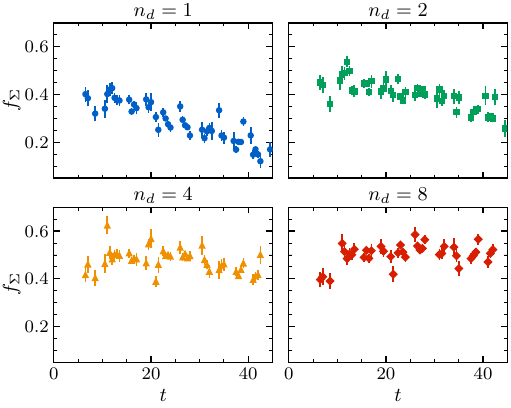}
\caption{
The recovered fraction of errors in the chiral condensate as a function of time for increasing numbers 
of error-detection layers, defined in Eq.~(\ref{eq:frac}).
The Trotterized simulations were performed using $1.2$M shots
with $p_2=0.001$.
The blue, green, orange and red points correspond to $n_{d}=1, 2, 4$ and 8 layers respectively.
Results are shown for $f_\chi(t,n_d)$ for which the associated uncertainty is less than $\delta f < 0.04$.
The numerical values displayed can be found in Table~\ref{tab:1082layers} (App.~\ref{app:1082}).
}
    \label{fig:Leq41082N1248}
\end{figure}
We define the  fraction of recovered errors by
\begin{align}
    f_\chi(t,n_d) & = \frac{|\chi(t,n_d)-\chi_{\rm Q8ps}(t)|}{\chi_{\rm exact}(t)-\chi_{\rm Q8ps}(t)} 
    \ ,
    \label{eq:frac}
\end{align}
where 
$\chi_{\rm exact}(t)$ is the chiral condensate from noiseless Trotterized simulations,
$\chi_{\rm Q8ps}(t)$ is the 8-qubit post-selected chiral condensate computed with $p_2=0.001$ (the result obtained from noisy physical qubits with postselection onto the Gauss's law preserving final states),
and 
$\chi(t,n_d)$ is the chiral condensate computed from the [[10,8,2]] encoding with $n_d$ layers of error detection and $p_2=0.001$.
Results obtained from the {\tt qiskit} {\tt AerSimulator} simulator are shown in Fig.~\ref{fig:Leq41082N1248}.
Because the difference between the post-selected and exact-Trotterized values can coincide during the evolution, there are regions of time for which the denominator in Eq.~(\ref{eq:frac}) becomes small, and the uncertainties in $f_\chi(t,n_d)$ become large.  
To account for this, a cut  of 0.04 has been placed on this uncertainty.

As expected, the temporal density of the detection layers impacts 
improvements in the chiral condensate, and  
at all times considered, the error detection layers improve the results. 
However,  it is found that the improvement is saturated to $\sim 0.55$ 
for 8 layers of error detection.
\begin{figure}[ht!]
    \centering
\includegraphics[width=0.95\linewidth,alt={FT rotations}]{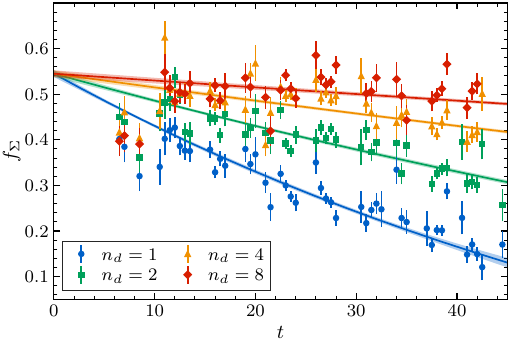}
\caption{
A correlated $\chi^2$-minimization
fit to the recovered fraction of errors in the chiral condensate as a function of time for increasing numbers 
of error-detection layers, defined in Eq.~(\ref{eq:frac}).
The Trotterized simulations were performed using $1.2$M shots
with $p_2=0.001$.
The blue, green, orange and red points correspond to $n_d=1, 2, 4$ and 8 layers respectively.
Results are shown for $f_\chi(t,n_d)$ for which the associated uncertainty is less than $\delta f < 0.04$.
The form of the correlated weighted $\chi^2$-minimization fit is
given in Eq.~(\ref{eq:fitform}), 
with the values of fit parameters and uncertainties displayed in 
Table~\ref{tab:fitparams} and Eq.~(\ref{eq:covfit}).
The fits with associated uncertainties are shown as the colored bands.
The numerical values displayed can be found in Table~\ref{tab:1082layers}.
}
    \label{fig:Leq41082N1248fit}
\end{figure}
A correlated $\chi^2$-minimization fit has been performed on the results shown in Fig.~\ref{fig:Leq41082N1248} of the form
\begin{align}
    f^{\rm fit}_\chi(t,n_d; A, C, \gamma) & = A e^{-\gamma t /n_d} + (C-A)
    \ ,
    \label{eq:fitform}
\end{align}
with the results displayed in Fig.~\ref{fig:Leq41082N1248fit}.
This functional form describes the results  well given their precision.
\begin{table}[htbp]
  \centering
  \begin{tabular}{lll}
    \hline\hline
    Parameter & Value & Description \\
    \hline
    $A$      & $0.92(24)$   & Amplitude \\
    $\gamma$ & $0.0133(46)$ & Decay rate \\
    $C$      & $0.545(7)$   & Value at $t=0$ \\
    \hline 
    \multicolumn{2}{l}{$\chi^2/\mathrm{dof}$} & $1.19$ \\
    \hline\hline
  \end{tabular}
    \caption{Correlated fit of the recovered fraction of the chiral condensate to
           the form given in Eq.~(\ref{eq:fitform}).
           A single parameter set $\{A, \gamma, C\}$ describes all four curves;
           the decay rate scales as $\gamma/n_d$, so each doubling of detection
           layers halves the decay rate.}
  \label{tab:fitparams}
\end{table}
The covariance matrix resulting from the fit is shown in Eq.~(\ref{eq:covfit}), 
\begin{equation}
\mathrm{Var}(A, \gamma, C) = 10^{-2}
\begin{pmatrix}
 5.751 & -0.1100 & -0.1232 \\
-0.1100 &  0.002122 &  0.002549 \\
-0.1232 &  0.002549 &  0.005312
\end{pmatrix}
\ .
\label{eq:covfit}
\end{equation}
A further reduction in the uncertainties of the results is required to 
scrutinize the limitations of the fit form we have assumed.
As expected, the value of A differs significantly from unity, due to undetectable errors introduced by the time evolution operator.
It is the case that this form is not applicable at late times, where it is expected that the recovered fraction vanishes.
These results imply that there exists an optimal frequency of syndrome measurements for a fixed error rate, code, and target evolution time, beyond which the improvements from increased syndrome extraction frequency are diminished.

\subsection{Encoding into $[[N+2,N,2]]$ with Partitions}
\label{sec:partitions}
\noindent
The $N=8$ system can be represented with multiple Iceberg encodings.
The eight sites can occupy one large code block ($[[10,8,2]]$), or be split into two ($[[6,4,2]]^{\otimes 2}$ and $[[8,6,2]]\otimes[[4,2,2]]$), three ($[[6,4,2]]\otimes[[4,2,2]]^{\otimes2}$), or four ($[[4,2,2]]^{\otimes4}$)
smaller blocks, down to the finest tiling the code family allows. 
Each block
carries its own pair of stabilizer checks, so a finer split detects errors at a higher density.

In order to make more robust numerical comparisons among encodings, 
we will work with variations of the metrics in previous sections.
In the following, $S(k)$ denotes 
acceptance rate after $k$ Trotter steps: the fraction
of shots that pass every check (mid-circuit ancilla flags, final-state
stabilizer checks, and the Gauss-law ($Q{=}0$) postselection).
To quantify the accuracy and performance of the different encodings in evaluating the chiral condensate, 
we measure its systematic error, 
denoted by ``SysErr'', 
and its statistical uncertainty, 
by combining them in quadrature,
\begin{itemize}
 \item $\mathrm{SysErr}\equiv\langle\Sigma\rangle_{\rm accepted}-\Sigma_{\rm ref}$:
    the systematic offset of the accepted-population mean from the
    reference. This is the error that postselection \emph{cannot} remove.
  \item $\sigma_1$: the per-accepted-shot standard deviation of
    $\Sigma$ (the spread of single kept shots).
\end{itemize}
The fixed-budget accuracy metric assembles bias, $\sigma_1$, $S$, 
and the number of shots, $n_s$,  as 
\begin{equation}
    \mathrm{RMSE}(n_s)=\sqrt{\mathrm{SysErr}^2+\sigma_1^2/(n_s\,S)}
    \ .
    \label{eq:rmse}
\end{equation}
\begin{figure}[ht!]
\centering
\includegraphics[width=0.95\linewidth]{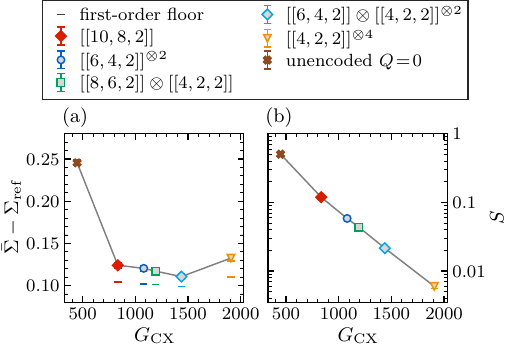}
\caption{Partitioning cost at $N=8$ ($14$ Trotter steps,
$p_2=0.003 $, $n_d=4$) as a function of  two-qubit gate count
$G_{\rm CX}$. 
The filled markers are measurements, open markers the exact
single-fault prediction.  
The brown cross, labeled unencoded, corresponds to postselection only (no encoding) 
(a) SysErr falls steeply from the unencoded results once any
encoding is applied, then plateaus. 
Within the encoded family it is not
monotonic in $G_{\rm CX}$, falling to a minimum at the intermediate-density
$[[6,4,2]]\otimes[[4,2,2]]^{\otimes2}$ and rising again at the finest tiling.   
(b) The acceptance rate falls exponentially with
$G_{\rm CX}$ (log axis), from $50\%$ unencoded through $12\%$ at the monolith
to $0.6\%$ at the finest tiling.  
The error bars denote $\sigma_1/\sqrt{n_{\rm kept}}$.
}
\label{fig:mechanism-partition}
\end{figure}
\begin{figure}[ht!]
\centering
\includegraphics[width=0.95\linewidth]{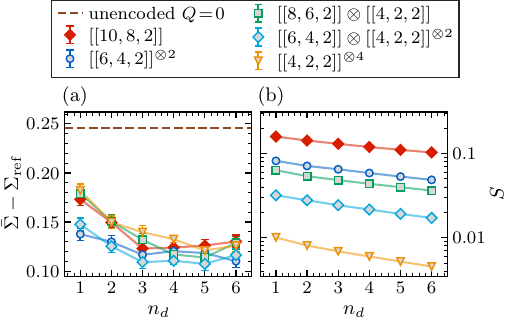}
\caption{The effects of the number of interior error-detection rounds $n_d$ at $N=8$ 
($14$ Trotter steps, $p_2=0.003 $). 
Each $n_d=4$ point pools $n_{\rm kept}\approx 2.3\times10^5$
accepted shots; the $n_d{=}5,6$ points pool $5.7-7.1\times10^4$.  
(a) SysErr against $n_d$ (error bars
$\sigma_1/\sqrt{n_{\rm kept}}$): 
it falls through about $n_d=3$ for most codes, then flattens
onto the single-fault floor. 
The finest tiling improves up to $n_d=5$ at this fixed number of Trotter steps.  The
intermediate-density $[[6,4,2]]\otimes[[4,2,2]]^{\otimes2}$ is lowest across
the sweep.  
(b) Acceptance rate as a function of $n_d$ (log axis): 
it falls with each added round, most steeply for the
finest tiling.}
\label{fig:round-scan}
\end{figure}
\begin{figure}[ht!]
\centering
\includegraphics[width=0.95\linewidth]{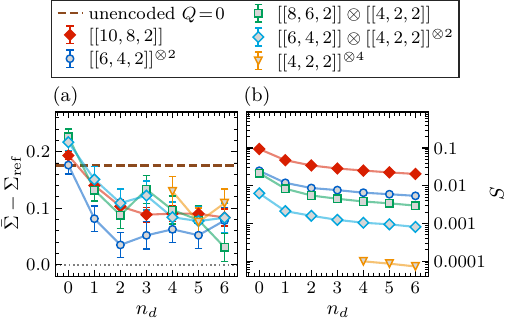}
\caption{
Deviations on the chiral condensate as a function of the number of rounds of error detection, 
$n_d$, for different encodings.
The systematic error (a) 
and postselection acceptance rate $S$ (b) 
versus $n_d$ for
$N{=}8$ with $k=28$ Trotter steps ($t=14$) and two-qubit gate error rate  $p_2=0.003 $, 
using the reference value of the chiral condensate
$\Sigma_{\rm ref}=3.6706$. 
The dashed line is the
unencoded results with postselection, 
$+0.175\pm0.004$ at acceptance $0.353$. 
(a) At $n_d{=}0$ no encoding outperforms the unencoded simulation.
(b) The acceptance rate falls with increasing $n_d$, 
and is ordered by block count: 
the monolithic $[[10,8,2]]$  code retains the most, $0.094$ at $n_d{=}0$, 
falling to $0.021$ at $n_d{=}6$.
The error bars are
$\sigma_1/\sqrt{n_{\rm kept}}$, with $n_{\rm kept}$ from $3.1\times10^3$ to
$4.3\times10^4$. 
Resource requirements are $1.5\times 10^8$ raw shots over 31 encoded cells and the
unencoded baseline.
}
\label{fig:allcode-vsm}
\end{figure}

The naive expectation is that the code with the finest error-detection should perform the best. 
However, as illustrated in Fig.~\ref{fig:mechanism-partition} the two-qubit gate count and hence the overhead increases significantly.~\footnote{The circuits in this section and all remaining sections use {\tt qiskit} {\tt AerSimulator} with \texttt{optimization\_level=0}.} 
Fig.~\ref{fig:mechanism-partition} (a) 
shows that all encodings significantly improve the accuracy (by a factor of 2) compared to the unencoded simulation after 14 Trotter steps (with four midcircuit measurements). 
This is consistent with the $L=2$ results in Sec.~\ref{sec:SM}.
Further, all encodings perform comparably.
However, as Fig.~\ref{fig:mechanism-partition} (b) shows, the acceptance rate drops significantly for the finer codes. 
Section~\ref{sec:SM1082}
showed that additional midcircuit measurements improve the performance at later times significantly. 
However, above a certain measurement frequency, 
the accuracy improvement saturates while the acceptance rate drops with any additional measurement. 
In Figs.~\ref{fig:round-scan} and \ref{fig:allcode-vsm}, 
we compare the systematic error in $\Sigma$ obtained 
from the different encodings after 14 and 28 Trotter steps, respectively, as a function of number of rounds of error detection.
We find that at 14 trotter steps, the  [6,4] and [4,4] partitions
have slight advantages over the others, 
but are not significantly better with sufficiently many 
rounds of error detection.
After 3 rounds most codes are saturated except for the finest [2,2,2,2], 
which keeps improving to 5 rounds (but has the lowest acceptance rate). 
Conclusions after 28 Trotter steps  are similar. 
We considered two scenarios: measuring only $\hat S_x$ and measuring $n_d$ rounds of $\hat S_x$ with a single midcircuit $\hat S_z$. 
Both performed worse compared to measuring both stabilizers throughout.

\begin{figure*}[ht!]
\centering
\includegraphics[width=0.95\linewidth]{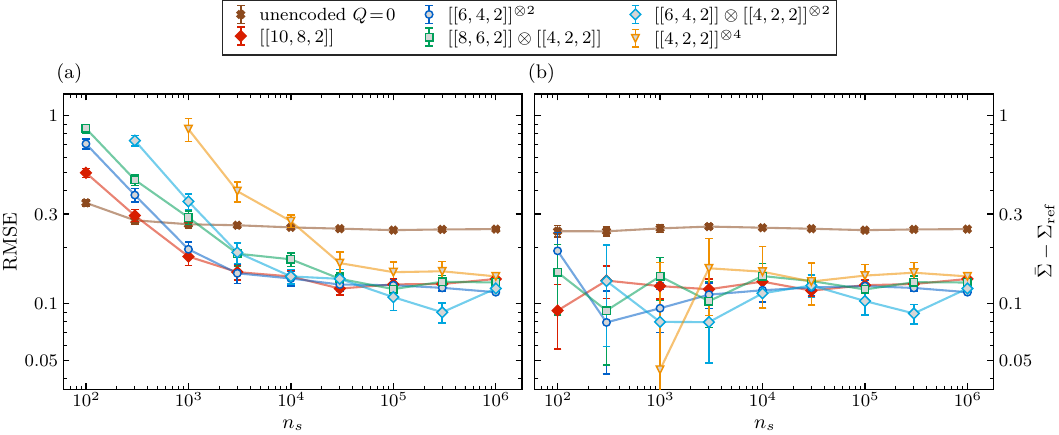}
\caption{Accuracy of the five $N=8$ partitions as a function of the number of shots at fixed resources
(condensate evaluated after $14$ Trotter steps at $p_2=0.003 $, $n_d=4$). 
The left panel (a) shows the accuracy
defined using Eq.~(\ref{eq:rmse}) and the right panel (b) shows the absolute deviation from the noiseless reference.
The monolithic
$[[10,8,2]]$ encoding is superior at a small number of shots.
The intermediate-density $[[6,4,2]]\otimes[[4,2,2]]^{\otimes2}$ 
is slightly favored 
above  $\sim 10^5$ shots and achieves the smallest error near $n_s=3\times 10^5$. The lowest $n_s$ of the highest density codes $[[4,2,2]]^{\otimes2}$ and $[[6,4,2]]\otimes[[4,2,2]]^{\otimes2}$ were removed since their total number of accepted shots was too low to be representative.}
\label{fig:crossover-shots}
\end{figure*}

Fig.~\ref{fig:crossover-shots} compares the accuracy of the encodings as a function of the number of shots.   We find that for a small number, the monolithic code performs the best,
while for a large number of shots,  [6,4] has a slight advantage.

\section{The Scaling of $[[N+2,N,2]]$ Encodings}
\label{sec:SMice}
\noindent
The previous sections discuss the performance and accuracy of Iceberg codes for small systems.
In order to quantify the scaling of the monolithic Iceberg code with system size $L$, 
we introduce two failure horizons:
\begin{itemize}
  \item $\Tshot$ (the shot horizon): the time at which shot acceptance falls
    below 1\%, which is a property of the circuit and is generally independent of the total number of shots. 
  \item $\Tacc$ (the accuracy horizon): the time at which the
    depolarizing noise has introduced a systematic error that is a fixed fraction of the chiral condensate. 
    We fit the noisy condensate to
    \begin{equation}
         \Sigma(t) \;=\; \frac{N}{2}
        \;+\; \Bigl(\Sigma_{\rm ref}(\kappa t)-\frac{N}{2}\Bigr)\,
              e^{-(a t + b t^{2})} \ ,
      \label{eq:envelope}
    \end{equation}
    where $\Sigma_{\rm ref}$ is the noiseless Trotter reference, $\kappa$ a phase rescaling, and $a$ and $b$ the damping rates. We fix $\epsilon_{\rm acc} = 0.10$.
    The exponential is exactly the fraction of the distance to the maximally mixed state with chiral condensate equal to $N/2$: 
    $= 1$ at $t=0$, 
    and $= 0$ once the state is fully mixed. 
    $\Tacc$
    is the earliest time at which that factor has fallen to
    $1-\epsilon_{\rm acc}$
      \begin{align}
     & a\,\Tacc + b\,\Tacc^{2} \;=\;
     -\ln(1-\epsilon_{\rm acc})
     \ .
      \label{eq:tacc}
    \end{align}
    
\end{itemize}

Fig.~\ref{fig:Leq2422422Chiral} and~\ref{fig:Leq41082Chiral} illustrate that the magnitude and shape of the 
chiral condensate changes significantly with system size. 
Moreover, the deviations of the noisy values from the noiseless reference are influenced by local maxima and minima (whose location in time changes with $N$). 
This motivated our choice of $\Tacc$ which is a phase adjusted, damping horizon.

\begin{figure}[ht!]
\centering
\includegraphics[width=0.95\columnwidth]{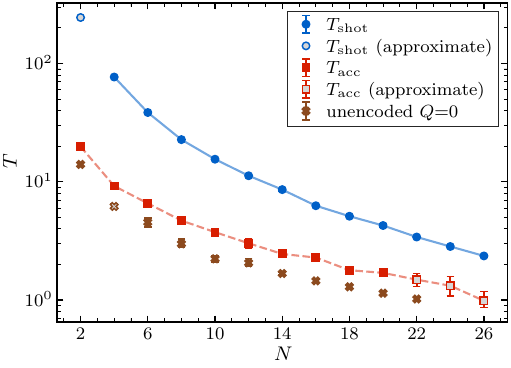}
\caption{Failure horizons of the monolithic $[[N{+}2,N,2]]$ code at $p_2=0.003 $,
for $N=2$ to $N=26$ ($n_d=2$). A median of $10^4$ shots was used. Small and large $N$ required a larger number of shots (up to $2.5\cdot 10^7$).}
\label{fig:tfail}
\end{figure}

Fig.~\ref{fig:tfail} shows the $T_\text{shot}$ and $T_\text{acc}$ as a function of 
system size at a fixed error rate $p_2=0.003 $. 
The first Trotter step is excluded from the $\Tacc$ fit since it only allows for one 
layer of error detection.
The accuracy horizon for the largest considered system sizes falls within 1-2 Trotter steps, and hence the determination has large extrapolation uncertainties.
The affected points are indicated by open markers, and are labeled approximate in the figure. 
The accuracy horizon of the encoded system (red squares) is always reached later than for the unencoded one (purple triangles), 
{\it i.e.}, 
the deviations from the noiseless simulations occur at later times for the encoded system.  
As expected, all curves fall monotonically with increasing  $N$.

The decay rate of the number of accepted shots as a function of time increases significantly 
with increasing $N$. 
For the smallest system, $N=2$,   only 0.004 of the shots per Trotter step  are lost. 
At $N=8$,
0.094 are lost (24× steeper); 
at $N=26$,  0.76 are lost. 
The $N=2$ point is labeled as approximate since the uncertainty of the crossing time $t^\star$ at which the acceptance rate drops to $S(t^\star)=1\%$ is inversely proportional to the decay rate.

\subsubsection{Resource Requirements and Scalings with System Size}
\noindent
In this section, we detail the scalings of the encoded circuits, 
which are split into two parts. 
The two-qubit gate count associated with a single Trotter step is $\Gstep$. 
The ``fixed'' part of the circuit refers to the constant overhead that does not depend on the number of Trotter steps: 
the encoding, midcircuit measurements and decoding which contribute $\Gfixed$ two-qubit gates. Together, they determine $\Gtot= \Gfixed+ k\,\Gstep$,
where $k$ is the number of Trotter steps, 
which is the total two-qubit-equivalent gate count. 
The gate-count error budget $x\equiv p_2\,\Gtot$ is the defining characteristic of the circuit that captures the noise impact of the circuit on the acceptance rate. 
We will see that it reveals a universal scaling.

For the monolithic $[[N{+}2,N,2]]$ code (with $n_d{=}2$ and at least two Trotter steps):
\begin{align}
&\Gfixed(N,n_d) = 6N + 18+(2N+8)n_d,\\ 
&\Gstep(N) = \tfrac{N^2}{2} + 2N.
\end{align}
For a general partitioning,
\begin{align}
  \Gfixed(N,P,n_d)
  &=(2N+2P)+(n_d+2)(2N+8P)
    \nonumber\\
 & =  (6+2n_d)N
    +(18+8n_d)P
    \ ,\nonumber\\
\Gstep(N,P,X) &= \tfrac{N^2}{2} + 2N + 14(P{-}1) + 4\tilde X, 
\label{eq:gsteppart}
\end{align} 
where $\tilde X$ depends on the location of the boundary
(see App.~\ref{app:RRscale} for more details).
It counts the nonzero electric 
$\widehat{\overline{Z}}\widehat{\overline{Z}}$
pairs that a boundary cuts, 
raising the requirements from 2 to 6 CNOT gates. 
With $L=N/2$, 
the nonzero contributions are exactly those with both sites below $L$ and those with both sites above it (with coefficients proportional to $L-{\rm max}(l,n)$ to ${\rm min}(l,n)-L$, respectively). 
Since no nonzero pair crosses the middle at $L$ (which belongs to the left half), a boundary can only split pairs whose two sites lie in the same half of the bipartition.
Writing $s_p^<$ and $s_p^>$ for the number of sites of block $p$ in the left and right half,~\footnote{Note that $s_p^<$ is strictly in the left half ($i<L$) and $s_p^>$ strictly in the right $(i>L)$.}
\begin{equation}
\tilde X = \Bigl[\tbinom{L}{2}-\sum_p\tbinom{s^{<}_p}{2}\Bigr]
  + \Bigl[\tbinom{L-1}{2}-\sum_p\tbinom{s^{>}_p}{2}\Bigr].
\label{eq:xcut}
\end{equation}

A boundary at the midpoint therefore does not generate a split, which makes equal halves the
one free two-block partition:~\footnote{The notation $[n_1,n_2]$ and $[n_1,n_2, n_3]$ is shorthand for $[[n_1+2,n_1,2]]\otimes[[n_2+2,n_2,2]]$ and $[[n_1+2,n_1,2]]\otimes[[n_2+2,n_2,2]]\otimes[[n_3+2,n_3,2]]$, respectively.} at $N{=}8$, $[4,4]$ has $\tilde X{=}0$ and $\Gstep{=}62$,
against $[6,2]$ with $\tilde X{=}2$ and $70$ and $[2,6]$ with $\tilde X{=}4$ and $78$, all
three sharing the same $\Gfixed$.
In particular, this implies that the  $[6,2]$ encoding 
requires fewer resources
than the $[2,6]$ encoding, even though both are based on the same combination of Iceberg codes.
Similarly [4,2,2] requires fewer resources than [2,2,4] and [2,4,2] which is the 
most resource heavy partition.

\begin{figure*}[ht!]
\centering
\includegraphics[width=\textwidth]{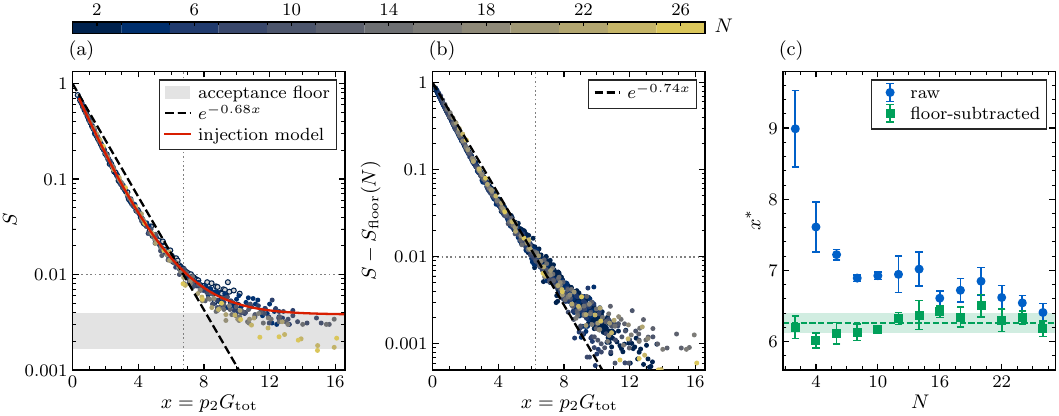}
\caption{
Acceptance rates for the monolithic $[[N{+}2,N,2]]$ code
at $p_2=0.003$.
(a)~Acceptance rate vs.\ the gate requirements $x = p\,\Gtot(N,t)$. 
Colored points:
$N=6$--$26$. The red
curve is the injection-based model of Section~\ref{subsec:inj}.
The shaded band spans the measured acceptance floors for $N=6$--$12$; dotted
lines indicate the acceptance rate falling below the 1\% threshold at the mean crossing budget $x^*=6.75$.
(b) The floor-subtracted acceptance rate collapses the lattice sizes $N=2$ to $26$
onto a single curve; the single curve decays exponentially as indicated by the dashed line corresponding to $e^{-0.74x}$. The dotted lines mark
the 1\% threshold $x^*=6.23$.
(c) $x^*$ against $N$. The raw crossing (open circles, taken from (a))
is inflated at small $N$, where the floor is a large fraction of the threshold,
and drifts down at large $N$; the floor-subtracted crossing (filled circles, taken from (b)) is
flat, $x^*=6.23$ and roughly independent of $N$.}
\label{fig:collapse}
\end{figure*}

We use these scaling considerations to understand aspects of 
the structure of the monolithic Iceberg code. 
In particular, the
acceptance re-scaling onto a single, universal curve in the resource variable
$x = p_2\,\Gtot(N,t) = p_2\,[\Gfixed(N) + k\,\Gstep(N)]$ for all $N \geq 6$ 
(see Fig.~\ref{fig:collapse}a). 
The predefined cutoff criterion is the time/depth at which the acceptance rate
first reaches 1\%. This crossing is reached for: $x^* = 6.75 \pm 0.2$.
The decay is not exactly an exponential, 
as is indicated by the black dashed line. 
Above $x\sim 7$, the curves have reached their acceptance floor and the points are scattered around a common mean. 
The two smallest $N$ are excluded from the plot since they reach their floor too early. The post-selected acceptance rate does not decay to zero since a corrupted state still passes every check with nonzero probability. 
$S_{\rm floor}(N)\approx\alpha\,
f_{\rm Gauss}(N)$ with $f_{\rm Gauss}=\binom{N}{N/2}/2^N$. 
The prefactor $\alpha\sim0.01$ depends on $n_d$ and slightly on $N$. 
For the $n_d=2$ simulation, 
it is of order $\alpha\approx0.01$ and varies slightly as a function of $N$.

If the floor is subtracted, as shown in Fig.~\ref{fig:collapse} (b), the decay is exponential $\sim e^{-0.74x}$ and the crossing sits at the universal value of  $x^*=6.23\pm0.16$. 
After floor subtraction,  any considered system size can be included, see Fig.~\ref{fig:collapse} (c). The blue dots show the unsubtracted data which collapses to the red band when the floor is subtracted.

\subsubsection{Injection model}\label{subsec:inj}
\noindent
In each post-selected circuit, the acceptance rate curve $S(x)$ is the fraction of shots retained after postselection at a fixed number of shots.
To understand the structure and scaling of  the encoded time evolution 
circuits better, we compute the probability that 
one-qubit and two-qubit
qubit errors are detected, and how many errors can remain in accepted shots.  
The errors are assumed to follow independent Poisson processes. 

Starting from the same circuit used in the noisy simulations, the depolarizing noise is removed 
and 
specific Pauli errors
are inserted after selected one-qubit and two-qubit gates.
By injecting one error at a time, 
the probability that each error is rejected by postselection or stabilizer measurements
can be measured. 
In this noise model, the expected numbers of one-qubit and two-qubit errors are
\begin{equation}
    \mu_1=\frac{3}{40}p\,G_1+\mathcal O(p^2), \qquad \mu_2=\frac{15}{16}p\,G_2+\mathcal O(p^2)
    \ ,
\end{equation}
where $G_1$ and $G_2$ denote the number of noisy one-qubit and two-qubit gates, respectively.
The acceptance probability of a shot with Pauli error $P$ inserted after gate location $i$ is $a_{iP}$. Hence, $f_{iP}\equiv 1-a_{iP}$ is the corresponding detection probability.

This injection experiment determines how often such faults are detected, 
given by $f_1$ (one-qubit errors) and $f_2$ (two-qubit errors), 
which are the averages of $f_{iP}$ over locations, 
and errors restricted to the Trotter part of the circuit. 
The two-qubit error in the fixed region are denoted by $\bar f_2$ and the total number in the circuit $\tilde f_2$ is the gate-weighted average of $f_2$ and $\bar f_2$, i.e. $\tilde f_2 = \frac{\Gfixed}{\Gtot}\,\bar f_2
   + \Bigl(1-\frac{\Gfixed}{\Gtot}\Bigr) f_2$.~\footnote{The fault ratio of one-qubit gates to two-qubit gates in the fixed region is 
   $\frac{2}{25}\frac{N + 36}{20N + 68}$ and hence negligible so we omit them for brevity. } 
The expected total number of detected errors is therefore 
$\lambda_\text{det}=\mu_1f_1+\mu_2 \tilde f_2 $ 
resulting in an acceptance probability of $S=e^{-\lambda_{\rm det}}$.

In the following, we fix $N=8$. 
Denoting the probability that a shot  contains exactly $n$ faults 
by $P(\mu,n)=\mu^ne^{-\mu}/n!$ and  its probability to be accepted  by $a_n$, 
then the acceptance rate can be decomposed as $S=\sum_n P(\mu,n)a_n$. 
The $a_n$ are determined by:
\begin{align}
   a_1 &= \frac{\mu_2\,(1-\tilde f_2) + \mu_1\,(1-f_1)}{\mu}, \qquad a_0 = 1,
\label{eq:a1}\\
a_n& = S_{\rm floor} + (a_2-S_{\rm floor})
      \left(\frac{a_2}{a_1}\right)^{n-2} \quad (n\geq2).
\end{align}
We  verify  this ansatz for $a_n$ by computing the first 13 $a_n$'s 
directly in the injection model.

The measured error detection probabilities are~\footnote{The one-qubit errors in the non Trotter step part of the circuit are only caught with $\bar f_1=0.73$.}
\begin{equation}
    f_1\sim0.90-0.15/k, \, f_2\sim0.88-0.06/k, \, \bar{f}_2=0.95
\end{equation}
indicating that errors in the encoding, decoding and syndrome measurement part are caught at a higher rate. In practice, this allows to use qubits with a higher error rate as ancilla qubits.
The acceptance rate curve for the injection model at $N=8$ is shown in red in Fig.~\ref{fig:collapse}(a) and matches our simulated data.

\subsubsection{Extrapolations to Large-$L$}
\noindent
The  $x^\star$ crossing point, where the acceptance rate drops to 1\%, 
is scale invariant information about a curve we find to be universal for all considered $N=[2,26]$ (see {\it e.g.}, Fig~\ref{fig:collapse}). 
To extrapolate  $x^\star$ to 100 qubits, 
we take its universal value and add an extrapolated floor  value 
$S_{\rm floor}(N=100)$ back in resulting in a crossing at $x^\star(100)=6.47\pm0.21$. 
For our given parameters, the shot horizon is reached within 
\begin{align}
\Tshot(100) \;&=\; dt\,\frac{x^*/p - \Gfixed(100)}{\Gstep(100)}
  \;\\&=\; \frac{6.47/0.003 - 1034}{2\cdot5200}\sim0.1,
  \end{align}
which is much less than one Trotter step at $dt=0.5$.
To estimate the precision required to stay within the shot horizon for $k$ Trotter steps, we can solve this equation for $p$ instead. 
The corresponding estimates are listed in Table~\ref{tab:100qest}.

To estimate the accuracy horizon for $100$ qubits, the data displayed in 
Fig.~\ref{fig:tfail}  must be extrapolated. 
The $T_{\rm acc}(N)$ for systems with $N\ge20$ 
are excluded
as they reach their accuracy horizon within 2-3 Trotter steps and carry non-negligible uncertainties. 
The nine smaller $N$ follow a power law with exponent $-1.06$ 
(with $\chi^2/\text{dof}=0.95$). 
Hence, the precision required to perform $k$ Trotter steps with 
$N=100$ within the accuracy horizon may be estimated as:
\begin{equation}
p_{\rm acc, req} 
      \;=\; \frac{p\,\Tacc(100)}{dt\,k}
      \;=\; \frac{0.003\times0.312}{k/2}.
\label{eq:preqest}
\end{equation}
The corresponding values for 20, 40 and 60 steps are listed in Table~\ref{tab:100qest}.

\begin{table}[htbp]
  \centering
\setlength{\tabcolsep}{4pt}
\begin{tabular}{rrrrrrr}
\hline
 & \multicolumn{3}{c}{monolithic $[[102,100,2]]$} & \multicolumn{2}{c}{unencoded} \\ \\
\cmidrule(lr){2-4}\cmidrule(lr){5-6}
 steps & $\Gtot$ & $p_{\rm shot, req}$ & $p_{\rm acc, req}$& $\Gtot$ & $p_{\rm acc, req}$
\\
\hline 
\rule{0pt}{2.5ex}
$20$ & $105{,}034$ & $6.2\times10^{-5}$ & $9.4\times10^{-5}$ & $100{,}000$ & $5.9\times10^{-5}$  \\
$40$ & $209{,}034$ & $4.7\times10^{-5}$ & $1.2\times10^{-5}$ & $200{,}000$ & $2.9\times10^{-5}$  \\
$60$ & $313{,}034$ & $3.1\times10^{-5}$ & $7.0\times10^{-6}$ & $300{,}000$ & $2.5\times10^{-5}$ \\
\hline
\end{tabular}
\caption{Extrapolated error rate estimates required to evolve 100 logical qubits 
(describing a $L=50$ lattice) 
through 20, 40 and 60 Trotter steps. 
$p_{\rm acc, req}$ denotes the accuracy required to stay within the accuracy horizon given in Eq.~(\ref{eq:tacc}). 
$p_{\rm shot, req}$ is the required accuracy for an acceptance rate larger than $1\%$. The unencoded reference value is given in the right column. }
\label{tab:100qest}
\end{table}

The results of this section indicate that with $N\sim 100$ and two-qubit gate fidelities 
at the $10^{-6}$ level on a quantum computer with all-to-all connectivity, 
monolithic Iceberg encodings appear to be able to improve the quality of (some) local observables,
based on this noise model.

\section{Encodings in Hypercube Codes: $[[2^N,N,2]]$  }
\label{sec:SMHC3}
\noindent
The $[[4,2,2]]$ code is the smallest member of the Iceberg code family, $[[N+2, N,2]]$, with $N$ an even integer. 
It is also the smallest member of another code family, the Hypercube code family, $[[2^N, N, 2]]$, where $N$ is any integer. 
These code families prioritize different aspects of quantum error detection: Iceberg codes maximize the ratio of logical to physical qubits for distance-two codes while Hypercube codes contain ladders of transversal logical entangling gates.
Therefore, it is worthwhile to explore how the Hypercube code family performs, compared to Iceberg codes, when carrying out quantum simulations of Trotterized time evolution of the Schwinger model.

\subsection{The $[[8, 3, 2]]$ Hypercube Code}
\begin{figure}[h]
\centering
\includegraphics[width=0.3\linewidth]{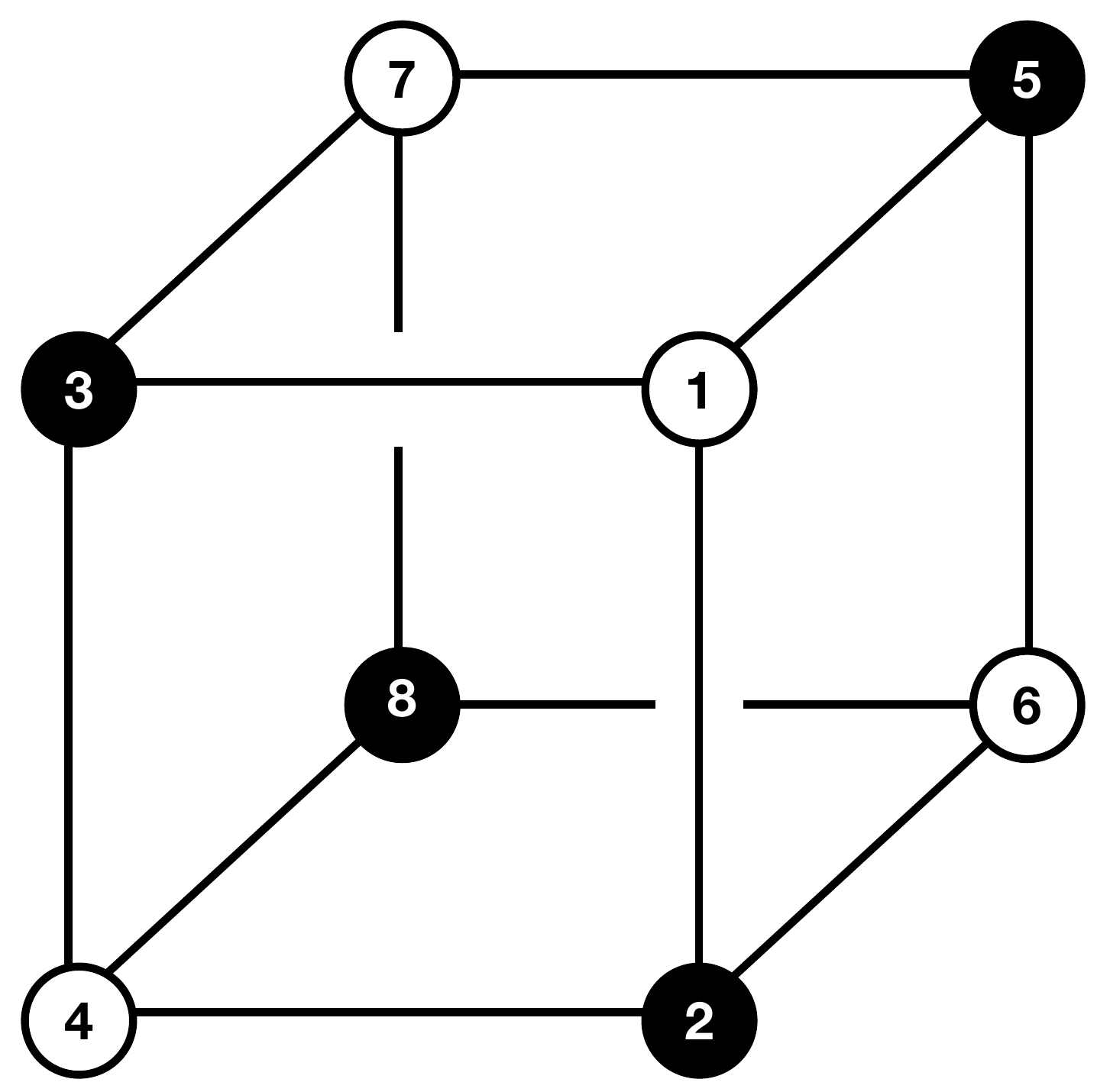}\\
\includegraphics[width=0.9\linewidth]{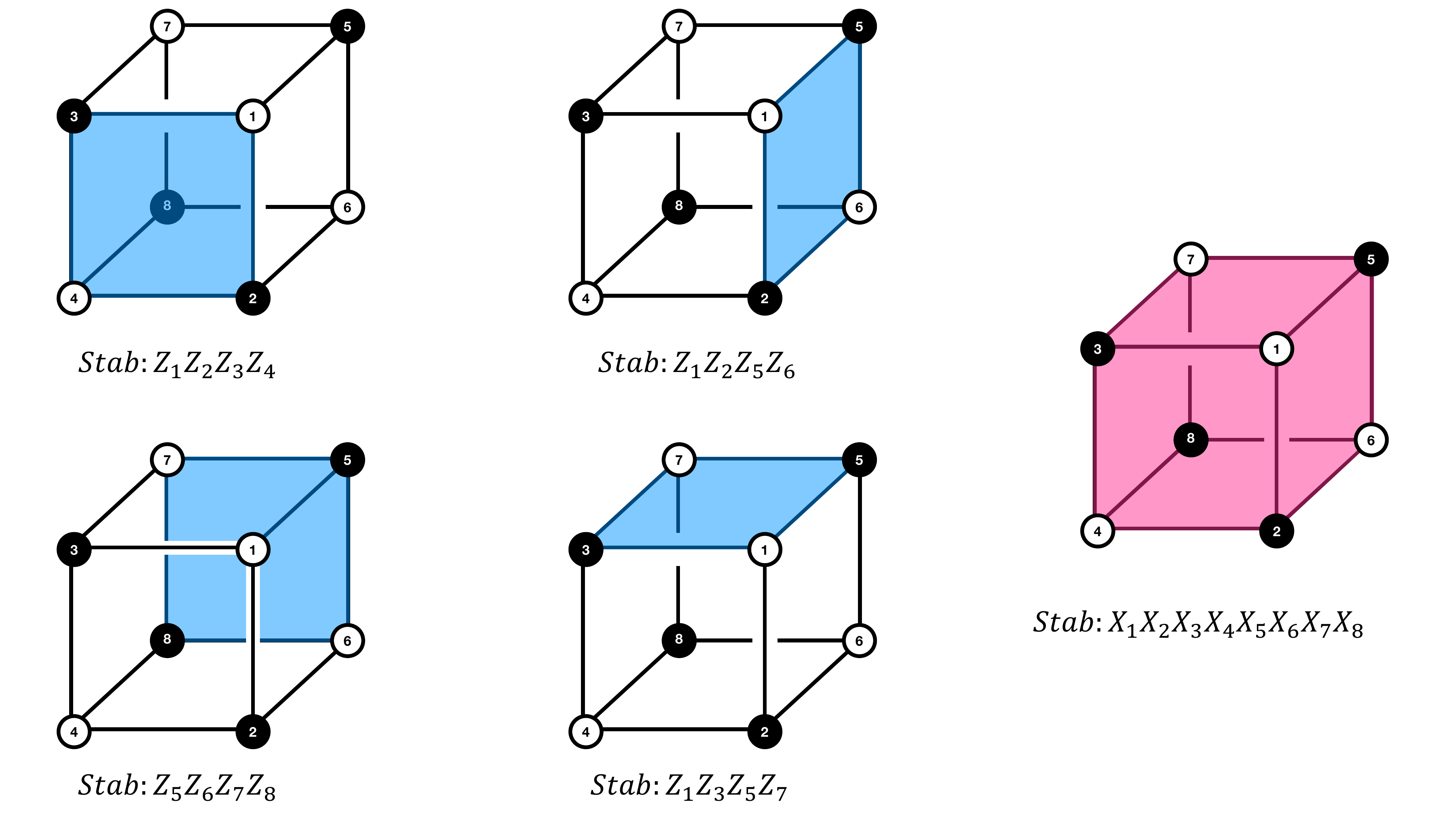}\\
\includegraphics[width=0.9\linewidth]{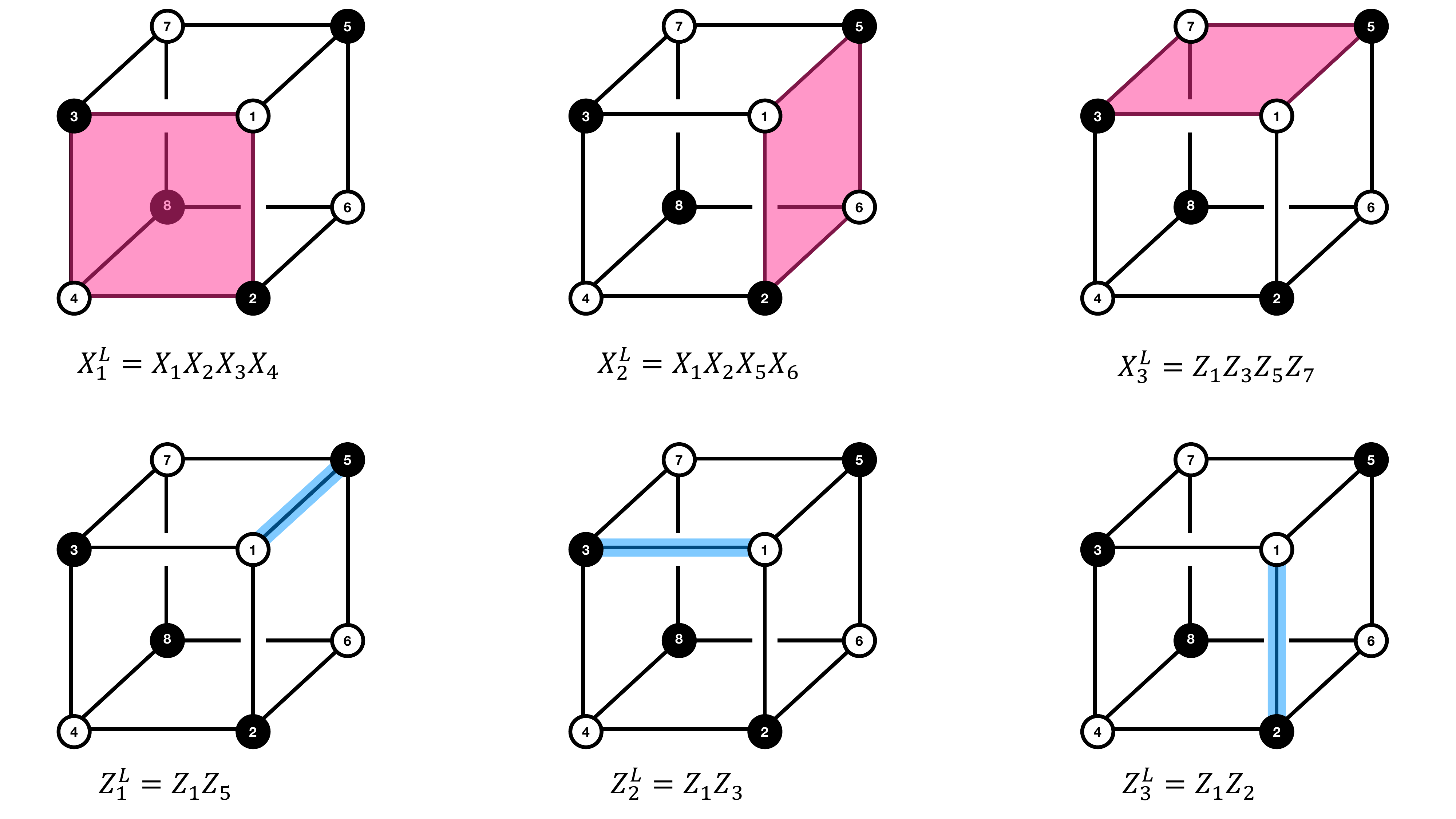}
\caption{
Geometric representation of the physical qubits, stabilizers and logical $\hat X$ and $\hat Z$ operators for the [[8,3,2]] quantum error detecting code. }
\label{fig:832GeoRep}
\end{figure}

\noindent
To understand the Hypercube code family, it is beneficial to first describe its next smallest member: $[[8,3,2]]$~\cite{Menendez:2023veg}. This code uses eight physical qubits to encode three logical qubits and therefore has five syndrome generators. 
Its key defining characteristics are that both the three logical CZ gates and the logical CCZ gate are all transversal.
The logical Pauli operators are defined by 
\begin{align}
\widehat{\overline{X}}_1 = \hat X_1 \, \hat X_2\, \hat X_3\, \hat X_4\ ,
\qquad \qquad \widehat{\overline{Z}}_1 = \hat Z_1 \, \hat Z_5 
\ ,
\nonumber \\
\widehat{\overline{X}}_2 = \hat X_1 \, \hat X_2\, \hat X_5\, \hat X_6\ ,
\qquad \qquad \widehat{\overline{Z}}_2 = \hat Z_1 \, \hat Z_3 \ ,
\nonumber \\
\widehat{\overline{X}}_3 = \hat X_1 \, \hat X_3\, \hat X_5\, \hat X_7\ ,
\qquad \qquad \widehat{\overline{Z}}_3 = \hat Z_1 \, \hat Z_2 
\ ,
\label{eq:832logicals}
\end{align}
where $\hat X_i$ is a Pauli X acting on the $i^\text{th}$ qubit. 
Analogous to the  [[4,2,2]] code, 
the $\ket{\bar{0} \bar{0} \bar{0}}$ state is the $\ket{\text{GHZ}(8)}$ state, 
\begin{align}
\ket{\bar{0}\bar{0}\bar{0}} &=\left[\frac{\ket{00000000}+\ket{11111111}}{\sqrt{2}}\right]
\ ,
\end{align}
these logical operators 
in Eq.~(\ref{eq:832logicals})
acting on the $\ket{\bar{0} \bar{0} \bar{0}}$ give the full dimension eight logical basis. The set of stabilizers is comprised of four weight-four Z stabilizers and one weight-eight X stabilizer,
\begin{align}
\hat S_Z &=\left\{\begin{array}{c} \hat Z_1 \, \hat Z_2\, \hat Z_3 \hat Z_4\\
\hat Z_5 \, \hat Z_6\, \hat Z_7 \hat Z_8\\
\hat Z_1 \, \hat Z_2\, \hat Z_5 \hat Z_6\\
\hat Z_1 \, \hat Z_3\, \hat Z_5 \hat Z_7 \end{array}\right\} 
\ ,
\nonumber \\
\hat S_X &=\hat X_1 \, \hat X_2 \, \hat X_3 \,\hat X_4 \,\hat X_5 \,\hat X_6 \,\hat X_7 \,\hat X_8 \, .
\end{align}
To better understand the structure of $[[8,3,2]]$ and generalize it to higher members of the Hypercube family, it is helpful to use its geometric representation, shown in Fig.~\ref{fig:832GeoRep}. 
The eight physical qubits  are represented as the vertices of a 3D cube. 
The Z stabilizers are associated with the six faces of the cube, of which only four are independent operators, and the X stabilizer acts on all eight physical qubits and is therefore associated with the entire cube. 
The logical X operators are associated with three faces of the cube  and the corresponding logical Z operators are associated with edges of the cube. 
The potential benefit of utilizing the $[[8,3,2]]$ code (or any Hypercube code) is that the multi-qubit entangling gates $C^{(n)} Z$, with $1\leq n\leq D-1$, are transversal.
An example of such a transversal gate within a $[[8,3,2]]$ codeblock is
\begin{align}
\widehat{\overline{CZ}}_{12} &= \hat S^{\phantom{\dagger}}_{1} \,\hat S^{\dagger}_{3} \, \hat S^{\dagger}_{5} \, \hat S^{\phantom{\dagger}}_{7}
\ .
\end{align}
Notice that in Fig.~\ref{fig:832GeoRep}, the undaggered operators correspond to white vertices, while daggered ones correspond to black vertices. 
The benefit of using this encoding may not be manifest using Trotter time evolution, 
and 
post-Trotter methods, {\it e.g.}, Ref.~\cite{Kubica2015},
are better suited for this code family.
Further details about this code can be found in App.~\ref{app:832}.

The generalization to larger members of this code  is carried out by utilizing this same geometric picture, but with higher dimensional codes. The generalization is:
\begin{itemize}
\item Physical qubits are associated with the corners of an $N$-dimensional Hypercube.
\item X stabilizer is weight $2^N$ and acts on all $2^N$ physical qubits.
Z stabilizers are weight-four and associated with the $\binom{N}{2} 2^{N-2}$ two-dimensional faces of the Hypercube. Since there are only $2^N-N-1$ Z stabilizers,   there is a convention choice for which faces to choose. 
\item Logical $\ket{\bar{0}}^{\otimes N}$ state is $\ket{\text{GHZ}(2^N)}$.
\item Logical Z operators are weight-2 and are associated with the edges of the Hypercube.
Logical X operators are weight $2^{N-1}$, associated with the facets (co-dimension 1 faces) of the Hypercube and share one physical qubit with the corresponding logical Z operator.
\item The $C^nZ$ gates within a codeblock 
are transversal, and are associated with $(n+1)$-dimensional faces of the Hypercube. 
\end{itemize}
Note that for $N = 3$, a two-dimensional face is also a facet of co-dimension one. 
This is why, for $N=3$, both the Z stabilizers and logical X operators are associated with the faces of the cube. For more details see App.~\ref{app:HGcodes}.
As the computations in this work are performed using classical simulators, the only two Hypercube codes that  are considered are $[[8,3,2]]$ and $[[16, 4, 2]]$.

\subsection{The [[16, 4, 2]] Hypercube Code}
\noindent
The Hypercube code corresponding to $D=4$ is the $[[16, 4, 2]]$ code, using sixteen physical qubits to encode four logical qubits with a distance of two. Continuing with the geometric picture for the [[8, 3, 2]] code,  the [[16, 4, 2]] code  is represented by a 4D Hypercube.
Representative logical Pauli operators are (see App.~\ref{app:1642} for more details),
\begin{align}
\widehat{\overline{X}}_1 &= \hat X_{9} \, \hat X_{10} \, \hat X_{11} \, \hat X_{12} \, \hat X_{13} \, \hat X_{14} \, \hat X_{15} \, \hat X_{16} 
\ ,
\nonumber \\
\widehat{\overline{Z}}_1 & = \hat Z_{1} \, \hat Z_{9} 
\ .
\end{align}
Each set of logical X and Z operators share only one physical qubit, 
ensuring the correct commutation relations are maintained.  
The sixteen codewords can be generalized as follows:
\begin{align}
|\bar{l}\rangle = \frac{1}{\sqrt{2}}\left( \bigotimes_{q=1}^{16} |l \cdot x(q)\rangle
 \; + \; \bigotimes_{q=1}^{16} |l \cdot x(q) \oplus 1\rangle \right)
\end{align}
where $|\bar{l}\rangle = |\bar{l}_1 \, \bar{l}_2 \, \bar{l}_3 \, \bar{l}_4\rangle$, 
and $x(q) = (x_1, x_2, x_3, x_4) \in \mathbb{F}_2^4$ is the coordinate vector of qubit $q$, 
defined by $q = 8x_1 + 4x_2 + 2x_3 + x_4 + 1$. 
This is the superposition of $l \cdot x = l_1 x_1 \oplus l_2 x_2 \oplus l_3 x_3 \oplus l_4 x_4$ evaluated over the vertices of the Hypercube, together with its complement. 
Specifically, $|\bar{0}\bar{0}\bar{0}\bar{0}\rangle$ is the 16-qubit GHZ state. 
Further details about this code can be found in App.~\ref{app:1642}.

\begin{figure}[ht!]
\centering
\includegraphics[width=0.9\linewidth]{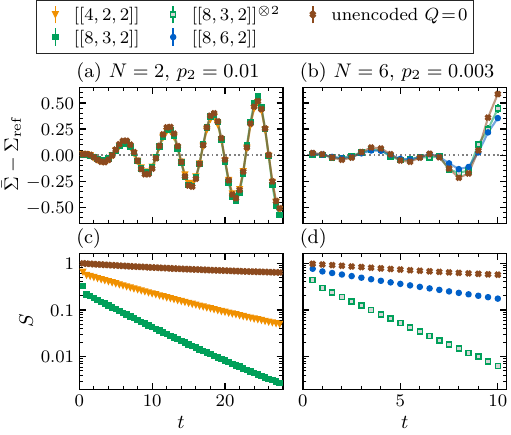}
\caption{The chiral condensate in the (a): $L=1$ and (b): $L=3$ Schwinger model 
embedded into the $[[4,2,2]]$ (orange), $[[8,3,2]]$ (green), $[[8,3,2]]^{\otimes 2}$ (green, open) and $[[8,6,2]]$ (blue) code with Gauss law postselection. The postselected, unencoded curve is drawn in brown. Shown is the deviation from the noiseless, unencoded reference
as a function of time starting from the Neel state. The encoded points are using two layers of FT error detection at the $1/3$ and $2/3$ points during the evolution. The Trotterized simulations were performed using  $10^5$ shots and depolarizing noise $p_2=0.01$ ($L=1$) and $p_2=0.003\,(L=3)$, respectively.
The bottom panel shows the corresponding acceptance rate as a function of time.
The $L=1$ simulation used $3\cdot10^6$ shots per point; $L=3$ used up to $5\cdot 10^5$.}
\label{fig:crossover832}
\end{figure}

\begin{figure}[ht!]
\centering
\includegraphics[width=0.9\linewidth]{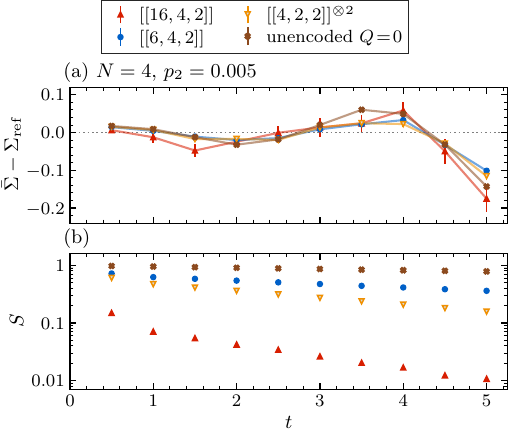}
\caption{The chiral condensate in the $L=2$ Schwinger model 
embedded into the $[[16,4,2]]$ (red), $[[4,2,2]]^{\otimes 2}$ (orange) and $[[6,4,2]]$ (blue) code with Gauss law postselection. The postselected, unencoded curve is drawn in brown. Shown is the deviation from the noiseless, unencoded reference
as a function of time starting from the Neel state. The encoded points are using two layers of FT error detection at the $1/3$ and $2/3$ points during the evolution. The Trotterized simulations were performed using  $10^5$ shots and depolarizing noise $p_2=0.005$.
The bottom panel (b) shows the corresponding acceptance rate as a function of time.}
\label{fig:1642comp}
\end{figure}

\subsection{Iceberg vs Hypercube Comparison}
\noindent
The encoding of the Schwinger model Hamiltonian into the Hypercube codes works analogously to the Iceberg encoding that we discussed in detail in previous sections. 
The circuits of the Trotterized time evolution introduce non-FT rotations, as it was the case for the Iceberg encoding. The state preparation circuits to encode the initial Neel state for the Hypercube codes match the ones outlined in Fig.~\ref{fig:422NeelFT}. For larger systems, such as the one considered in this section, we found it to be beneficial to skip the $\widehat{\overline Z}$ which adds $2N$ CNOT gates.  
Since the stabilizer structure is similar, i.e. $\hat S_X$ is a product of an even number of $\hat X_i$ and $\hat S_Z$ of $\hat Z_i$, we can use the same FT flagging for the stabilizer checks as for the Iceberg codes~\ref{fig:422NeelFT}. After non-FT time evolution, we use the same decoding procedure as outlined in Fig~\ref{fig:422Final} for the Iceberg codes including the syndrome measurement and final Gauss law postselection. For a comparison of gate counts between the Iceberg and Hypercube codes see App.~\ref{app:gate}. The quantum simulations are performed using the same mass $m$, coupling $g$, Trotter step $dt$ and isotropic depolarizing noise model that was used throughout the previous sections.

A point of comparison between the  Iceberg codes and Hypercube codes is the accuracy of the chiral condensate expectation value after a given number of Trotter steps, 
initialized on the Neel state for $L=1, 2$ and 3 of the Schwinger model, as discussed above.

Due to the staggering of the lattice, the Schwinger model requires an even number of lattice sites which is at odds with the three logical sites of the  $[[8,3,2]]$ code. Hence, we can only encode the $L=1$ system and the implementation of the $[[8,3,2]]$ keeps one logical qubit idle. The corresponding Iceberg code is the $[[4,2,2]]$ code. The mismatch in encoding overhead is immediately obvious: The $[[4,2,2]]$ code requires 4 physical qubits and has two stabilizers. The FT flags~\ref{fig:422NeelFT} require 2 ancillas for measuring a stabilizer. The $[[8,3,2]]$ code requires 8 physical and has five stabilizers.

The results of the simulation are presented in Fig.~\ref{fig:crossover832} and Fig.~\ref{fig:1642comp}. Both Figures display the time evolution of the chiral condensate (with the noiseless reference value subtracted) as a function of time. 
While both the Iceberg code and the Hypercube code have similar levels of accuracy, the acceptance rate of the Hypercube code is found to decay significantly faster.

\section{Summary and Outlook}
\label{sec:conc}
\noindent
Optimizing the scientific impact of near-term quantum computers requires   tuning 
all aspects of simulations to minimize a 
combination of systematic and statistical errors
in target observables.
With continually improving physical qubit fidelities and gate operations, 
simulations with partial FT and error detection provide a near-term step
away from the NISQ era toward fully error-corrected simulations.
Our recent work showing a substantial reduction in simulation errors
through encoded simulations with error detection
using IBM's quantum computers, 
along with a number of other works and  foundational theoretical papers, 
provides motivation for further exploring this path at this time.

In this work, we have explored small instances of the Schwinger model 
encoded into logical qubits with Hamming-distance-two error-detecting codes
using 
classical simulations  under single-qubit and two-qubit depolarizing noise
and all-to-all connectivity. 
Error detection was implemented by a small number of layers of FT stabilizer measurements. 
Post-processing error mitigation was not implemented to clearly 
isolate the role of error-detection.
We anticipate that the combination of error detection removing leading-order errors with error mitigation suppressing higher-order errors will further reduce the uncertainties in observables of interest.
We have quantified the performance of both Iceberg codes ($[[N+2,N,2]]$) and Hypercube codes
($[[2^N,N,2]]$) for $L=2,4,6$, with different degrees of error resolution, but without including spacetime optimizations in event rejection. 
Imposing limits on the quantum resources available for simulation means that tunings are required for optimizing the quality of target observables, particularly in situations where the total shot-rejection rate is high, resulting in small ensembles for post processing.
While an obvious point, embedding the codespace into a much larger Hilbert space usefully reduces 
${\cal O}(p^2)$ errors down to some easily calculable floor.
This is helpful for signal extraction at late times through high statistics production runs, 
when the shot rejection rate remains exponential in time, increasing the time before the codespace becomes uniformly populated by shots.

As the Schwinger model Hamiltonian conserves electric charge, final-state postselection 
into the same charge sector as the initial state removes the charge-violating 
${\cal O}(p)$ errors,
without encoding.
It is in the size of the sub-leading ${\cal O}(p^2)$ errors 
where encoding into logical qubits and layers of FT error-detection 
is shown to reduce deviations from exact results, limited by the circuit depth of one Trotter step and by the number of shots surviving at the end of the simulation.
While the encoding  overhead degrades the quality of observables due to the increased noise per unit time, this is compensated for by a modest number of layers of error detection, out to large numbers of Trotter steps for the considered error rates.
The $[[10,8,2]]$ code provides comparable results to the other possible partitions for $L=4$,  with $[[6,4,2]]^{\otimes 2}$  being optimal.  
For a fixed Trotter-step size, there is an optimal (uniform) density of detection layers 
interspersed between steps that minimizes the ${\cal O}(p^2)$ errors (including postselection).
For the $[[10,8,2]]$ code, the effectiveness of a fixed number of error-detection layers decreases with increasing number of Trotter steps, and for the selected couplings in the Hamiltonian and noise parameters, evolution through 90 Trotter steps remains 
essentially optimal for 8 layers.
This effect is analogous to the error in Trotterized time evolution and is expected to be a generic feature of quantum simulations with sparse error detection.
Simulations with the same parameters and constraints 
were performed using Hypercube encodings, and compared with the results of Iceberg encodings.
The quality of the results in reducing systematic uncertainties is found to be comparable.
However, the shot rejection rate is higher, due to the larger ratio of physical to logical qubits,
and corresponding deeper circuit depths, relegating them to second place behind Iceberg codes in resource constrained scenarios.
Different simulation parameters will require a new tuning to identify the optimal density of error-detection layers.

Simulations on present-day and near-term quantum computers will likely benefit from 
utilizing light-weight logical encodings, such as Iceberg and Hypercube codes, 
with sparse FT error-detection layers implemented during time evolution.
This holds for simulations performed with sufficient resources to furnish post-selected ensembles 
that yield statistical errors within desired tolerances for target observables within the required time interval.
Even prior to full FT powered by magic state injection or code switching, partially FT schemes such as those described in this work will prove advantageous. 
While these schemes require careful tuning to balance resource overhead with accuracy gains, their benefit to quantum simulations of fundamental physics can likely be realized on the current generation of quantum hardware.

\section*{Acknowledgments}
\noindent
This work was supported 
by U.S. Department of Energy, Office of Science, Office of Nuclear Physics, InQubator for Quantum Simulation (IQuS)\footnote{\url{https://iqus.uw.edu}} under Award Number DOE (NP) Award DE-SC0020970 via the program on Quantum Horizons: QIS Research and Innovation for Nuclear Science\footnote{\url{https://science.osti.gov/np/Research/Quantum-Information-Science}}.
Support is also acknowledged from the U.S. Department of Energy, Office of Science, National Quantum Information Science Research Centers, Quantum Systems Accelerator (Award No. DE-SCL0000121). This work was also supported by the U.S. Department of Energy, Office of Science, Office of Nuclear Physics, Grant No. DE-FG02-97ER-41014 (UW Nuclear Theory). 
S.G. was supported in part by a Feodor Lynen Research fellowship of the Alexander von Humboldt foundation. This work was also supported, in part,
through the Department of Physics\footnote{\url{https://phys.washington.edu}}
and the College of Arts and Sciences\footnote{\url{https://www.artsci.washington.edu}} at the University of Washington. 
This work was enabled, in part, by the use of advanced computational, storage and networking infrastructure provided by the Hyak supercomputer system at the University of Washington.
We acknowledge the use of {\tt Claude} for code development.

\appendix

\section{Coherent Error Propagation}
\label{app:CEP}
\begin{figure}[ht!]
    \centering
\includegraphics[width=0.45\linewidth]{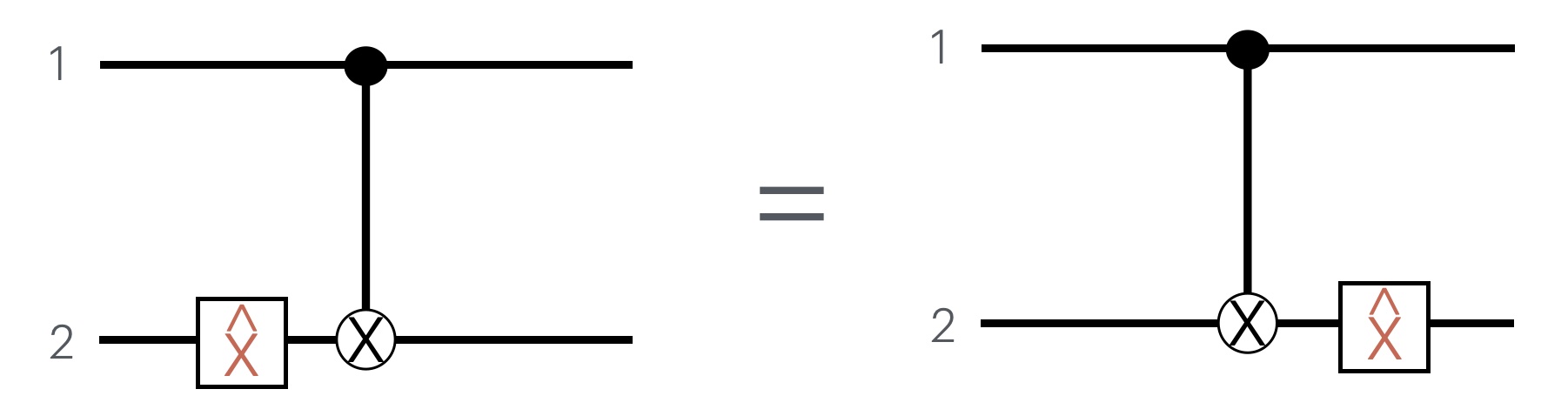}
\  , \ \ \ 
\includegraphics[width=0.45\linewidth]{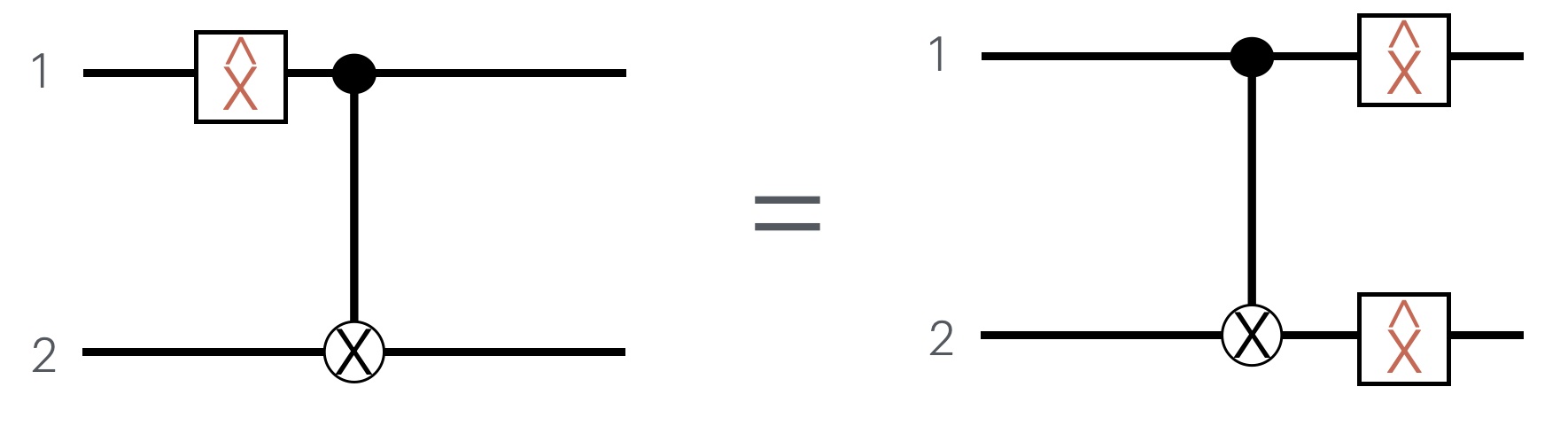}
\caption{
The propagation  of single-qubit $\hat X$ errors via  a CNOT gate.
}
    \label{fig:Xcnot}
\end{figure}

\begin{figure}[ht!]
    \centering
\includegraphics[width=0.45\linewidth]{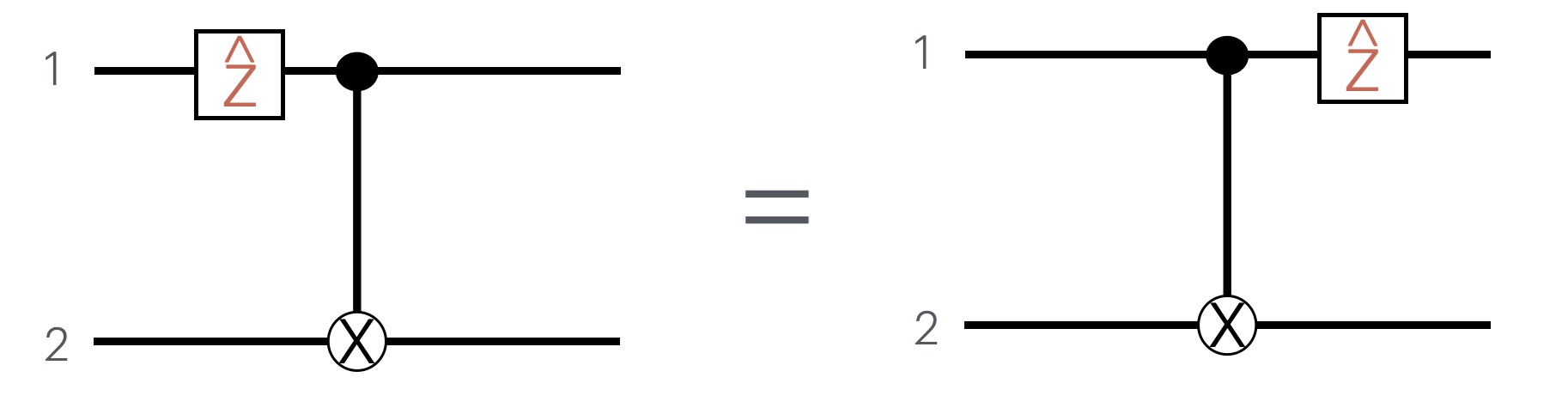}\ \ \ \ 
\includegraphics[width=0.45\linewidth]{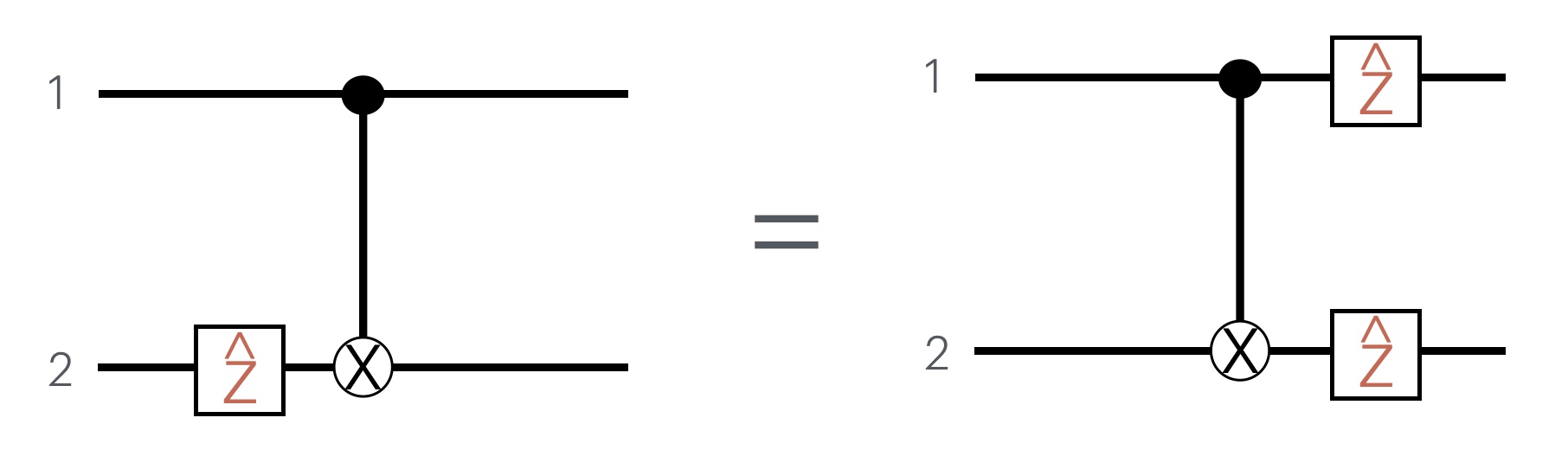}
\caption{
The propagation of single-qubit $\hat Z$ errors via  a CNOT gate.
}
    \label{fig:Zcnot}
\end{figure}

\noindent
FT is central to parametric error suppression in quantum simulation.
Specifically, as discussed in the main text, 
naively detectable single-qubit errors can become multi-qubit coherent errors 
with non-FT circuit designs, 
leaving the systems in the codeword space but with incorrect quantum amplitudes at 
${\cal O}(p)$.
The simplest examples of such error proliferation are shown in Figs.~\ref{fig:Xcnot} and \ref{fig:Zcnot},
where $\hat X$ and $\hat Z$ Pauli errors become correlated multi-Pauli errors via the CNOT gate.


\section{Details of $[[4,2,2]]^{\otimes 2}$}
\label{app:422sq}
\noindent
The $Q=0$ codewords of the $[[4,2,2]]^{\otimes 2}$ encoding, 
and the physical states they correspond to, are given by 
\begin{align}
    |\_\ \_\ \_\ \_\ \rangle & = |\bar{0} \bar{1}\rangle\otimes |\bar{0} \bar{1}\rangle
    \nonumber\\
    & = 
    \frac{1}{2}\left( 
    | \ 0101\ 0101\ \rangle + |\ 0101\ 1010\ \rangle 
\right. \nonumber\\ & \left. \qquad    
    + |\ 1010\ 0101\ \rangle + |\ 1010\ 1010\ \rangle
    \right)
    \ ,
\nonumber
\end{align}
\begin{align}
     |e^+e^-\_\ \_\rangle & = |\bar{1} \bar{0}\rangle\otimes |\bar{0} \bar{1}\rangle
    \nonumber\\
    & = 
    \frac{1}{2}\left( 
    | \ 1100\ 0101\ \rangle + |\ 1100\ 1010\ \rangle 
\right. \nonumber\\ & \left. \qquad    
    + |\ 0011\ 0101\ \rangle + |\ 0011\ 1010\ \rangle
    \right)
    \ ,
\nonumber
\end{align}
\begin{align}
    |\ \_\ \_\ e^+e^-\rangle & = |\bar{0} \bar{1}\rangle\otimes |\bar{1} \bar{0}\rangle
    \nonumber\\
    & = 
    \frac{1}{2}\left( 
    | \ 0101\ 1100\ \rangle + |\ 0101\ 0011\ \rangle 
\right. \nonumber\\ & \left. \qquad    
    + |\ 1010\ 1100\ \rangle + |\ 1010\ 0011\ \rangle
    \right)
    \ ,
\nonumber
\end{align}
\begin{align}
    |\ \_\ e^-e^+\ \_\ \rangle & = |\bar{0} \bar{0}\rangle\otimes |\bar{1} \bar{1}\rangle
    \nonumber\\
    & = 
    \frac{1}{2}\left( 
    | \ 0000\ 0110\ \rangle + |\ 0000\ 1001\ \rangle 
\right. \nonumber\\ & \left. \qquad    
    + |\ 1111\ 0110\ \rangle + |\ 1111\ 1001\ \rangle
    \right)
    \ ,
\nonumber
\end{align}
\begin{align}
    |e^+\ \_\ \_\ e^-\rangle & = |\bar{1} \bar{1}\rangle\otimes |\bar{0} \bar{0}\rangle
    \nonumber\\
    & = 
    \frac{1}{2}\left( 
    | \ 0110\ 0000\ \rangle + |\ 0110\ 1111\ \rangle 
\right. \nonumber\\ & \left. \qquad    
    + |\ 1001\ 0000\ \rangle + |\ 1001\ 1111\ \rangle
    \right)
    \ ,
\nonumber
\end{align}
\begin{align}
    |e^+e^-e^+e^-\rangle & = |\bar{1} \bar{0}\rangle\otimes |\bar{1} \bar{0}\rangle
    \nonumber\\
    & = 
    \frac{1}{2}\left( 
    | \ 1100\ 1100\ \rangle + |\ 1100\ 0011\ \rangle 
\right. \nonumber\\ & \left. \qquad    
    + |\ 0011\ 1100\ \rangle + |\ 0011\ 0011\ \rangle
    \right)
    \ ,
\end{align}

\section{Details of [[6,4,2]]}
\label{app:642}
\noindent
In this appendix, we provide details of the quantum circuits, codewords and observables used in simulations of the $[[6,4,2]]$ embedding of the $L=2$ Schwinger model, with two and three stabilizer checks.

\begin{figure}[ht!]
    \centering
\includegraphics[width=0.45\linewidth]{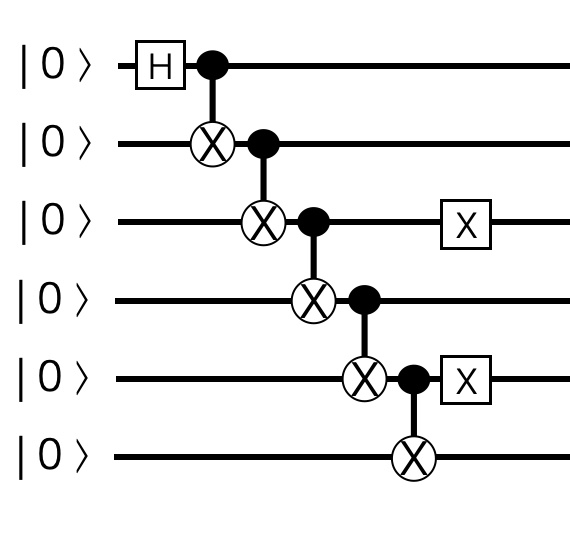}\qquad
\includegraphics[width=0.4\linewidth]{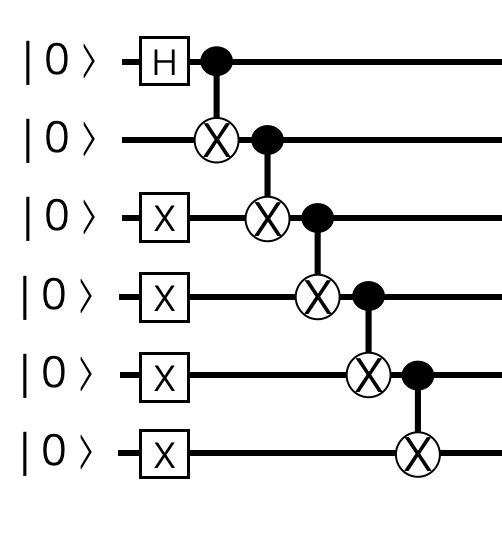}
\caption{
Two quantum circuits for preparing the Neel state in [[6,4,2]].
}
    \label{fig:642Neel}
\end{figure}

We start by considering the evolution of this system without any flags or FT/EC.
There are at least two ways we have identified for preparing the Neel state, as shown in Fig.~\ref{fig:642Neel}.
One may be more readily made FT than the other, depending on the nature of the errors.

\begin{figure}[ht!]
    \centering
\includegraphics[width=0.75\linewidth]{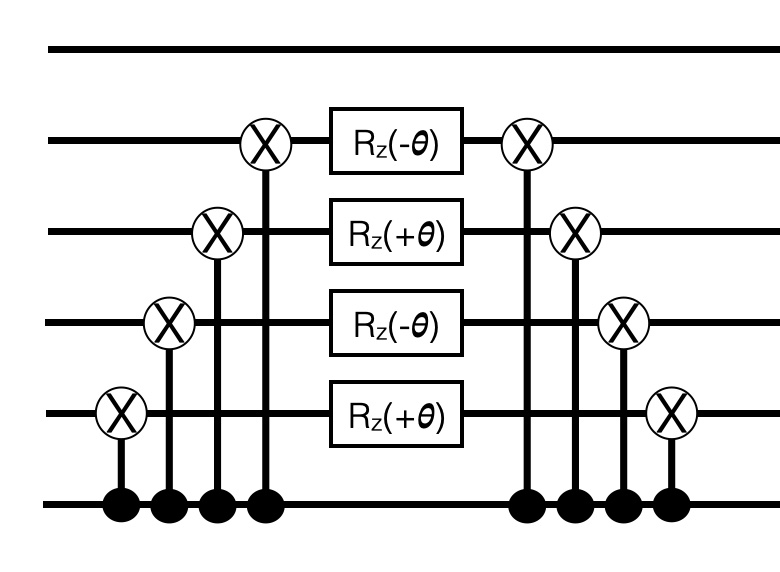}
\caption{
The quantum circuit implementing the mass term, corresponding to 
$e^{-i{\theta\over 2} (-\hat Z_{L1}+\hat Z_{L2}-\hat Z_{L3}+\hat Z_{L4})}$.
Matching to the Hamiltonian gives $\theta = t m$.
}
    \label{fig:642mass}
\end{figure}

\begin{figure}[th!]
    \centering
    \includegraphics[width=0.9\linewidth]{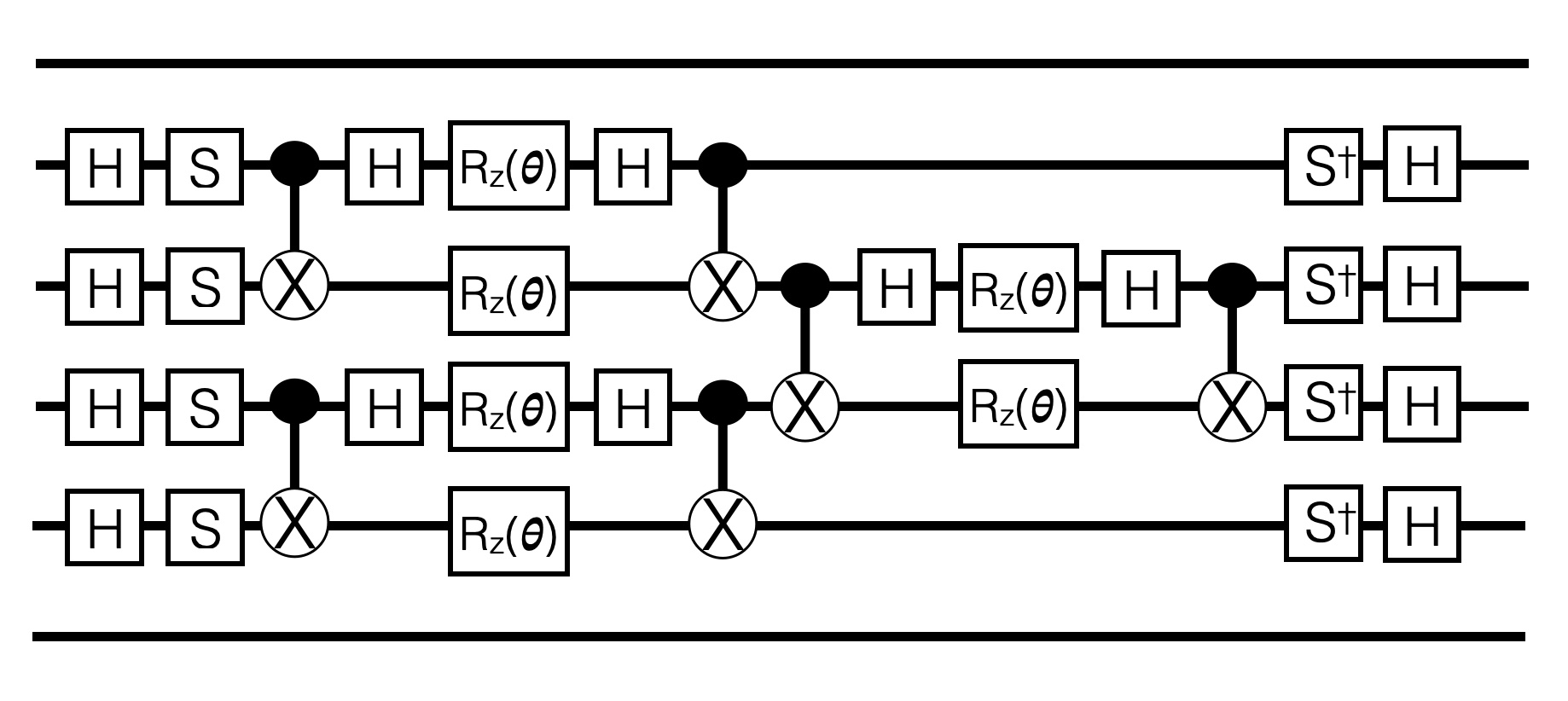} 
    \caption{
    The quantum circuit implementing  
    $e^{-i \theta (\sigma^+_2\sigma^-_3 +{\rm h.c.})}$
    $e^{-i \theta (\sigma^+_1\sigma^-_2 +{\rm h.c.})}$
    $e^{-i \theta (\sigma^+_3\sigma^-_4 +{\rm h.c.})}$,
    from the Hamiltonian given in Eq.~(\ref{eq:HamLeq2phys}).
    Matching to the Hamiltonian gives $\theta = t/2$.
    }
    \label{fig:Ukin6}
\end{figure}

\begin{figure}[th!]
    \centering
    \includegraphics[width=0.85\linewidth]{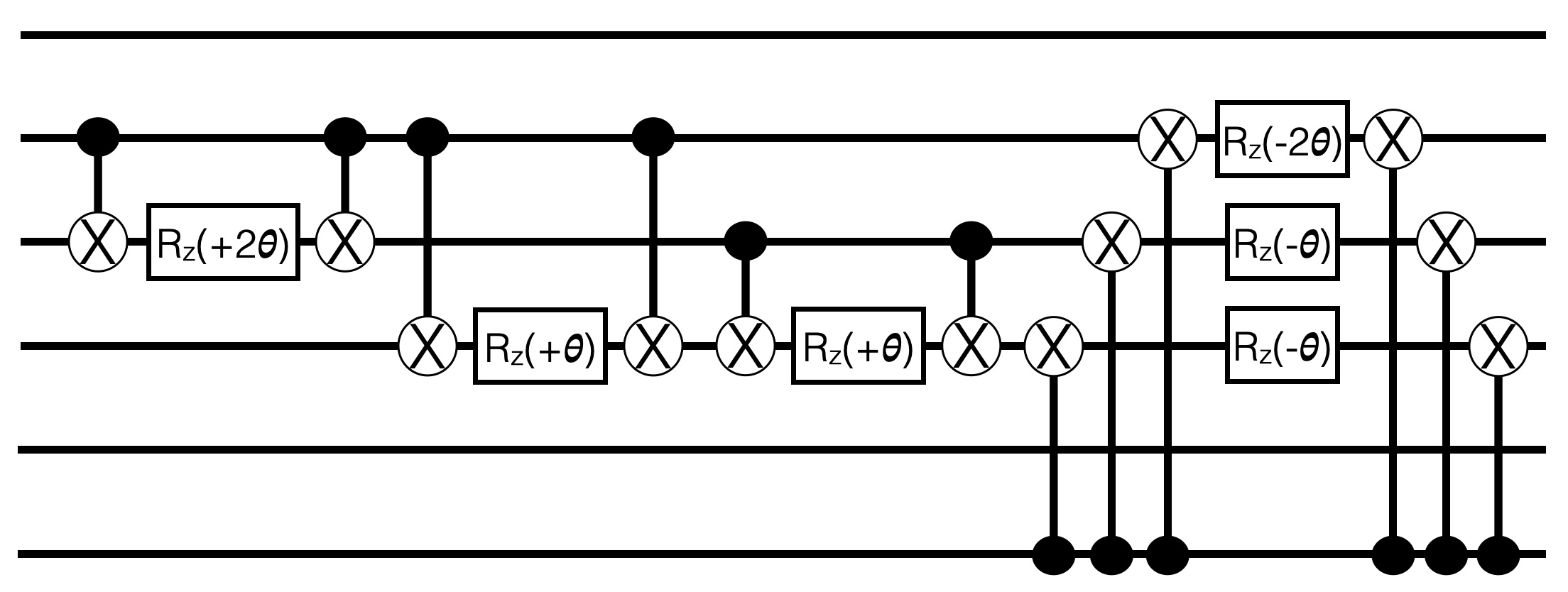} 
    \caption{
    The quantum circuit implementing  
    $e^{-i \theta (\hat Z_{L1}\hat Z_{L2} + {1\over 2}\hat Z_{L1}\hat Z_{L3} + {1\over 2}\hat Z_{L2}\hat Z_{L3} )} $
    $e^{-i \theta (-\hat Z_{L1} - {1\over 2}\hat Z_{L2}- {1\over 2}\hat Z_{L3})}$
    from the Hamiltonian given in Eq.~(\ref{eq:HamLeq2phys}).
    Matching to the Hamiltonian gives $\theta = g^2 t/2$.
    }
    \label{fig:Ugluon6}
\end{figure}

The mass term given in Eq.~(\ref{eq:HamLeq2phys}) can be implemented with the circuit shown in Fig.~\ref{fig:642mass}.
The kinetic term, after canceling contributions to the top and bottom qubits, is implemented by the circuit in Fig.~\ref{fig:Ukin6}.
The contribution to the evolution operator from the gauge field, given  in Eq.~(\ref{eq:HamLeq2phys}), is shown in Fig.~\ref{fig:Ugluon6}.
The single-qubit phase contributions to the gluon and mass circuits can be combined using the circuit in Fig.~\ref{fig:642mass} with the replacement
\begin{eqnarray}
    \theta_1 & = & t  m \rightarrow t (m+g^2)
    \ \ ,\ \ 
    \theta_2\ =\ t m \rightarrow t (m-{g^2\over 2})
    \nonumber\\
    \theta_3 & = & t  m \rightarrow t (m+{g^2\over 2})
    \ \ ,\ \ 
    \theta_4\ =\ t m \rightarrow t m 
    \ .
\end{eqnarray}
and without including the corresponding gates in Fig.~\ref{fig:Ugluon6}.
In the case of the symmetric gauge circuit, the single-qubit phases can be combined to give
\begin{eqnarray}
    \theta_1 & = & t  m \rightarrow t (m+{g^2\over 2})
    \ \ ,\ \ 
    \theta_2\ =\ t m \rightarrow t m 
    \nonumber\\
    \theta_3 & = & t  m \rightarrow t m 
    \ \ ,\ \ 
    \theta_4\ =\ t m \rightarrow t (m+{g^2\over 2}) 
    \ .
\end{eqnarray}

Preparing the circuit for measurement, to reduce the post-processing overhead, 
the superposition logical states is transformed back to a single state.  
The states in the $Q=0$ sector are of the form 
\begin{eqnarray}
    |\psi\rangle & \sim & |0abcd0\rangle + |1\tilde{a}\tilde{b}\tilde{c}\tilde{d}1\rangle
    \ ,
    \label{eq:Ftrans642}
\end{eqnarray}
\begin{figure}[ht!]
    \centering
\includegraphics[width=0.3\linewidth]{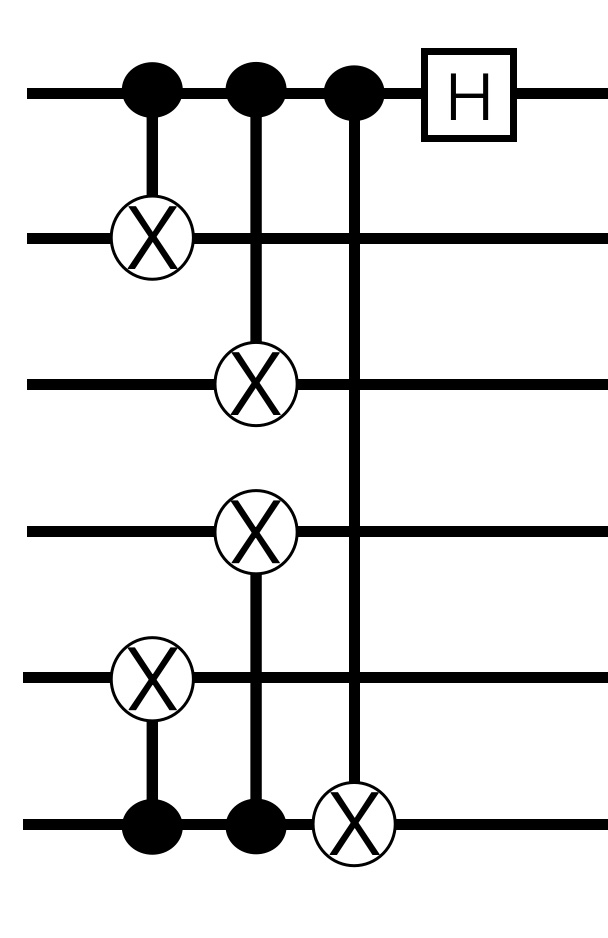}
\caption{
The final-stage circuit element, to be included immediately prior to measurement, to bring the wavefunction into a simpler form for post-processing, see Eq.~(\ref{eq:Ftrans642}).  
}
    \label{fig:642Ftrans}
\end{figure}
and Fig.~\ref{fig:642Ftrans} shows the final circuit element added to transform states 
in the $Q=0$ sector (more precisely even-charge sectors) back to physical basis for 
post-processing,
\begin{eqnarray}
    |\psi\rangle & \sim & |0abcd0\rangle + |1\tilde{a}\tilde{b}\tilde{c}\tilde{d}1\rangle
    \rightarrow |0abcd0\rangle
    \ .
    \label{eq:FlagA642}
\end{eqnarray}
This also has the nice feature of eliminating single-$\hat Z$ errors, as 
\begin{eqnarray}
    |\psi\rangle & \sim & |0abcd0\rangle - |1\tilde{a}\tilde{b}\tilde{c}\tilde{d}1\rangle
    \rightarrow |1abcd0\rangle
    \ ,
    \label{eq:FlagB642}
\end{eqnarray}
which is projected out in post-processing.

Counting CNOTs and circuit depth, there are
6-CNOTs with depth-4 for preparing the Neel state,  
20-CNOTs with depth-18 for one Trotter step,
and 5-CNOTs of depth-3 for the final transformation.
This gives a total of $11+20 N_{\rm Trott}$ CNOTs with depth
$7+18 N_{\rm Trott}$.

\section{Results for the [[10,8,2]] Encoding}
\label{app:1082}
\noindent

The numerical values for the post-selected chiral condensate as a function of time in the [[10,8,2]] encoding for up to $n_d=8$ are listed in Table~\ref{tab:1082layers}.

\begin{table}[htbp]
  \centering
  \begin{tabular}{c cccc}
    \hline\hline
    time & $n_{d}=1$ & $n_{d}=2$ & $n_{d}=4$ & $n_{d}=8$ \\
    \hline
6.5 & $0.410(40)$ & $0.460(40)$ & $0.430(40)$ & $0.410(40)$ \\
    11.5 & $0.401(34)$ & $0.471(35)$ & $0.490(40)$ & $0.500(40)$ \\
    12.5 & $0.382(26)$ & $0.496(27)$ & $0.498(27)$ & $0.501(29)$ \\
    13.5 & $0.367(29)$ & $0.406(30)$ & $0.490(31)$ & $0.518(32)$ \\
    15.5 & $0.380(21)$ & $0.448(21)$ & $0.508(22)$ & $0.491(22)$ \\
    16.5 & $0.354(19)$ & $0.406(19)$ & $0.482(20)$ & $0.483(21)$ \\
    19.5 & $0.342(26)$ & $0.422(26)$ & $0.543(28)$ & $0.513(28)$ \\
    21.5 & $0.242(35)$ & $0.390(40)$ & $0.450(40)$ & $0.410(40)$ \\
    22.5 & $0.328(21)$ & $0.468(22)$ & $0.518(23)$ & $0.511(24)$ \\
    23.5 & $0.285(13)$ & $0.383(14)$ & $0.506(14)$ & $0.518(15)$ \\
    26.5 & $0.276(17)$ & $0.412(18)$ & $0.477(18)$ & $0.525(20)$ \\
    27.5 & $0.263(13)$ & $0.424(15)$ & $0.487(15)$ & $0.528(15)$ \\
    30.5 & $0.270(40)$ & -- & -- & -- \\
    31.5 & $0.244(20)$ & $0.371(22)$ & $0.458(23)$ & $0.505(26)$ \\
    34.5 & $0.230(23)$ & $0.327(25)$ & $0.456(29)$ & $0.497(29)$ \\
    37.5 & $0.172(16)$ & $0.306(18)$ & $0.432(21)$ & $0.486(23)$ \\
    38.5 & $0.206(12)$ & $0.341(13)$ & $0.441(15)$ & $0.515(16)$ \\
    40.5 & $0.240(40)$ & -- & -- & -- \\
    41.5 & $0.165(15)$ & $0.304(17)$ & $0.407(19)$ & $0.504(21)$ \\
    42.5 & $0.106(29)$ & $0.380(35)$ & $0.490(40)$ & -- \\
    44.5 & $0.150(32)$ & $0.240(40)$ & -- & -- \\    \hline\hline
  \end{tabular}
  \caption{Post-selected chiral condensate as a function of time for the [[10,8,2]] Iceberg code
           with $N_{\mathrm{layers}}=1,2,4,8$ layers of error detection from 1.2M shots. Values are
           bootstrap mean with the standard deviation on the last digits in parentheses;
           ``--'' denotes data with errors exceeding $0.04$.}
   \label{tab:1082layers}
\end{table}

\section{Resource Requirements for Scaling}
\label{app:RRscale}
\noindent
In this appendix, we generalize the counts for the circuit of the monolithic code to different partitions and detail their decomposition. 
We can partition $N$ into $P$ blocks ($P=1$ is the monolith Iceberg). 
For a block containing $n_p$ logical sites, 
the encoding and decoding cost $n_p+1$ CNOT gates each 
\begin{equation}
G_{\mathrm{encode+decode}}
  =2\sum_{p=1}^P(n_p+1)
  =2N+2P
\end{equation}
assuming one round of syndrome extraction each.
The stabilizer of block $p$ acts on
\begin{equation}
  w_p=n_p+2
  \ ,
\end{equation}
physical code qubits.  The flagged $S_X$ extraction uses $w_p$
CNOT gates touching the code qubits and two flag CNOT gates, for a total
$w_p+2=n_p+4$.  The linked $S_Z$ extraction has the same CNOT count.  One full
$S_X+S_Z$ round therefore requires
$  2(n_p+4)=2n_p+8$
for block $p$, or
\begin{equation}
  G_{\mathrm{1 round}}=2N+8P
  \ ,
\end{equation}
over the complete partition. 
There are $n_d$ interior rounds and two bracket rounds. Hence
\begin{align}
  \Gfixed(N,P,n_d)
  &=(2N+2P)+(n_d+2)(2N+8P)\notag\\
  &=(6+2n_d)N
    +(18+8n_d)P
    \ .                 
\label{eq:gfixed}
\end{align}
The monolithic per step CNOT count is $2(N-1)+ 2\left(\frac{N}{2}-1\right)^2+2N=N^2/2+2N$ corresponding to kinetic and the electric $ZZ$ and $Z$ contribution.
Taking into account the cross partition $ZZ$ terms for general partitions yields the general per-step count
\begin{equation}
  \Gstep(N,P,\widetilde X)
  =\frac{N^2}{2}+2N+14(P-1)+4\widetilde X.  
\end{equation}
The $14(P-1)$ term depends only on the number of blocks.  The
$4\widetilde X$ term contains all dependence on the positions of their
boundaries.

\section{Hypercube Codes}
\label{app:HGcodes}
\noindent
In this appendix, we provide details supporting the discussion of Hypercube codes in main text.

\subsection{[[8,3,2]]}
\label{app:832}
The complete set of logical states is
\begin{align}
\ket{\bar{0}\bar{0}\bar{0}} &=\left[\frac{\ket{00000000}+\ket{11111111}}{\sqrt{2}}\right]
\ ,\nonumber \\
\ket{\bar{1}\bar{0}\bar{0}} &=\left[\frac{\ket{11110000}+\ket{00001111}}{\sqrt{2}}\right] \ ,
\nonumber \\
\ket{\bar{0}\bar{1}\bar{0}} &=\left[\frac{\ket{11001100}+\ket{00110011}}{\sqrt{2}}\right] \ ,
\nonumber \\
\ket{\bar{1}\bar{1}\bar{0}} &=\left[\frac{\ket{00111100}+\ket{11000011}}{\sqrt{2}}\right]\ ,
\nonumber \\
\ket{\bar{0}\bar{0}\bar{1}} &=\left[\frac{\ket{10101010}+\ket{01010101}}{\sqrt{2}}\right]\ ,
\nonumber \\
\ket{\bar{1}\bar{0}\bar{1}} &=\left[\frac{\ket{01011010}+\ket{10100101}}{\sqrt{2}}\right]\ ,
\nonumber \\
\ket{\bar{0}\bar{1}\bar{1}} &=\left[\frac{\ket{01100110}+\ket{10011001}}{\sqrt{2}}\right] \ ,
\nonumber \\
\ket{\bar{1}\bar{1}\bar{1}} &=\left[\frac{\ket{10010110}+\ket{01101001}}{\sqrt{2}}\right] 
\, .
\end{align}

Examples of transversal gates within a $[[8,3,2]]$ codeblock  are
\begin{align}
\widehat{\overline{CZ}}_{12} &= \hat S^{\phantom{\dagger}}_{1} \,\hat S^{\dagger}_{3} \, \hat S^{\dagger}_{5} \, \hat S^{\phantom{\dagger}}_{7} \ ,
\nonumber \\
\widehat{\overline{CZ}}_{13} &= \hat S^{\phantom{\dagger}}_{1} \,\hat S^{\dagger}_{2} \, \hat S^{\dagger}_{5} \, \hat S^{\phantom{\dagger}}_{6} \ ,
\nonumber \\
\widehat{\overline{CZ}}_{23} &= \hat S^{\phantom{\dagger}}_{1} \,\hat S^{\dagger}_{2} \, \hat S^{\dagger}_{3} \, \hat S^{\phantom{\dagger}}_{4} \ ,
\nonumber \\
\widehat{\overline{CCZ}}_{123} &= \hat T^{\phantom{\dagger}}_{1} \, \hat T^{\dagger}_{2} \, \hat T^{\dagger}_{3} \, \hat T^{\phantom{\dagger}}_{4} \, \hat T^{\dagger}_{5} \, \hat T^{\phantom{\dagger}}_{6} \, \hat T^{\phantom{\dagger}}_{7} \, \hat T^{\dagger}_{8}
\ .
\end{align}
%

\subsection{[[16,4,2]]}
\label{app:1642}
\noindent
One choice for the logical operators is given by
\begin{align}
\widehat{\overline{X}}_1 &= \hat X_{9} \, \hat X_{10} \, \hat X_{11} \, \hat X_{12} \, \hat X_{13} \, \hat X_{14} \, \hat X_{15} \, \hat X_{16} \ ,
\nonumber \\
\widehat{\overline{X}}_2 &= \hat X_{5} \, \hat X_{6} \, \hat X_{7} \, \hat X_{8} \, \hat X_{13} \, \hat X_{14} \, \hat X_{15} \, \hat X_{16} \ ,
\nonumber \\
\widehat{\overline{X}}_3 &= \hat X_{3} \, \hat X_{4} \, \hat X_{7} \, \hat X_{8} \, \hat X_{11} \, \hat X_{12} \, \hat X_{15} \, \hat X_{16} \ ,
\nonumber \\
\widehat{\overline{X}}_4 &= \hat X_{2} \, \hat X_{4} \, \hat X_{6} \, \hat X_{8} \, \hat X_{10} \, \hat X_{12} \, \hat X_{14} \, \hat X_{16}
\ ,
\end{align}
and
\begin{align}
\widehat{\overline{Z}}_1 &= \hat Z_{1} \, \hat Z_{9} \ ,
\nonumber \\
\widehat{\overline{Z}}_2 &= \hat Z_{1} \, \hat Z_{5} \ ,
\nonumber \\
\widehat{\overline{Z}}_3 &= \hat Z_{1} \, \hat Z_{3} \ ,
\nonumber \\
\widehat{\overline{Z}}_4 &= \hat Z_{1} \, \hat Z_{2}\, .
\end{align}

The twelve stabilizers of the $[[16, 4, 2]]$ code are given by:
\begin{align}
\hat S_Z &=\left\{\begin{array}{c} \hat Z_{13} \, \hat Z_{14} \, \hat Z_{15} \, \hat Z_{16}\\
\hat Z_{11} \, \hat Z_{12} \, \hat Z_{15} \, \hat Z_{16}\\
\hat Z_{10} \, \hat Z_{12} \, \hat Z_{14} \, \hat Z_{16}\\
\hat Z_7 \, \hat Z_8 \, \hat Z_{15} \, \hat Z_{16}\\
\hat Z_6 \, \hat Z_8 \, \hat Z_{14} \, \hat Z_{16}\\
\hat Z_4 \, \hat Z_8 \, \hat Z_{12} \, \hat Z_{16}\\
\hat Z_9 \, \hat Z_{10} \, \hat Z_{11} \, \hat Z_{12}\\
\hat Z_5 \, \hat Z_6 \, \hat Z_7 \, \hat Z_8\\
\hat Z_3 \, \hat Z_7 \, \hat Z_{11} \, \hat Z_{15}\\
\hat Z_2 \, \hat Z_6 \, \hat Z_{10} \, \hat Z_{14}\\
\hat Z_1 \, \hat Z_2 \, \hat Z_3 \, \hat Z_4
\end{array}\right\}  \nonumber \\
\hat S_X &=\hat X^{\otimes 16} = \hat X_1 \, \hat X_2 \, \hat X_3 \, ... \, \hat X_{16}.
\end{align}
Lastly, the tower of transversal logical entangling gates is built from the single-qubit phase gates
\begin{align}
\hat R_n = \begin{pmatrix}
1&0\\
0&e^{i\pi/2^n}
\end{pmatrix} \, ;
\end{align}
note that $\hat R_1 = \hat S$ and $\hat R_2 = \hat T$. Defining $\hat R_{n,q}$ for $\hat R_n$ acting on qubit $q$, the tower is
\begin{align}
\widehat{\overline{CZ}}_{12} &= \hat R^{\phantom{\dagger}}_{1,1} \, \hat R^{\dagger}_{1,5} \, \hat R^{\dagger}_{1,9} \, \hat R^{\phantom{\dagger}}_{1,13} \nonumber \\
\widehat{\overline{CZ}}_{13} &= \hat R^{\phantom{\dagger}}_{1,1} \, \hat R^{\dagger}_{1,3} \, \hat R^{\dagger}_{1,9} \, \hat R^{\phantom{\dagger}}_{1,11} \nonumber \\
\widehat{\overline{CZ}}_{14} &= \hat R^{\phantom{\dagger}}_{1,1} \, \hat R^{\dagger}_{1,2} \, \hat R^{\dagger}_{1,9} \, \hat R^{\phantom{\dagger}}_{1,10} \nonumber \\
\widehat{\overline{CZ}}_{23} &= \hat R^{\phantom{\dagger}}_{1,1} \, \hat R^{\dagger}_{1,3} \, \hat R^{\dagger}_{1,5} \, \hat R^{\phantom{\dagger}}_{1,7} \nonumber \\
\widehat{\overline{CZ}}_{24} &= \hat R^{\phantom{\dagger}}_{1,1} \, \hat R^{\dagger}_{1,2} \, \hat R^{\dagger}_{1,5} \, \hat R^{\phantom{\dagger}}_{1,6} \nonumber \\
\widehat{\overline{CZ}}_{34} &= \hat R^{\phantom{\dagger}}_{1,1} \, \hat R^{\dagger}_{1,2} \, \hat R^{\dagger}_{1,3} \, \hat R^{\phantom{\dagger}}_{1,4}
\end{align}
for two-qubit gates, 
\begin{align}
\widehat{\overline{CCZ}}_{123} &= \hat R^{\phantom{\dagger}}_{2,1} \, \hat R^{\dagger}_{2,3} \, \hat R^{\dagger}_{2,5} \, \hat R^{\phantom{\dagger}}_{2,7} \, \hat R^{\dagger}_{2,9} \, \hat R^{\phantom{\dagger}}_{2,11} \, \hat R^{\phantom{\dagger}}_{2,13} \, \hat R^{\dagger}_{2,15} \nonumber \\
\widehat{\overline{CCZ}}_{124} &= \hat R^{\phantom{\dagger}}_{2,1} \, \hat R^{\dagger}_{2,2} \, \hat R^{\dagger}_{2,5} \, \hat R^{\phantom{\dagger}}_{2,6} \, \hat R^{\dagger}_{2,9} \, \hat R^{\phantom{\dagger}}_{2,10} \, \hat R^{\phantom{\dagger}}_{2,13} \, \hat R^{\dagger}_{2,14} \nonumber \\
\widehat{\overline{CCZ}}_{134} &= \hat R^{\phantom{\dagger}}_{2,1} \, \hat R^{\dagger}_{2,2} \, \hat R^{\dagger}_{2,3} \, \hat R^{\phantom{\dagger}}_{2,4} \, \hat R^{\dagger}_{2,9} \, \hat R^{\phantom{\dagger}}_{2,10} \, \hat R^{\phantom{\dagger}}_{2,11} \, \hat R^{\dagger}_{2,12} \nonumber \\
\widehat{\overline{CCZ}}_{234} &= \hat R^{\phantom{\dagger}}_{2,1} \, \hat R^{\dagger}_{2,2} \, \hat R^{\dagger}_{2,3} \, \hat R^{\phantom{\dagger}}_{2,4} \, \hat R^{\dagger}_{2,5} \, \hat R^{\phantom{\dagger}}_{2,6} \, \hat R^{\phantom{\dagger}}_{2,7} \, \hat R^{\dagger}_{2,8}
\end{align}
for three-qubit gates and 
\begin{align}
\widehat{\overline{CCCZ}}_{1234} 
&= \hat R^{\phantom{\dagger}}_{3,1} \, \hat R^{\dagger}_{3,2} \, \hat R^{\dagger}_{3,3} \, \hat R^{\phantom{\dagger}}_{3,4} \, \hat R^{\dagger}_{3,5} \, \hat R^{\phantom{\dagger}}_{3,6} \, \hat R^{\phantom{\dagger}}_{3,7} \, \hat R^{\dagger}_{3,8} \nonumber \\
&\qquad\;\, 
\hat R^{\dagger}_{3,9} \, \hat R^{\phantom{\dagger}}_{3,10} \, 
\hat R^{\phantom{\dagger}}_{3,11} \, \hat R^{\dagger}_{3,12} \, \hat R^{\phantom{\dagger}}_{3,13} \, \hat R^{\dagger}_{3,14} 
\nonumber \\
&\qquad\;\, 
\hat R^{\dagger}_{3,15} \, \hat R^{\phantom{\dagger}}_{3,16}
\end{align}
for the four-qubit gate.

\subsection{Comparison of gate counts: Hypercubic and Iceberg}\label{app:gate}
\noindent
The two codes considered here are
 $ D=3: [[8,3,2]],\, D=4: [[16,4,2]]$.
Their minimum logical weights are asymmetric:
\begin{align}
  [[8,3,2]]&:\ (d_X,d_Y,d_Z)=(4,5,2),\\
  [[16,4,2]]&:\ (d_X,d_Y,d_Z)=(8,9,2).
\end{align}
Thus increasing the Hypercube dimension raises the protection against
$X$-type logical faults but leaves $d_Z=2$.
For both codes, $\widehat{\overline{Z}}_i$ has weight two.  Consequently, an electric
$\widehat{\overline{Z}}_i\widehat{\overline{Z}}_j$ rotation costs two CNOT gates when its sites share a block and
six when they occupy different blocks, exactly as for the Iceberg codes.
The same cross-block electric-pair count $\widehat{\overline{X}}$ of
Eq.~(\ref{eq:xcut}) therefore produces the same additional cost of $4\widehat{\overline{X}}$.
The difference appears in the kinetic operators.  
After optimization, the physical supports and CNOT costs of one logical $XX+YY$ bond,
as given in Table~\ref{tab:HCencode}.
\begin{table}[htbp]
  \centering
\setlength{\tabcolsep}{4pt}
\begin{tabular}{lcccc}
\toprule
code & location & $w(XX)$ & $w(YY)$ & CNOT per bond \\
\midrule
Iceberg & within a block & $2$ & $2$ & $2$ \\
Iceberg & across blocks & $4$ & $6$ & $16$ \\
$[[8,3,2]]$ & within a block & $4$ & $4$ & $6$ \\
$[[8,3,2]]$ & across blocks & $8$ & $10$ & $32$ \\
$[[16,4,2]]$ & within a block & $8$ & $8$ & $14$ \\
$[[16,4,2]]$ & across blocks & $16$ & $18$ & $64$ \\
\bottomrule
\end{tabular}
\caption{Gate counts for Hypercube encodings.}
\label{tab:HCencode}
\end{table}
The within-block $XX$ and $YY$ rotations commute and share one support.  
Hence, a weight-$w$ pair costs
$2(w-1)$ CNOT gates.   The cross-block rows contain two distinct
supports and their CNOT costs are
$2(8-1)+2(10-1)=32$ and $2(16-1)+2(18-1)=64$, respectively.
\bibliography{main}

\end{document}